\documentclass[11pt]{article}
\usepackage{jheppub}
\usepackage{rotating}
\usepackage{enumitem}
\usepackage[left=5cm,right=0cm,top=5cm,bottom=1cm,footskip=1cm]{geometry}
\usepackage{amsfonts}
\usepackage{amsmath}
\usepackage{slashed}
\usepackage{amssymb}
\usepackage{latexsym}
\usepackage{multirow}
\usepackage{hyperref}
\usepackage{mathtools}
\hypersetup{colorlinks=true,citecolor=red,linkcolor=magenta,urlcolor=blue}
\usepackage{relsize}
\usepackage{booktabs}
\usepackage{cleveref}
\usepackage{fancyhdr}
\usepackage{float}
\usepackage{graphicx}
\usepackage{microtype}
\usepackage{tabularx}
\usepackage[dvipsnames]{xcolor}
\usepackage{orcidlink}
\usepackage{wrapfig}
\usepackage{siunitx,adjustbox,bm}
\usepackage[bottom]{footmisc}
\usepackage{placeins}

\usepackage{pdflscape}
\usepackage{caption}
\usepackage{arydshln}
\usepackage{tikz}
\usetikzlibrary{arrows.meta,positioning}
\usepackage{tikz-feynman}
\usetikzlibrary{decorations.pathmorphing}
\usepackage[most]{tcolorbox}    
\usepackage{array}      
\newcommand{\Qchunk}[3]{%
\begin{tabular}[t]{@{}p{0.175\textwidth}r@{}}
\multicolumn{2}{@{}c@{}}{\textbf{\(#1\)}}\\
\midrule
#2
\midrule
\textbf{Total} & \(\mathbf{#3}\)
\end{tabular}%
}

\newcommand{\cA}{\mathcal{A}}
\newcommand{\hQ}{\widehat{Q}}
\newcommand{\delmu}{\overset{\leftrightarrow}{\partial}_\mu}

\definecolor{oceanblue}{HTML}{0077BE}

\def\blue{\color{blue}}

\def\red{\color{red}}
\def\black{\color{black}}

\def\purp{\color{purple}}
\def\magenta{\color{magenta}}
\def\Green{\color{OliveGreen}}

\def\black{\color{black}}

\newcommand{\nn}{\nonumber \\}

\title{Hilbert Series of Pseudoscalar Mesons: Operators and Sum Rules}
\author[a]{Joydeep Chakrabortty\orcidlink{0000-0001-8709-916X},}
\author[a]{Sabyasachi Chakraborty\orcidlink{0000-0001-5356-7607},} 
\author[a]{Siddhartha Karmakar\orcidlink{0009-0003-0609-9689}} 

\affiliation[a]{Indian Institute of Technology Kanpur, Kanpur-208016, Uttar Pradesh, India}

\emailAdd{joydeep@iitk.ac.in}
\emailAdd{sabyac@iitk.ac.in}
\emailAdd{siddharthak@iitk.ac.in}

\abstract{
We construct the complete non-redundant basis of operators built from pseudoscalar mesons using the Hilbert series method, subject to the global flavor symmetry $SU(3)_L \times SU(3)_R \times SU(2)_{L_H} \times SU(2)_{R_H}$, together with the underlying shift symmetry. The SM weak interactions are incorporated systematically by promoting the CKM matrix elements to spurion fields transforming under the global symmetry. The resulting operator basis provides a model-independent parametrization of two- and three-body weak decay amplitudes of heavy mesons, with the corresponding Wilson coefficients serving as reduced hadronic amplitudes. We demonstrate that all amplitude sum rules follow directly from the structure of the operator basis and propose an unfolding procedure that connects the effective operators to their possible ultraviolet origins. Our framework reproduces the known flavor relations while predicting additional identities arising from the symmetry-constrained operator basis. 
}

\begin{document}
\allowdisplaybreaks
\flushbottom
%%%%%%%%%%%%%%%%%%%%%%%%%%%%%%%%%%%%%%%%%%%%%%%%%%%%%
\maketitle
\allowdisplaybreaks
%%%%%%%%%%%%%%%%%%%

%%%%%%%%%%%%%%%%%%%%%%%%%%%%%%%%%
\section{Introduction}
%%%%%%%%%%%%%%%%%%%%%%%%%%%%%%%%%%%%%%%%%%%%%%%%%%%%%%%%%
Over the past decades, advances in lattice QCD algorithms and numerical resources have played a crucial role in determining hadronic observables such as mass, decay constants, form factors, and matrix elements, where non-perturbative strong interactions play the most dominant role. Lattice QCD provides a first principle approach for studying strongly coupled QCD and has achieved significant success in describing various hadronic phenomena. Nevertheless, lattice calculations can become inefficient or computationally demanding in  physical problems with multiple scales or processes with large recoils, etc. This is where effective theories are indispensable as they systematically exploit the hierarchy, i.e., separation of scales and the symmetries of the problem, thus providing a complementary framework. This also enables improved predictions in regions where lattice calculations are rather difficult to perform.

Prominent examples of such effective theories include Chiral Perturbation Theory ($\chi$PT)~\cite{Weinberg:1978kz,Gasser:1983yg,Gasser:1984gg}, Heavy Quark Effective Theory (HQET)~\cite{Isgur:1989vq,Eichten:1989zv, Isgur:1990yhj,Georgi:1990um,Grinstein:1990mj,Neubert:1993mb,Manohar:2000dt}, and Soft-Collinear Effective Theory (SCET)~\cite{Bauer:2000yr,Bauer:2000ew,Bauer:2001ct, Bauer:2001yt, Beneke:2002ph} etc. Chiral Perturbation Theory provides a systematic low-energy description of the interactions among the pseudo-Goldstone bosons of spontaneously broken chiral symmetry, namely the pions, kaons, and eta mesons. Organized as an expansion in external momenta and light-quark masses over the chiral symmetry breaking scale, $\chi$PT has become the standard framework for studying low-energy hadronic processes~\cite{Gasser:1984ux}, pion-pion~\cite{Colangelo:2001df} and pion-kaon~\cite{Bernard:1990kw} scattering, and the quark-mass and finite-volume effects of lattice QCD calculations~\cite{Gasser:1987zq,Colangelo:2005gd}.

Heavy Quark Effective Theory (HQET) exploits the hierarchy $m_Q \gg \Lambda_{\rm QCD}$ to describe hadrons containing a heavy quark~\cite{Isgur:1989vq,Georgi:1990um,Neubert:1993mb,Manohar:2000dt}. In the heavy-quark limit, i.e., $m_Q\to\infty$, the heavy quark behaves as a static color source, rendering the dynamics of the light degrees of freedom insensitive to the flavor and spin of the heavy quark. This gives rise to heavy-quark flavor and spin symmetries, which results in non-trivial relations among decay constants, form factors, and transition amplitudes. Corrections to these symmetry relations are systematically incorporated through an expansion in powers of $1/m_Q$. HQET has found widespread applications in the study of heavy-flavor phenomenology, particularly in semileptonic decays such as $\bar{B}\to D^{(*)}\ell\bar{\nu}$ and $B\to\pi\ell\nu$~\cite{Burdman:1993es,Charles:1998dr}, rare decays including $B\to K\ell^+\ell^-$~\cite{Grinstein:2004vb}, and heavy-meson spectroscopy~\cite{Eichten:1989zv}. The heavy-quark symmetry predicts that, in the infinite-mass limit, all form factors governing $B\to D^{(*)}$ transitions are determined by a single universal Isgur–Wise function~\cite{Isgur:1989vq,Isgur:1990yhj}. These features have made HQET an indispensable framework for precision determinations of the Cabibbo–Kobayashi–Maskawa (CKM) matrix elements $|V_{cb}|$~\cite{Caprini:1997mu}, as well as for interpreting lattice QCD calculations of heavy-hadron observables.

Finally, SCET is formulated to describe processes involving highly energetic particles moving close to the light cone, where several widely separated momentum scales coexist~\cite{Bauer:2000yr,Bauer:2000ew,Bauer:2001ct,Beneke:2002ph}. By introducing distinct hard, hard-collinear, collinear, and soft degrees of freedom together with a manifest power-counting scheme, SCET systematically factorizes short-distance perturbative effects from long-distance non-perturbative hadronic dynamics, usually included in wave functions. This factorization enables the resummation of large logarithms through renormalization-group evolution, thereby improving the convergence of perturbative calculations and enhancing the precision of theoretical predictions. SCET has found widespread applications in exclusive and inclusive $B$-meson decays, including $B\to K^{(*)}\ell^+\ell^-$~\cite{Beneke:2001at}, $B\to K^{(*)}\gamma$~\cite{Beneke:2004dp}, $B\to\pi\ell\nu$~\cite{Bauer:2002aj}, and $B\to X_s\gamma$~\cite{Bauer:2000yr}, as well as in jet production and event-shape observables at high-energy colliders~\cite{Becher:2008cf}. SCET has also been used to probe spectator and soft form factor contributions in the presence of light new physics such as axions~\cite{Bisht:2026vcw}. More generally, the framework provides a systematic expansion in powers of the small ratio of momentum scales ($\lambda$), allowing both leading- and subleading-power contributions to be incorporated in a controlled manner. As a result, SCET has become an indispensable tool for precision flavor physics and collider phenomenology.

Each of these effective theories is tailored to a specific kinematic regime. Their explicit mode separation provides a manifest power-counting scheme, allowing operators and observables to be organized systematically and distinguishing them from generic Wilsonian EFTs. However, these same features make it challenging to construct a unified description spanning multiple EFTs. A notable step in this direction is heavy meson chiral perturbation theory (HM$\chi$PT)~\cite{Wise:1992hn,Burdman:1992gh,Yan:1992gz, Casalbuoni:1996pg}, which combines the chiral and heavy-quark symmetries to describe the interactions of heavy-light mesons with soft Goldstone bosons, and whose multiplet structure underlies the field content adopted below. Even there, however, the weak interactions must be introduced process by process through matching. Motivated by these considerations, a general framework was recently developed~\cite{Chakraborty:2026vdz}, in which effective operators for charged-current processes are constructed directly from the underlying symmetries and low-energy degrees of freedom. A key ingredient of this formalism is the systematic incorporation of weak interactions by promoting the CKM matrix elements to spurion fields with prescribed transformation properties under the global symmetry. In the heavy-light sector, this provides a unified, symmetry-based description that avoids the ambiguities of combining different power-counting schemes.

Once the relevant symmetries and field content are specified, the resulting effective theory generally admits a large number of invariant operators. A systematic construction therefore requires an efficient method to enumerate all independent operators while eliminating redundancies arising from symmetry identities, integration by parts, and equations of motion. The Hilbert series provides precisely such a framework, yielding a complete and non-redundant operator basis consistent with the underlying symmetries. In this work, we employ the Hilbert series (HS) method to construct composite operators built from pseudoscalar mesons, following the general strategy developed for effective field theories~\cite{Hanany:2010vu,  Lehman:2015via,Henning:2015alf,Henning:2015daa,Graf:2020yxt,Banerjee:2020bym, Wang:2021wdq, Chakrabortty:2026swu}. Rather than organizing the basis by canonical mass dimension, as is customary in SMEFT, we classify operators containing four meson fields and two derivatives. This defines the effective theory considered in the present work, where the relevant expansion is determined by the mesonic building blocks and derivative structure rather than by the canonical dimensions of the constituent fields.

The fundamental degrees of freedom of our effective theory consist of the light pseudo-Goldstone bosons together with the heavy-heavy and heavy-light mesons. As in $\chi$PT, the octet of light pseudoscalar mesons is encoded in the nonlinear field $U_\pi$~\cite{Coleman:1969sm, Callan:1969sn}, which realizes the spontaneously broken chiral symmetry $SU(3)_L\times SU(3)_R$ in a manifestly invariant manner within a vector-like multiplet $\Sigma$. The heavy-heavy pseudoscalar mesons are collected in a vector-like multiplet $\Sigma_H$ under $SU(2)_{L_H}\times SU(2)_{R_H}$, while the heavy-light pseudoscalar mesons are assembled in a doublet representation $\Phi(\Phi_c)$ of  $SU(2)_{L_H}(SU(2)_{R_H})$. This arrangement makes the heavy- and light-flavor symmetry structures of the theory explicit and provides a convenient basis for constructing the most general effective Lagrangian consistent with the underlying symmetries. We also take spurion representations of $\widehat{Q}_{ij}\; \forall (i,j)\in \{Q,q \}$, charged under the above mentioned global flavor symmetries. This construction allows us to encapsulate all the viable interactions among light, heavy-light, and heavy-heavy mesons  within a single unified framework. 

The paper is organized as follows. Sec.~\ref{sec:HS-OP} presents the Hilbert series construction of the flavor-invariant heavy-light meson operator basis. Secs.~\ref{sec:3} and \ref{sec:4} apply the formalism to the two-body decays $B\to P_1P_2$ and $B\to DP$, where $P$ refers to a light pseudoscalar meson and $D$ is a heavy pseudoscalar meson with a charm quark. We derive the corresponding amplitudes and flavor sum rules for both the cases. Sec.~\ref{sec:3body} discusses the extension to three-body decays, while Secs.~\ref{sec:unfolding} and \ref{sec:comparison} relate the EFT to the underlying Standard Model description and compare our approach with existing methods. We conclude in Sec.~\ref{sec:conclusion}. Additional operator tables and explicit expansions are collected in the appendices \ref{app:oplist} and \ref{app:explcit_amp}, respectively.
%%%%%%%%%%%%%%%%%%%%%%%%%%%%%%%%%%%%%%%%%%%%%%%%%%%%%%%%%%%%%%%
\section{Hilbert Series and Counting of Effective Operators}\label{sec:HS-OP}

\begin{table}[H]
    \centering
    \renewcommand{\arraystretch}{1.2}
    \begin{minipage}{0.48\textwidth}
        \centering
        \resizebox{.99\textwidth}{!}{
        \begin{tabular}{|c|c|c|c|c|}
            \hline
            Field & $SU(3)_L$ & $SU(3)_R$ & $SU(2)_{L_H}$ & $SU(2)_{R_H}$ \\
            \hline
             $\Phi$ & 1 & $\bar 3$ & 2 & 1 \\
             $\Phi^\dagger$ & 1 & $3$ & 2 & 1 \\
             $\Phi_c$ & 3 & 1 & 1 & 2 \\
             $\Phi_c^\dagger$ & $\bar 3$ & 1 & 1 & 2 \\
             \hline
             $\Sigma$ & 3 & $\bar 3$ & 1 & 1 \\
             $\Sigma^\dagger$ & $\bar 3$ & $3$ & 1 & 1 \\
             \hline
             $\Sigma_H$ & 1 & 1 & 2 & 2 \\
             $\Sigma_H^\dagger$ & 1 & 1 & 2 & 2 \\
            \hline
        \end{tabular}
        }
        \caption*{Meson field content.}
    \end{minipage}
    \hfill
    \begin{minipage}{0.48\textwidth}
        \centering
        \renewcommand{\arraystretch}{1.2}
        \resizebox{.99\textwidth}{!}{
        \begin{tabular}{|c|c|c|c|c|}
            \hline
            Spurion & $SU(3)_L$ & $SU(3)_R$ & $SU(2)_{L_H}$ & $SU(2)_{R_H}$ \\
            \hline
             $\hQ_{QQ}$ & 1  & 1 & 3 & 1 \\
             $\hQ_{QQ}^\dagger$ & 1  & 1 & 3 & 1 \\
             \hline
             $\hQ_{qQ}$ & 3  & 1 & 2 & 1 \\
             $\hQ_{qQ}^\dagger$ & $\bar 3$  & 1 & 2 & 1 \\
             \hline
             $\hQ_{Qq}$ & $\bar 3$  & 1 & 2 & 1 \\
             $\hQ_{Qq}^\dagger$ & $3$  & 1 & 2 & 1 \\
             \hline
            $\hQ_{qq}$ & $8$  & 1 & 1 & 1 \\
            $\hQ_{qq}^\dagger$ & $8$  & 1 & 1 & 1 \\
            \hline
        \end{tabular}
        }
        \caption*{Spurion for CKM insertion.}
    \end{minipage}
    \caption{Field content and representations.}
    \label{tab:meson-qn} 
\end{table}
%%%%%%%%%%%%%%%%%%%%%%%%%%%%%%%%%
\subsection{A Review of Hilbert Series Method}\label{subsec:HS}

The Hilbert Series (HS) is a symmetry-based technique for systematically constructing the complete set of group-invariant polynomials formed from spurion fields transforming in representations \((\mathbb{R})\) of a symmetry group. In the context of particle physics, the interactions among the relevant degrees of freedom at a given energy scale are described by an effective Lagrangian. The construction of such a Lagrangian is governed by two key ingredients: the symmetries of the theory and the on-shell degrees of freedom that remain dynamical at that scale. The effective Lagrangian is then organized as a sum of independent local operators that are invariant under the underlying symmetry group and are built from these degrees of freedom.   

HS, defined as the integral of plethystics ($\textbf{PE} (\Phi, \mathbb{R})$) in the group-space $\mathcal{G}_i$ ~\cite{Henning:2015daa, Hanany:2010vu, Lehman:2015via, Feng:2007ur}
\begin{equation}
    \mathbb{H} (\Phi) = \prod_{i=1}^{N} \int_{\mathcal{G}_i} d\mu_i \; \textbf{PE} (\Phi, \mathbb{R})\;,
\end{equation}
allows us to compute such invariant local independent operators up to any mass dimensions. Here, $d\mu_i $ depicts the Haar-measure for group $\mathcal{G}_i$, $\Phi$ is the representative spurion of the representation $\mathbb{R}$. In this paper, we work with only scalar degrees of freedom, i.e., bosonic spurions $(\varphi)$, and the respective plethystic is given as~~\cite{Henning:2015daa, Hanany:2010vu, Lehman:2015via, Feng:2007ur}
\begin{align}
    \textbf{PE} (\varphi, \mathbb{R}) &= {\rm exp}\Big[ \sum_{n=1}^{\infty}   \frac{\varphi^n\; \chi[z_j^n, \mathbb{R}]}{n} \Big]\;,
\end{align}
where $\chi[z_j^n, \mathbb{R}]$ is the Weyl character of the representation $\mathbb{R}$ of the group $\mathcal{G}$. We identify the maximal torus corresponding to the group $G$ of rank $k$, and $\chi$ is written in terms of the maximal torus coordinates $z_j, \forall j\in \{1,\cdots,k\}$. 

We work with the following internal global symmetry $SU(3)_L\times SU(3)_R \times SU(2)_{L_H} \times SU(2)_{R_H}\equiv \mathcal{G}_{3322}$, and the fields are light mesons $\Phi, \Sigma, \Sigma_H$ that transform under these symmetries and also being Goldstone bosons, they respect the shift symmetry. We provide the CKM-matrix elements $\hQ_{ij}, \; \forall i,j \in \{Q, q \}$ the same status as the scalar fields, but unlike the mesons, they are not dynamical, i.e., they do not possess any equation of motion. The quantum numbers of these fields under the mentioned symmetries are summarized in Table~\ref{tab:meson-qn}. The Weyl characters for these fields can be written as a function of the coordinates $(z_i, i=\{1,\cdots, N-1\})$ of a maximal torus $\mathbb{T}^{N-1} \equiv \underbrace{ U(1)\otimes \cdots \otimes U(1)}_{N-1} $ corresponding to the internal symmetry group $SU(N)$~\cite{brocker2003representations}. $\mathbb{T}^{N-1}$ is depicted as
$\text{diag}(e^{i\theta_1,\;e^i(\theta_2-\theta_1),\; \cdots,\; e^{i\theta_{N-1}}})$, with $z_i = e^{i \theta_i}$.

The character formula for the irreducible representation $\mathbb{R}$ of $SU(N)$ is given as \cite{Balantekin:2001id, Plymen:1976su, koike1987littlewood, littlewood1977theory, rossmann2006lie}
\begin{align}
    \chi[\mathbb{R}] = \frac{|\epsilon^{r_1},\cdots, \epsilon^{r_{N-1}},1|}{|\epsilon^{N-1},\cdots, \epsilon^{1},1|}\;,
\end{align}
with $\prod_{i=1}^N \epsilon_i=1$, and $\epsilon_i = \epsilon_i(z_1, \cdots, z_{N-1})$. The relation among $\epsilon_i$ and $z_j$ can be computed employing the weight tree of the lowest dimensional fundamental representation, see Ref.~\cite{foster2016lie} which is unique for a given group and is independent of the choice of representations of the group, for example, for $SU(2): \epsilon_1=z_1, \epsilon_2=1/z_1, $ and $SU(3): \epsilon_1=z_2, \epsilon_2=z_3/z_2, \epsilon_3= 1/z_3$.

Here, we note down the characters of the relevant representations~\cite{slansky1981group, Yamatsu:2015npn}, see Table~\ref{tab:meson-qn}, as
\begin{align}
 \chi_{SU(2)}[2] &= z_1+\frac{1}{z_1};\;\; \chi_{SU(2)}[3] = z_1^2+\frac{1}{z_1^2}+1\;, \nonumber \\
    \chi_{SU(3)}[3] &= z_2+\frac{z_3}{z_2}+\frac{1}{z_3}; \;\;\;  \chi_{SU(3)}[\overline{3}] = z_3+\frac{z_2}{z_3}+\frac{1}{z_2}\;, \nonumber \\
    \chi_{SU(3)}[8] &= z_2 z_3+\frac{1}{z_2 z_3}+ \frac{z_2^2}{z_3} + \frac{z_3^2}{z_2}+ \frac{z_2}{z_3^2} + \frac{z_3}{z_2^2}+2\;.
\end{align}

The Haar measures for the $SU(2)$ and $SU(3)$ groups are given as
\begin{align}
    d\mu_{SU(2)} &\equiv  \frac{1}{2 (2\pi i )} 
    \frac{d{z_1}}{z_1} \Big[  2 - z_1^2 -\frac{1}{z_1^2}\Big]\;, \nonumber\\
      d\mu_{SU(3)} &\equiv  \frac{1}{3! (2\pi i )^2} \frac{d{z_2}}{z_2} \frac{d{z_3}}{z_3} \Big[ 2-z_2 z_3 -\frac{1}{z_2 z_3}\Big] \Big[ 2- \frac{z_2}{z_3^2} -\frac{z_3^2}{z_2}\Big]  \Big[ 2- \frac{z_2^2}{z_3} -\frac{z_3}{z_2^2}\Big]\;.
      \label{eq:Haar-SU2-SU3}
\end{align}

We work with a connected compact internal symmetry group, e.g., $SU(N)$.  The space-time is described by non-compact Lorentz symmetry $SO(3,1) \simeq SO(4, \mathbb{C})$ which does not have unitary finite dimensional representations. Thus, for the sake of HS construction, we work in Weyl representations, i.e., the space-time symmetry is realized in terms of compact $SU(2)_L\times SU(2)_R$. We assume that the unimodular torus coordinates corresponding to the $SU(2)_L$ and  $SU(2)_R$ groups are $\alpha$ and $\beta$, respectively.  Within this framework, we define the characters for the scalar fields as follows
\begin{equation}
    \chi_s^4 \Big[ \mathcal{D},\alpha, \beta\Big] = \mathcal{D}\; (1-\mathcal{D}^2)\; \mathbb{P}^{4}\;,
\end{equation}
where $\mathbb{P}^{4}=\Big[ \Big(1- \mathcal{D} \alpha \beta \Big) \Big(1- \frac{\mathcal{D}}{ \alpha \beta}\Big)\Big(1- \frac{\mathcal{D} \alpha}{ \beta} \Big) \Big(1- \frac{\mathcal{D} \beta}{ \alpha}\Big) \Big]^{-1}$ is the momentum generating function that takes care of the constraints due to integration by parts (IBP) \cite{Henning:2015alf}. As the $\widehat{Q}$'s are non-dynamical scalars, so their character under space-time symmetry is identity, i.e., 1. The Haar measure for Lorentz symmetry (in Weyl representation) reads~\cite{Henning:2015alf, Hanany:2014dia}
\begin{eqnarray}
d\mu_{LG} &=& \Big[ \mathbb{P}^{4}\Big]^{-1} d\mu_{SU(2)_L} d\mu_{SU(2)_R}\;, \nonumber \\
&=& \Big[ \mathbb{P}^{4}\Big]^{-1} \Big[ \frac{1}{2(2\pi i)}\Big]^2 \frac{d\alpha}{\alpha}\frac{d\beta}{\beta}\; (2-\alpha^2 -\alpha^{-2}) (2-\beta^2 -\beta^{-2})\;.  
\end{eqnarray}

Now, we are ready to write down the characters of the relevant representations as follows:
\begin{align}
	\Phi&\rightarrow \chi(\Phi) = \chi^{(4)}_{s}\left(D^r,\alpha^r,\beta^r\right) \chi_{(SU(3)_{R})_{\bar 3}}(z^{\prime r}_2,z^{\prime r}_3)\;\chi_{(SU(2)_{L_H})_2}(z_1^{r})
	\;,\nn
	\Phi_c&\rightarrow \chi(\Phi_c) = \; \chi^{(4)}_{s}\left(D^r,\alpha^r,\beta^r\right) \chi_{(SU(3)_{L})_{3}}(z^{r}_2,z^{r}_3)\;\chi_{(SU(2)_{R_H})_{2}}(z_1^{\prime r})
	\;,\nn
	\Sigma&\rightarrow \chi(\Sigma) = \chi^{(4)}_{s}\left(D^r,\alpha^r,\beta^r\right) \chi_{(SU(3)_{L})_{3}}(z^{r}_2,z^{r}_3)\;\chi_{(SU(3)_{R})_{\bar 3}}(z^{\prime r}_2,z^{\prime r}_3)
	\,,\nn
	\Sigma_H&\rightarrow \chi(\Sigma_H) =  \chi^{(4)}_{s}\left(D^r,\alpha^r,\beta^r\right) \chi_{(SU(2)_{L_H})_2}(z_1^{r})\;\chi_{(SU(2)_{R_H})_{2}}(z_1^{\prime r})
	\,,\nn
	\hQ_{QQ}&\rightarrow \chi(\hQ_{QQ}) =  \,  \chi_{(SU(2)_{L_H})_3}(z_1^{r})
	\,,\nn
	\hQ_{qQ}&\rightarrow \chi(\hQ_{qQ}) = \,  \chi_{(SU(3)_{L})_{3}}(z^{r}_2,z^{r}_3)\;\chi_{(SU(2)_{L_H})_{2}}(z_1^{r})
	\,,\nn
	\hQ_{Qq}&\rightarrow \chi(\hQ_{Qq}) = \,  \chi_{(SU(3)_{L})_{\bar 3}}(z^{r}_2,z^{r}_3)\;\chi_{(SU(2)_{L_H})_2}(z_1^{r})
	\,,\nn
	\hQ_{qq}&\rightarrow \chi(\hQ_{qq}) = \,  \chi_{(SU(3)_{L})_{8}}(z^{r}_2,z^{r}_3)\;.
	\,
\end{align}

Thus, the complete HS can be expressed as 
\begin{align}
	\mathbb{H} (\Phi) &= \int_{\mathcal{LG}}  d\mu_{SU(2)}\times d\mu_{SU(2)}\frac{1}{P^{(4)}({\cal D}, \alpha, \beta)} \left[ \int_{\mathcal{G}} d\mu_{SU(3)_L}  d\mu_{SU(3)_R}  d\mu_{SU(2)_{L_H}}  d\mu_{SU(2)_{R_H}} \right. \nn
	&\left.\exp\Big[ \sum_{r=1}^{\infty}   \frac{\Phi^r\; \chi(\Phi)}{r}\Big]\,\exp\Big[ \sum_{r=1}^{\infty}   \frac{\Phi_c^r\; \chi(\Phi_c)}{r}\Big] \,\exp\Big[ \sum_{r=1}^{\infty}   \frac{\Sigma^r\; \chi(\Sigma)}{r}\Big] \, \exp\Big[ \sum_{r=1}^{\infty}   \frac{\Sigma_H^r\; \chi(\Sigma_H)}{r}\Big] \right.\nn
	&\left. \exp\Big[ \sum_{r=1}^{\infty}   \frac{\hQ_{QQ}^r\; \chi(\hQ_{QQ})}{r}\Big]\,\exp\Big[ \sum_{r=1}^{\infty}   \frac{\hQ_{qQ}^r\; \chi(\hQ_{qQ})}{r}\Big]\,\exp\Big[ \sum_{r=1}^{\infty}   \frac{\hQ_{Qq}^r\; \chi(\hQ_{Qq})}{r}\Big]\exp\Big[ \sum_{r=1}^{\infty}   \frac{\hQ_{qq}^r\; \chi(\hQ_{qq})}{r}\Big]\right]\;.\label{HSfinal}
\end{align}
This plethystic exponents contain all possible combinations of fields. We use the character orthogonality theorem
\begin{equation}
    \int_\mathcal{G}\; d\mu_{\mathcal{G}}\;\; \chi_i[\mathbb{R}_i] \;\;\chi_j[\mathbb{R}_j] = \delta_{ij}\;,
\end{equation}
to project invariant operators of different dimensions. We are interested in operators $(D6)$ that consist of four meson fields and two derivatives and are invariant under internal symmetry $SU(3)_L\times SU(3)_R \times SU(2)_{L_H} \times SU(2)_{R_H}$. In the next section, we tabulate all such relevant operators. 

\subsection{Haar projection and invariant polynomials: an example}
Here, we demonstrate how the Hilbert series selects invariant polynomials for an example scenario having scalar fields transform under global symmetry $SU(2)\times SU(3)$
\begin{align}
	\Phi \sim (2,\bar 3)\;,
	\qquad
	\Phi_d \sim (2,3)\;.
\end{align}
The plethystic exponential then takes the form
\begin{align}
	\mathrm{PE}
	&=
	\exp\left[
	\sum_{r=1}^{\infty}
	\frac{\Phi^r}{r}
	\chi_\Phi(z_1^r,z_2^r,z_3^r)
	\right]
	\exp\left[
	\sum_{r=1}^{\infty}
	\frac{\Phi_d^r}{r}
	\chi_{\Phi_d}(z_1^r,z_2^r,z_3^r)
	\right]\;,
\end{align}
where $[\{z_1\},\{z_2,z_3\}]$ are the maximal torus coordinates corresponding to the $[\{SU(2)\},\{ SU(3)\}]$ groups, respectively.
The plethystic exponential can be written as a Laurent series in these variables as
\begin{align}
	\mathrm{PE}
	=
	\sum_{m,n}
	\sum_{a,b,c}
	N_{m,n}^{a,b,c}\,
	\Phi^m\Phi_d^n\,
	z_1^a z_2^b z_3^c\;,
\end{align}
with $N_{m,n}^{a,b,c}$ containing the corresponding numerical factors after the expansion. The powers of $z_1,z_2,z_3$ label the weights of the corresponding tensor structures under the maximal torus, i.e., representations under $SU(2)\times SU(3)$. Thus, a specific choice of $\{a,b,c\}$ corresponds to the symmetry invariant terms where others do not. Character functions $\chi$ are orthonormal , and employing this property we can project out invariants, i.e., singlets. This is equivalent to finding the residues of the simple poles while performing the group-space integral using the Haar measure: 
\begin{align}
	\mathrm{HS}(\Phi,\Phi_d)
	&=
	\oint d\mu_{SU(2)}(z_1)
	\oint d\mu_{SU(3)}(z_2,z_3)\,
	\mathrm{PE}\;, \nonumber \\
	&=
	\oint \frac{dz_1}{2\pi i} \oint \frac{dz_2}{2\pi i} \oint \frac{dz_3}{2\pi i}\;
	T(z_1, z_2, z_3)\;,
\end{align}
where the complex integrand
\begin{align}
    T(z_1, z_2, z_3) \equiv \mathrm{Haar}_{SU(2)}(z_1)\;
	\mathrm{Haar}_{SU(3)}(z_2,z_3)\;
	\mathrm{PE}\;.
\end{align}
As the Haar measures are of the form $1/z_1$, $1/z_2$, $1/z_3$, see eq.~\eqref{eq:Haar-SU2-SU3},  the character orthonormality extracts the residue proportional to simple poles in $(z_1, z_2, z
_3)$ as
\begin{align}
	\frac{1}{z_1z_2z_3}\;.
\end{align}
After expanding the plethystics, $T(z_1, z_2, z_3)$ takes the following form
\begin{align}
    T(z_1, z_2, z_3)& = \frac{1}{z_1 z_2 z_3}(
	\Phi\Phi_d
	+
	2\Phi^2\Phi_d^2
	+
	2\Phi^3\Phi_d^3
	+
	3\Phi^4\Phi_d^4
	+\cdots )\\
    &+ \frac{z_3}{z_1 z_2} \left(-\frac{\Phi^2}{3}
	-\frac{\Phi^4\Phi_d^2}{3}
	+\frac{\Phi_d^4}{3} + \cdots\right) 
     + {z_1 z_2 z_3} \left(
	-\frac{\Phi^2\Phi_d^2}{12}
	-\frac{\Phi^4 \Phi_d^4}{6} + \cdots\right) 
     + \cdots ~. \nonumber
\end{align}
Employing the contour integrations corresponding to these three terms, we note
\begin{equation}
\begin{aligned}
   &\frac{1}{(2\pi i)^3} \oint dz_1 \oint dz_2 \oint dz_3 \frac{1}{z_1 z_2 z_3} = 1~,\\
%\end{align}
%\begin{align}
  &\frac{1}{(2\pi i)^3}  \oint dz_1 \oint dz_2 \oint dz_3 \frac{z_3}{z_1 z_2} = 0~,\\
%\end{align}
%and
%\begin{align}
 &\frac{1}{(2\pi i)^3}   \oint dz_1 \oint dz_2 \oint dz_3 \, (z_1 z_2 z_3) = 0~.
\end{aligned}
\end{equation}
Now it is evident how the residue of simple poles, i.e., $(\Phi\Phi_d + 2\Phi^2\Phi_d^2 + 2\Phi^3\Phi_d^3+ 3\Phi^4\Phi_d^4 + \cdots)$ are related to the polynomials invariant under $SU(2)\times SU(3)$. We employ this algorithm to construct the HS and compute the invariant effective operators for our scenario. The Hilbert series computation is performed using \texttt{GrIP}, an automated package for group-integration-based invariant counting~\cite{Banerjee:2020bym}.

%%%%%%%%%%%%%%%%%%%%%%%%%%%%%%%%
\subsection{Class of operators}
%%%%%%%%%%%%%%%%%%%%%%%%%%%%%%%%
Having specified the transformation properties of the meson fields and weak spurions, we now extract the operator classes relevant for the present analysis from the Hilbert series in eq.~\eqref{HSfinal}. We focus on the section with two weak-spurion insertions, four meson fields, and two derivatives $\hQ_i \hQ_j^\dagger, D^2, \mathcal{M}_1\mathcal{M}_2\mathcal{M}_3\mathcal{M}_4$, where $\mathcal{M}_k\in\{{\Phi,\Phi^\dagger,\Phi_c,\Phi_c^\dagger,\Sigma,\Sigma^\dagger,\Sigma_H,\Sigma_H^\dagger}\}$. The Hilbert series counts the independent invariant contractions after accounting for internal symmetry relations, integration by parts, and equations of motion for the dynamical fields. The weak spurions are treated as non-dynamical background fields that encode the flavor and CKM structure of the underlying weak interaction. As a result, our operators would include all the charged and neutral current processes mediated by weak interaction. The resulting operator classes are summarized in Table~\ref{tab:full_table}. For each weak-spurion pair, the table lists the allowed meson field content together with the corresponding number of independent operators. At the end of each block, the complete Hilbert series count for that spurion sector at $\mathcal{O}(D^2)$ is given. Although Hermitian conjugate blocks are not independent in the Lagrangian, we retain them separately for convenience when projecting onto specific decay channels. Each block corresponds to a specific weak-spurion combination, while the rows enumerate the allowed meson field content together with the number of independent invariant contractions after accounting for the symmetry constraints. The table thus provides a complete, process-independent classification of all local operators permitted by the flavor symmetries at this order. For any given decay process, the relevant effective operator basis is obtained by selecting the appropriate operator classes from this catalog and projecting them onto the desired external meson states. In the following sections, we explicitly carry out this projection for charmless two-body decays $B\to P_1P_2$.

\begin{table}[h]
\centering
\renewcommand{\arraystretch}{1.08}
\setlength{\tabcolsep}{2pt}
\tiny
\resizebox{\textwidth}{!}{%
\begin{tabular}{@{}c@{\hspace{0.018\textwidth}}c@{\hspace{0.018\textwidth}}c@{\hspace{0.018\textwidth}}c@{}}
\toprule
%%%%%%%%%%%%%%%%%%%%%%%%%%%%%%%%%%%%%%%%%%%%%%%%%%%%%%%%%%%%
% Row 1: self hermitian
%%%%%%%%%%%%%%%%%%%%%%%%%%%%%%%%%%%%%%%%%%%%%%%%%%%%%%%%%%%%
\Qchunk{\hQ_{qQ}\hQ_{qQ}^{\dagger}\,D^2}
{
\((\Sigma^\dagger)^2\,\Sigma^2\) & \(2\)\\
\(\Sigma^\dagger\,\Sigma\,\Sigma_H^\dagger\,\Sigma_H\) & \(4\)\\
\(\Phi\,\Sigma\,\Phi^\dagger\,\Sigma^\dagger\) & \(8\)\\
\(\Phi_c\,\Sigma\,\Phi_c^\dagger\,\Sigma^\dagger\) & \(6\)\\
\(\Phi_c\,\Sigma_H\,\Phi_c^\dagger\,\Sigma_H^\dagger\) & \(8\)\\
\(\Phi\,\Phi_c\,\Phi^\dagger\,\Phi_c^\dagger\) & \(4\)\\
\(\Phi_c^2\,(\Phi_c^\dagger)^2\) & \(4\)\\
\(\Phi\,\Phi_c\,\Sigma^\dagger\,\Sigma_H^\dagger\) & \(4\)\\
\(\Phi\,\Sigma_H\,\Phi^\dagger\,\Sigma_H^\dagger\) & \(5\)\\
}
{45} 
&
\Qchunk{\hQ_{Qq}\hQ_{Qq}^{\dagger}\,D^2}
{
\((\Sigma^\dagger)^2\,\Sigma^2\) & \(2\)\\
\(\Sigma^\dagger\,\Sigma\,\Sigma_H^\dagger\,\Sigma_H\) & \(4\)\\
\(\Phi\,\Sigma\,\Phi^\dagger\,\Sigma^\dagger\) & \(8\)\\
\(\Phi_c\,\Sigma\,\Phi_c^\dagger\,\Sigma^\dagger\) & \(6\)\\
\(\Phi_c\,\Sigma_H\,\Phi_c^\dagger\,\Sigma_H^\dagger\) & \(8\)\\
\(\Phi\,\Phi_c\,\Phi^\dagger\,\Phi_c^\dagger\) & \(4\)\\
\(\Phi_c^2\,(\Phi_c^\dagger)^2\) & \(4\)\\
\(\Phi^\dagger\,\Phi_c^\dagger\,\Sigma\,\Sigma_H\) & \(4\)\\
\(\Phi\,\Sigma_H\,\Phi^\dagger\,\Sigma_H^\dagger\) & \(5\)\\
}
{45}
&
\Qchunk{\hQ_{QQ}\hQ_{QQ}^{\dagger}\,D^2}
{
\((\Sigma_H^\dagger)^2\,\Sigma_H^2\) & \(4\)\\
\(\Sigma_H^\dagger\,\Sigma_H\,\Sigma^\dagger\,\Sigma\) & \(2\)\\
\(\Sigma_H^\dagger\,\Sigma_H\,\Phi^\dagger\,\Phi\) & \(4\)\\
\(\Sigma_H^\dagger\,\Sigma_H\,\Phi_c^\dagger\,\Phi_c\) & \(4\)\\
\((\Phi^\dagger)^2\,\Phi^2\) & \(4\)\\
\(\Phi^\dagger\,\Phi\,\Sigma^\dagger\,\Sigma\) & \(4\)\\
\(\Phi^\dagger\,\Phi\,\Phi_c^\dagger\,\Phi_c\) & \(2\)\\
\(\Sigma_H^\dagger\,\Phi\,\Sigma^\dagger\,\Phi_c\) & \(2\)\\
\(\Sigma_H\,\Sigma\,\Phi^\dagger\,\Phi_c^\dagger\) & \(2\)\\
}
{28}
&
\Qchunk{\hQ_{qq}\hQ_{qq}^{\dagger}\,D^2}
{
\(\Sigma^2\,(\Sigma^\dagger)^2\) & \(4\)\\
\(\Sigma\,\Sigma^\dagger\,\Phi\,\Phi^\dagger\) & \(6\)\\
\(\Sigma\,\Sigma^\dagger\,\Phi_c\,\Phi_c^\dagger\) & \(13\)\\
\(\Phi\,\Phi^\dagger\,\Phi_c\,\Phi_c^\dagger\) & \(3\)\\
\(\Phi_c^2\,(\Phi_c^\dagger)^2 \) & \(4\)\\
\(\Phi_c\,\Phi_c^\dagger\,\Sigma_H\,\Sigma_H^\dagger\) & \(6\)\\
\(\Sigma^\dagger\,\Phi_c\,\Sigma_H^\dagger\,\Phi\) & \(3\)\\
}
{39}
\\
%%%%%%%%%%%%%%%%%%%%%%%%%%%
% Row 2
%%%%%%%%%%%%%%%%%%%%%%%%%%%
\Qchunk{\hQ_{qq}\hQ_{qQ}^{\dagger}\,D^2}
{
\(\Sigma_H\,\Phi_c^\dagger\,\Sigma\,\Sigma^\dagger\) & \(4\)\\
\(\Sigma_H^2\,\Phi_c^\dagger\,\Sigma_H^\dagger\) & \(2\)\\
\(\Sigma_H\,\Phi_c^\dagger\,\Phi\,\Phi^\dagger\) & \(2\)\\
\(\Sigma_H\,(\Phi_c^\dagger)^2\,\Phi_c\) & \(4\)\\
\(\Sigma^2\,\Sigma^\dagger\,\Phi^\dagger\) & \(4\)\\
\(\Phi\,\Sigma^\dagger\,\Sigma_H\,\Sigma_H^\dagger\) & \(2\)\\
\(\Phi^2\,\Sigma^\dagger\,\Phi^\dagger\) & \(2\)\\
\(\Phi\,\Sigma^\dagger\,\Phi_c\,\Phi_c^\dagger\) & \(4\)\\
\(\Sigma_H\,\Sigma^\dagger\,\Sigma\,\Phi_c^\dagger\) & \(4\)\\
\(\Sigma_H\,\Sigma^\dagger\,\Phi\,\Sigma_H^\dagger\) & \(2\)\\
\(\Phi\,\Phi_c^\dagger\,\Phi_c\,\Sigma^\dagger\) & \(4\)\\
\(\Phi\,\Phi_c^\dagger\,\Sigma_H\,\Phi^\dagger\) & \(2\)\\
}
{36}
&
{\purp\Qchunk{\hQ_{qQ}\hQ_{qq}^{\dagger}\,D^2}
{
\(\Sigma_H^\dagger\,\Phi_c\,\Sigma^\dagger\,\Sigma\) & \(4\)\\
\((\Sigma_H^\dagger)^2\,\Phi_c\,\Sigma_H\) & \(2\)\\
\(\Sigma_H^\dagger\,\Phi_c\,\Phi^\dagger\,\Phi\) & \(2\)\\
\(\Sigma_H^\dagger\,\Phi_c^\dagger\,\Phi_c^2\) & \(4\)\\
\((\Sigma^\dagger)\,\Sigma^2\,\Phi^\dagger\) & \(4\)\\
\(\Phi^\dagger\,\Sigma\,\Sigma_H^\dagger\,\Sigma_H\) & \(2\)\\
\((\Phi^\dagger)^2\,\Sigma\,\Phi\) & \(2\)\\
\(\Phi^\dagger\,\Sigma\,\Phi_c^\dagger\,\Phi_c\) & \(4\)\\
\(\Sigma_H^\dagger\,\Sigma\,\Sigma^\dagger\,\Phi_c\) & \(4\)\\
\(\Sigma_H^\dagger\,\Sigma\,\Phi^\dagger\,\Sigma_H\) & \(2\)\\
\(\Phi^\dagger\,\Phi_c\,\Phi_c^\dagger\,\Sigma\) & \(4\)\\
\(\Phi^\dagger\,\Phi_c\,\Sigma_H^\dagger\,\Phi\) & \(2\)\\
}
{36}}
&
\Qchunk{\hQ_{qq}\hQ_{Qq}^{\dagger}\,D^2}
{
\(\Sigma_H\,\Phi_c^\dagger\,\Sigma\,\Sigma^\dagger\) & \(4\)\\
\(\Sigma_H^2\,\Phi_c^\dagger\,\Sigma_H^\dagger\) & \(2\)\\
\(\Sigma_H\,\Phi_c^\dagger\,\Phi\,\Phi^\dagger\) & \(2\)\\
\(\Sigma_H\,(\Phi_c^\dagger)^2\,\Phi_c\) & \(4\)\\
\(\Sigma^2\,\Sigma^\dagger\,\Phi^\dagger\) & \(4\)\\
\(\Phi\,\Sigma^\dagger\,\Sigma_H\,\Sigma_H^\dagger\) & \(2\)\\
\(\Phi^2\,\Sigma^\dagger\,\Phi^\dagger\) & \(2\)\\
\(\Phi\,\Sigma^\dagger\,\Phi_c\,\Phi_c^\dagger\) & \(4\)\\
\(\Sigma_H\,\Sigma^\dagger\,\Sigma\,\Phi_c^\dagger\) & \(4\)\\
\(\Sigma_H\,\Sigma^\dagger\,\Phi\,\Sigma_H^\dagger\) & \(2\)\\
\(\Phi\,\Phi_c^\dagger\,\Phi_c\,\Sigma^\dagger\) & \(4\)\\
\(\Phi\,\Phi_c^\dagger\,\Sigma_H\,\Phi^\dagger\) & \(2\)\\
}
{36}
&
{\purp\Qchunk{\hQ_{Qq}\hQ_{qq}^{\dagger}\,D^2}
{
\(\Sigma_H^\dagger\,\Phi_c\,\Sigma^\dagger\,\Sigma\) & \(4\)\\
\((\Sigma_H^\dagger)^2\,\Phi_c\,\Sigma_H\) & \(2\)\\
\(\Sigma_H^\dagger\,\Phi_c\,\Phi^\dagger\,\Phi\) & \(2\)\\
\(\Sigma_H^\dagger\,\Phi_c^\dagger\,\Phi_c^2\) & \(4\)\\
\((\Sigma^\dagger)\,\Sigma^2\,\Phi^\dagger\) & \(4\)\\
\(\Phi^\dagger\,\Sigma\,\Sigma_H^\dagger\,\Sigma_H\) & \(2\)\\
\((\Phi^\dagger)^2\,\Sigma\,\Phi\) & \(2\)\\
\(\Phi^\dagger\,\Sigma\,\Phi_c^\dagger\,\Phi_c\) & \(4\)\\
\(\Sigma_H^\dagger\,\Sigma\,\Sigma^\dagger\,\Phi_c\) & \(4\)\\
\(\Sigma_H^\dagger\,\Sigma\,\Phi^\dagger\,\Sigma_H\) & \(2\)\\
\(\Phi^\dagger\,\Phi_c\,\Phi_c^\dagger\,\Sigma\) & \(4\)\\
\(\Phi^\dagger\,\Phi_c\,\Sigma_H^\dagger\,\Phi\) & \(2\)\\
}
{36}}
\\[1.2em]
%%%%%%%%%%%%%%%%%%%%%%%%%%%%%%%%%%%%%%%%%%%%%%%%%%%%%%%%%%%%
% Row 3: remaining medium chunks
%%%%%%%%%%%%%%%%%%%%%%%%%%%%%%%%%%%%%%%%%%%%%%%%%%%%%%%%%%%%

\Qchunk{\hQ_{Qq}\hQ_{QQ}^{\dagger}\,D^2}
{
\((\Sigma^\dagger)^2\,\Sigma\,\Phi\) & \(2\)\\
\(\Sigma^\dagger\,\Phi\,\Sigma_H^\dagger\,\Sigma_H\) & \(3\)\\
\(\Sigma^\dagger\,\Phi^2\,\Phi^\dagger\) & \(3\)\\
\(\Sigma^\dagger\,\Phi\,\Phi_c^\dagger\,\Phi_c\) & \(2\)\\
\(\Sigma^\dagger\,\Sigma\,\Phi_c^\dagger\,\Sigma_H\) & \(2\)\\
\(\Sigma_H^\dagger\,\Sigma_H^2\,\Phi_c^\dagger\) & \(3\)\\

\((\Phi_c^\dagger)^2\,\Sigma_H\,\Phi_c\) & \(2\)\\
\(\Phi_c^\dagger\,\Phi\,\Phi^\dagger\,\Sigma_H\) & \(3\)\\

}
{20}
&
{\purp \Qchunk{\hQ_{QQ}\hQ_{Qq}^{\dagger}\,D^2}
{
\(\Sigma^2\,\Sigma^\dagger\,\Phi^\dagger\) & \(2\)\\
\(\Sigma\,\Phi^\dagger\,\Sigma_H\,\Sigma_H^\dagger\) & \(3\)\\
\(\Sigma\,(\Phi^\dagger)^2\,\Phi\) & \(3\)\\
\(\Sigma\,\Phi^\dagger\,\Phi_c\,\Phi_c^\dagger\) & \(2\)\\
\(\Sigma^\dagger\,\Sigma\,\Phi_c\,\Sigma_H^\dagger\) & \(2\)\\
\(\Sigma_H\,(\Sigma_H^\dagger)^2\,\Phi_c\) & \(3\)\\
\(\Phi_c^2\,\Sigma_H^\dagger\,\Phi_c^\dagger\) & \(2\)\\
\(\Phi_c\,\Phi^\dagger\,\Phi\,\Sigma_H^\dagger\) & \(3\)\\
}
{20}
}
&

\Qchunk{\hQ_{QQ}\hQ_{qQ}^{\dagger}\,D^2}
{
\(\Sigma^2\,\Sigma^\dagger\,\Phi^\dagger\) & \(2\)\\
\(\Sigma\,\Phi^\dagger\,\Sigma_H\,\Sigma_H^\dagger\) & \(3\)\\
\(\Sigma\,(\Phi^\dagger)^2\,\Phi\) & \(3\)\\
\(\Sigma\,\Phi^\dagger\,\Phi_c\,\Phi_c^\dagger\) & \(2\)\\
\(\Sigma^\dagger\,\Sigma\,\Phi_c\,\Sigma_H^\dagger\) & \(2\)\\
\(\Sigma_H\,(\Sigma_H^\dagger)^2\,\Phi_c\) & \(3\)\\
\(\Phi_c^2\,\Sigma_H^\dagger\,\Phi_c^\dagger\) & \(2\)\\
\(\Phi_c\,\Phi^\dagger\,\Phi\,\Sigma_H^\dagger\) & \(3\)\\
}
{20}
&

{\purp\Qchunk{\hQ_{qQ}\hQ_{QQ}^{\dagger}\,D^2}
{
\((\Sigma^\dagger)^2\,\Sigma\,\Phi\) & \(2\)\\
\(\Sigma^\dagger\,\Phi\,\Sigma_H^\dagger\,\Sigma_H\) & \(3\)\\
\(\Sigma^\dagger\,\Phi^2\,\Phi^\dagger\) & \(3\)\\
\(\Sigma^\dagger\,\Phi\,\Phi_c^\dagger\,\Phi_c\) & \(2\)\\
\(\Sigma^\dagger\,\Sigma\,\Phi_c^\dagger\,\Sigma_H\) & \(2\)\\
\(\Sigma_H^\dagger\,\Sigma_H^2\,\Phi_c^\dagger\) & \(3\)\\
\((\Phi_c^\dagger)^2\,\Sigma_H\,\Phi_c\) & \(2\)\\
\(\Phi_c^\dagger\,\Phi\,\Phi^\dagger\,\Sigma_H\) & \(3\)\\

}
{20}}
\\[1.2em]
%%%%%%%%%%%%%%%%%%%%%%%%%%%%%%%%%%%%%%%%%%%%%%%%%%%%%%%%%%%%
% Row 4: 
%%%%%%%%%%%%%%%%%%%%%%%%%%%%%%%%%%%%%%%%%%%%%%%%%%%%%%%%%%%%
\Qchunk{\hQ_{qq}\hQ_{QQ}^{\dagger}\,D^2}
{
\(\Sigma\,\Sigma^\dagger\,\Phi\,\Phi^\dagger\) & \(2\)\\
\(\Sigma\,\Sigma^\dagger\,\Sigma_H\,\Sigma_H^\dagger\) & \(1\)\\
\(\Phi_c\,\Phi_c^\dagger\,\Phi\,\Phi^\dagger\) & \(1\)\\
\(\Phi_c\,\Phi_c^\dagger\,\Sigma_H\,\Sigma_H^\dagger\) & \(2\)\\
\(\Sigma^\dagger\,\Phi_c\,\Sigma_H^\dagger\,\Phi\) & \(1\)\\
}
{7}
&
{\purp \Qchunk{\hQ_{QQ}\hQ_{qq}^{\dagger}\,D^2}
{
\(\Sigma^\dagger\,\Sigma\,\Phi^\dagger\,\Phi\) & \(2\)\\
\(\Sigma^\dagger\,\Sigma\,\Sigma_H^\dagger\,\Sigma_H\) & \(1\)\\
\(\Phi_c^\dagger\,\Phi_c\,\Phi^\dagger\,\Phi\) & \(1\)\\
\(\Phi_c^\dagger\,\Phi_c\,\Sigma_H^\dagger\,\Sigma_H\) & \(2\)\\
\(\Sigma\,\Phi_c^\dagger\,\Sigma_H\,\Phi^\dagger\) & \(1\)\\
}
{7}}
&
\Qchunk{\hQ_{qQ}\hQ_{Qq}^{\dagger}\,D^2}
{
\((\Sigma^\dagger)^2\,\Phi^2\) & \(2\)\\
\((\Phi_c^\dagger)^2\,\Sigma_H^2\) & \(2\)\\
\(\Sigma^\dagger \Sigma_H\,\Phi_c^\dagger \Phi\) & \(4\)\\
}
{8}
&
{\purp\Qchunk{\hQ_{Qq}\hQ_{qQ}^{\dagger}\,D^2}
{
\(\Sigma^2\,(\Phi^\dagger)^2\) & \(2\)\\
\(\Phi_c^2\,(\Sigma_H^\dagger)^2\) & \(2\)\\
\(\Sigma^\dagger \Sigma_H\,\Phi_c^\dagger \Phi\) & \(4\)\\
}
{8}}
\\

\bottomrule
\end{tabular}%
}

\caption{\label{tab:full_table}Operator classes. Purple colored blocks are hermitian conjugates of the blocks just left to them. The operators in the first row of block are self hermitian. We have not mentioned the specific position of $D_\mu$ as it is not unique, rather can be attached with any fields respecting the shift symmetry.}
\label{tab:spurion_counts_four_chunks}
\end{table}

%%%%%%%%%%%%%%%%%%%%%%%%%%%%%%%%%%%%%%%%%%%%%%%%%%%%%%%%%
\section{Application to $B\to P_1P_2$}
\label{sec:3}
%%%%%%%%%%%%%%%%%%%%%%%%%%%%%%%%%%%%%%%%%%%%%%%%%%
We now apply the operator basis constructed in the previous section to charmless two-body decays of $B$ mesons, $B\to P_1P_2$, with $P_i\in\{\pi,K,\eta_8\}$. Physical decay amplitudes are obtained by inserting the weak spurion matrices and expanding the meson multiplets into their component fields. The resulting amplitudes are expressed as linear combinations of process-independent EFT Wilson coefficients multiplied by CKM factors and kinematic structures. Since the same finite set of Wilson coefficients contributes to several decay channels, the amplitudes are not independent but satisfy symmetry-imposed relations. These correlations are a direct consequence of the operator basis and lead to model-independent flavor sum rules. In the following, we identify the subset of operators relevant for $B\to P_1P_2$, specify the weak spurions and meson multiplets used in the flavor projection, illustrate the procedure with the decay $\bar B_d^0\to\pi^+\pi^-$, and finally derive the corresponding amplitude sum rules.

%%%%%%%%%%%%%%%%%%%%%%%%%%%%%%%%%
\subsection{Building blocks}\label{subsec:OP}
%%%%%%%%%%%%%%%%%%%%%%%%%%%%%%%%%
To project the invariant operators onto the physical decay amplitudes, we now specify the explicit flavor matrices used in the expansion, which are the building blocks of our construction. The weak spurions are treated as non-dynamical background fields. Their nonzero entries select the flavor flow of the underlying weak transition and carry the corresponding CKM factors. In the present convention, $\hQ_{qq}$ describes light-sector flavor transitions, $\hQ_{QQ}$ describes heavy-sector transitions, while $\hQ_{qQ}$ and $\hQ_{Qq}$ connect the light and heavy flavor sectors. Their explicit forms are obtained by matching with the SM currents and derived in~\cite{Chakraborty:2026vdz}
\begin{align}
\widehat{Q}_{qq} &=
\begin{pmatrix}
0 & 0 & 0 \\
V_{\bar d u} & 0 & 0 \\
V_{\bar s u} & 0 & 0
\end{pmatrix}\;,
&
\widehat{Q}_{QQ} &=
\begin{pmatrix}
0 & 0 \\
V_{\bar b c} & 0
\end{pmatrix}\;,
& \widehat{Q}_{qQ} &=
\begin{pmatrix}
0 & 0 \\
V_{\bar d c} & 0 \\
V_{\bar s c} & 0
\end{pmatrix}\;,
& \widehat{Q}_{Qq} &=
\begin{pmatrix}
0 & 0 & 0 \\
V_{\bar b u} & 0 & 0
\end{pmatrix}\;.
\end{align}
Here, the position of a nonzero entry fixes the flavor indices contracted in the invariant operators, while the entry itself keeps track of the CKM factor associated with that contraction.

The heavy-light meson multiplets are written as
\begin{align}
\Phi &=
\begin{pmatrix}
D_u & D_d^+ & D_s^+ \\
B_u^- & \bar B_d^0 & \bar B_s^0
\end{pmatrix}\;,
&
\Phi_c &=
\begin{pmatrix}
D_u & B_u^+ \\
D_d^- & B_d^0 \\
D_s^- & B_s^0
\end{pmatrix}\;.
\end{align}
Thus, selecting a particular entry of $\Phi$ or $\Phi_c$ fixes the external heavy meson that appears in the decay amplitude. For the charmless modes considered in this section, the relevant external heavy field is the initial $B$ meson, while the light pseudoscalars are generated by expanding the chiral field $\Sigma$ in powers of the light-meson matrix $M$. The light pseudoscalar fields are collected in the matrix
\begin{align}
\widetilde{M} = M + \frac{1}{\sqrt{3}}\eta_0 =
\begin{pmatrix}
\dfrac{\pi^0}{\sqrt{2}}+\dfrac{\eta_8}{\sqrt{6}}
& \pi^+ & K^+ \\
\pi^- &
-\dfrac{\pi^0}{\sqrt{2}}+\dfrac{\eta_8}{\sqrt{6}}
& K^0 \\
K^- & \bar K^0 &
-\dfrac{2\eta_8}{\sqrt{6}}
\end{pmatrix} \; + \;\frac{1}{\sqrt{3}}\eta_0,
\end{align}
and the adjoint $M$ enters the invariant operators through
\begin{align}
\Sigma = \frac{f_\pi}{2}\exp\left(\frac{2iM}{f_\pi}\right)\;.
\end{align}
For a two-body amplitude $B\to P_1P_2$, we expand the operators to second order in the light fields contained in $\Sigma$ and $\Sigma^\dagger$.

We also introduce the heavy-sector field
\begin{align}
\Sigma_H =
\begin{pmatrix}
\lambda_c+i\eta_c & iB_c^+ \\
iB_c^- & \lambda_b+i\eta_b
\end{pmatrix}\;,
\label{eq:sigmaH}
\end{align}
where the off-diagonal entries $B_c^{\pm}$ are the charged heavy-heavy pseudoscalars, $\eta_c$ and $\eta_b$ denote the neutral charmonium- and bottomonium-like states, and $\lambda_c$, $\lambda_b$ are real parameters proportional to the condensates $\langle \bar{c}c\rangle$ and $\langle \bar{b}b\rangle$. 

With the explicit definition of the meson multiplets, our choice of $SU(2)_{L_H}\times SU(2)_{R_H}$ symmetry for the heavy mesons becomes clear. Unlike the light chiral symmetry, it is not a dynamical symmetry of QCD. We use it as a spurion symmetry that the heavy meson masses break explicitly. Instead of a single $SU(2)$, the doubling lets the weak spurions track the flavor flow of the underlying transition: $\widehat Q_{qQ}$ transforms under $L\, \widehat Q_{qQ} L_{H}^\dagger$ and $\widehat Q_{Qq}$ as $L_H\, \widehat Q_{Qq} L_{}^\dagger$, so operators mediating $b\to c\bar u q$ and $b\to u\bar c q$ fall into distinct invariant classes, and $\Phi$ and $\Phi_c$ (hence $D$ and $\bar D$ final states) are distinguished by symmetry alone.

%%%%%%%%%%%%%%%%%%%%%%%%%%%%%%%%%%%%%%%%%%%%%%%%%%%
\subsection{Relevant operators for $B\to P_1P_2$}
%%%%%%%%%%%%%%%%%%%%%%%%%%%%%%%%%%%%%%%%%%%%%%%%%%%
For charmless two-body decays $B\to P_1P_2$, the external states contain one $B$ meson and two light pseudoscalars. We therefore select from Table~\ref{tab:full_table} only those four-meson, two-derivative operators that can produce this field content after expanding the light-meson matrices $\Sigma$ and $\Sigma^\dagger$. The first set, denoted by $O_{1a,b}$ and $O_{2a,b}$, arises from the spurion pair ${\hQ_{qQ},\hQ_{QQ}^\dagger}$ and contains the heavy-sector weak spurion together with a light-heavy spurion. The second set, denoted by $O_{3a,\ldots,d}$ and $O_{4a,\ldots,d}$, involves the light-sector spurion $\hQ_{qq}^\dagger$ together with $\hQ_{Qq}$ and accounts for the remaining allowed charmless flavor contractions. These operators form the complete basis contributing to $B\to P_1P_2$ in this order and are listed in Table~\ref{tab:BPP_oplist}.

%%%%%%%%%%%%%%%%%%%%%%%%%%%%%%%%%%%%%%%%%%%%%%%%%%
\begin{table}[h]
\centering
\renewcommand{\arraystretch}{1.25}
\resizebox{.9\textwidth}{!}{%
\setlength{\tabcolsep}{0pt}
\begin{tabular}{p{0.2\textwidth} p{0.2\textwidth} p{0.12\textwidth} p{0.50\textwidth}}
\toprule
\textbf{Field content} & \textbf{Non-redundant operators} & \textbf{Operator} & \textbf{Covariant form} \\
\midrule
\multirow{2}{=}{ \(\Sigma^\dagger,\Sigma,\Sigma^\dagger,\Phi\)\\
 \(\{\hQ_{qQ},\hQ_{QQ}^\dagger\}\)}
& \multirow{2}{*}{2}
& \(O_{1a}\) &
\((\Sigma^\dagger T_L^A \hQ_{qQ}\tau_L^I \delmu \Phi)
(\Sigma^\dagger \,\hQ_{QQ}^\dagger \tau_L^I \, T_L^A \delmu\Sigma)\)
\\[2ex]
&& \(O_{1b}\) &
\((\Sigma^\dagger T_L^A T_R^B \hQ_{qQ}\tau_L^I \delmu \Phi)
(\Sigma^\dagger \,\hQ_{QQ}^\dagger \tau_L^I \, T_L^A T_R^B \delmu\Sigma)\)
\\
\midrule
\multirow{2}{=}{ \(\Sigma^\dagger,\Sigma,\Phi_c^\dagger,\Sigma_H\)\\
 \(\{\hQ_{qQ},\hQ_{QQ}^\dagger\}\)}
& \multirow{2}{*}{2}
& \(O_{2a}\) &
\((\Sigma^\dagger \hQ_{qQ}\tau_L^I \delmu \Sigma_H)
(\Phi_c^\dagger \langle \hQ_{QQ}^\dagger \tau_L^I \rangle \delmu \Sigma)\)
\\[2ex]
&& \(O_{2b}\) &
\((\Sigma^\dagger T_L^A \hQ_{qQ}\tau_L^I \delmu \Sigma_H)
(\Phi_c^\dagger \langle \hQ_{QQ}^\dagger \tau_L^I \rangle T_L^A \delmu \Sigma)\)
\\
\midrule
\midrule
\multirow{4}{=}{ \(\Sigma_H^\dagger,\Phi_c,\Sigma^\dagger,\Sigma\)\\
 \(\{\hQ_{Qq},\hQ_{qq}^\dagger\}\)}
& \multirow{4}{*}{4}
& \(O_{3a}\) &
\((\Sigma^\dagger \,\hQ_{qq}^\dagger T_L^A\, T_L^A \delmu\Sigma)\,
(\Sigma_H^\dagger \hQ_{Qq}\delmu \Phi^c)\)
\\[2ex]
&& \(O_{3b}\) &
\((\Sigma^\dagger \,\hQ_{qq}^\dagger T_L^A\, \delmu\Sigma)\,
(\Sigma_H^\dagger\hQ_{Qq} T_L^A \delmu \Phi^c)\)
\\[2ex]
&& \(O_{3c}\) &
\(f^{ABC}(\Sigma^\dagger \,\hQ_{qq}^\dagger T_L^A\, T_L^B \delmu\Sigma)\,
(\Sigma_H^\dagger\hQ_{Qq} T_L^C \delmu \Phi^c)\)
\\[2ex]
&& \(O_{3d}\) &
\(d^{ABC}(\Sigma^\dagger \,\hQ_{qq}^\dagger T_L^A\, T_L^B \delmu\Sigma)\,
(\Sigma_H^\dagger\hQ_{Qq} T_L^C \delmu \Phi^c)\)
\\
\midrule
\multirow{4}{=}{ \(\Sigma^\dagger,\Sigma,\Phi^\dagger,\Sigma\)\\
 \(\{\hQ_{qq}^\dagger,\hQ_{Qq}\}\)}
& \multirow{4}{*}{4}
& \(O_{4a}\) &
\((\Sigma^\dagger T_R^D \,\hQ_{qq}^\dagger T_L^A \, T_L^A \delmu\Sigma)
(\Phi^\dagger T_R^D\hQ_{Qq} \delmu\Sigma)\)
\\[2ex]
&& \(O_{4b}\) &
\((\Sigma^\dagger T_R^D \,\hQ_{qq}^\dagger T_L^A \, \delmu\Sigma)
(\Phi^\dagger T_R^D\hQ_{Qq} T_L^A \delmu\Sigma)\)
\\[2ex]
&& \(O_{4c}\) &
\(f^{ABC}(\Sigma^\dagger T_R^D \,\hQ_{qq}^\dagger T_L^A \, T_L^B \delmu\Sigma)
(\Phi^\dagger T_R^D\hQ_{Qq} T_L^C \delmu\Sigma)\)
\\[2ex]
&& \(O_{4d}\) &
\(d^{ABC}(\Sigma^\dagger T_R^D \,\hQ_{qq}^\dagger T_L^A \, T_L^B \delmu\Sigma)
(\Phi^\dagger T_R^D\hQ_{Qq} T_L^C \delmu\Sigma)\)
\\
\bottomrule
\end{tabular}}
\caption{\label{tab:BPP_oplist}Operator list in covariant form for charmless $B$ decay modes: $B \to P_1 P_2$. In HS method we include derivative as covariant derivative in a generic sense, but, here, in absence of any underlying gauge symmetry $D_\mu$ reads $\partial_\mu$. Also, we define $\Phi_1^\dagger \delmu \Phi_2\equiv \Phi_1^\dagger (\partial_\mu \Phi_2)-(\partial_\mu \Phi_1^\dagger) \Phi_2$, with $d^{ABC}, f^{ABC}$ as symmetric tensors and anti-symmetric structure constants of $SU(3)$, respectively.}
\end{table}
%%%%%%%%%%%%%%%%%%%%%%%%%%%%%%%%%%%%%%%%%%%%%%%%%%

Note that the $\Sigma_H$ field, given in eq.~\eqref{eq:sigmaH}, plays a role analogous to the chiral field in the heavy-flavor sector. Its diagonal entries have $\lambda_c$ and $\lambda_b$ that denote possible heavy-quark condensate or background insertions in the charm and bottom directions. Therefore, terms proportional to them captures the {\it novel} and subleading contributions to the amplitudes. The operators containing $\Sigma_H$ can contribute not only to processes with explicit heavy-heavy mesons, such as $J/\psi\to D\pi$ or processes involving $B_c$-type states, but also to ordinary heavy-light decay amplitudes when $\Sigma_H$ is replaced by its diagonal background components. This is why terms proportional to $\lambda_b$ appear in the projected $B$-decay amplitudes even when no heavy-heavy meson is present as an external state.

With these conventions, the projection of an invariant operator onto a physical channel is straightforward. The relevant fields $\Sigma$, $\Sigma^\dagger$, and, when present, $\Sigma_H$ are expanded to the required order, the weak spurion matrices are inserted and the coefficient of the desired external-field product is selected. The derivatives acting on external fields are then replaced by the corresponding momenta. In this way, each invariant operator gives a contribution proportional to the CKM factors, Wilson coefficients, possible heavy-sector background insertions such as $\lambda_b$, and scalar products of external momenta.
%%%%%%%%%%%%%%%%%%%%%%%%%%%%%%%%%%%%%%%%%%%%%%%%
\subsection{Amplitudes in terms of Wilson coefficients}
\label{subsec:amplitudes-wcs}
%%%%%%%%%%%%%%%%%%%%%%%%%%%%%%%%%%%%%%%%%%%%%%%%

We now show how explicit decay amplitudes are extracted from invariant operators by expanding the meson multiplets and spurions
$\widehat{Q}_{ij}$.The derivatives acting on the
external fields supply the momentum structures and the $\widehat{Q}_{ij}$ spurions give the CKM elements. We illustrate this for
$\bar{B}^0_d \to \pi^+\pi^-$, which receives contributions from
$\mathcal{O}_{1b}$, $\mathcal{O}_{2b}$, $\mathcal{O}_{4a}$, $\mathcal{O}_{4c}$
and $\mathcal{O}_{4d}$. Upon expansion, the relevant operator contributions lead to the following contribution
\begin{align}
    {\cal A}(\bar B_d^0 \to \pi^+\pi^-)
    ={}&
    4\,C_{1b}\,V_{\bar bc}V_{d\bar c}
    \left(
    2\,p_B\!\cdot p_{\pi^+}
    -2\,p_{\pi^+}\!\cdot p_{\pi^-}
    \right)
    \notag\\
    &+
    \frac{4}{3}\,C_{2b}\,V_{\bar bc}V_{d\bar c}\lambda_b
    \left(
    2\,p_B\!\cdot p_{\pi^+}
    -5\,p_{\pi^-}\!\cdot p_{\pi^+}
    +3\,p_B\!\cdot p_{\pi^-}
    \right)
    \notag\\
    &+
    \frac{4}{3}\,C_{4a}\,V_{d\bar u}V_{\bar b u}
    \left(
    -p_{\pi^-}\!\cdot p_{\pi^+}
    +p_B\!\cdot p_{\pi^-}
    \right)
    \notag\\
    &+
    \frac{4i}{3}\,C_{4c}\,V_{d\bar u}V_{\bar b u}
    \left(
    p_B\!\cdot p_{\pi^-}
    +p_{\pi^-}\!\cdot p_{\pi^+}
    \right)
    \notag\\
    &+
    C_{4d}\,V_{d\bar u}V_{\bar b u}
    \left(
    \frac{4}{3}\,p_B\!\cdot p_{\pi^+}
    +
    \frac{16}{9}\,p_B\!\cdot p_{\pi^-}
    -
    \frac{28}{9}\,p_{\pi^-}\!\cdot p_{\pi^+}
    \right)\;,
    \label{eq:Bd-pipi-amplitude-expanded}
\end{align}
with $C_i$'s as the Wilson coefficients (WCs) corresponding to the effective operators $O_i$ of Table~\ref{tab:BPP_oplist}. Equivalently, collecting the independent momentum structures, one obtains the following
\begin{align}
    {\cal A}(\bar B_d^0 \to \pi^+\pi^-)
    ={}&
    \mathcal C_{B+}^{\pi\pi}\,p_B\!\cdot p_{\pi^+}
    +
    \mathcal C_{B-}^{\pi\pi}\,p_B\!\cdot p_{\pi^-}
    -
    \mathcal C_{+-}^{\pi\pi}\,p_{\pi^-}\!\cdot p_{\pi^+}\;,
    \label{eq:Bd-pipi-amplitude-compact}
\end{align}
where
\begin{align}
    \mathcal C_{B+}^{\pi\pi}
    ={}&
    8C_{1b}V_{\bar bc}V_{d\bar c}
    +
    \frac{8}{3}C_{2b}V_{\bar bc}V_{d\bar c}\lambda_b
    +
    \frac{4}{3}C_{4d}V_{d\bar u}V_{\bar b u}\;,
    \notag\\[0.4em]
    \mathcal C_{B-}^{\pi\pi}
    ={}&
    4C_{2b}V_{\bar bc}V_{d\bar c}\lambda_b
    +
    \frac{4}{3}C_{4a}V_{d\bar u}V_{\bar b u}
    +
    \frac{4i}{3}C_{4c}V_{d\bar u}V_{\bar b u}
    +
    \frac{16}{9}C_{4d}V_{d\bar u}V_{\bar b u}\;,
    \notag\\[0.4em]
    \mathcal C_{+-}^{\pi\pi}
    ={}&
    8C_{1b}V_{\bar bc}V_{d\bar c}
    +
    \frac{20}{3}C_{2b}V_{\bar bc}V_{d\bar c}\lambda_b
    +
    \frac{4}{3}C_{4a}V_{d\bar u}V_{\bar b u}
    -
    \frac{4i}{3}C_{4c}V_{d\bar u}V_{\bar b u}
    +
    \frac{28}{9}C_{4d}V_{d\bar u}V_{\bar b u}\;.
    \label{eq:Bd-pipi-coefficients}
\end{align}
This form makes the EFT interpretation transparent: the amplitude is a linear combination of a small number of universal Wilson coefficients, while the channel dependence is carried by the CKM factors, Clebsch coefficients, and external momenta.

For comparison, in a factorized QCD description, the same decay is usually written in terms of a decay constant and a $B\to\pi$ transition form factor as~\cite{Beneke:1999br, Beneke:2001ev}
\begin{align}
    \mathcal {\cal A}(\bar B_d^0 \to \pi^+\pi^-)
    =
    \frac{G_F}{\sqrt{2}}\,
    f_\pi\,
    \left(m_B^2-m_\pi^2\right)
    F_0^{B\to\pi}(m_\pi^2)
    \left[
    V_{\bar b u}V_{d \bar u}\,a_1
    -
    V_{\bar bt}V_{d\bar t}\,a_4^{\rm eff}
    +\cdots
    \right] .
    \label{eq:Bd-pipi-qcdf}
\end{align}
Here $F_0^{B\to \pi}$ is the scalar $B\to \pi$ transition form factor in BSW convention~\cite{Wirbel:1985ji, Bauer:1986bm}, and $a_1 = c_1 + c_2/N_c$, $a_4 = c_4 + c_3/N_c$ are the standard effective coefficients of naive factorization~\cite{Bauer:1986bm, Buchalla:1995vs, Beneke:2000ry}.
The factorized matrix element underlying this expression is
\begin{align}
    &\langle \pi^-|\,\bar d\gamma_\mu(1-\gamma_5)u\,|0\rangle
    \,
    \langle \pi^+|\,\bar u\gamma^\mu(1-\gamma_5)b\,|\bar B_d^0\rangle
    \notag\\
    &\hspace{3cm}
    =
    i f_\pi\,
    \left(m_B^2-m_\pi^2\right)
    F_0^{B\to\pi}(m_\pi^2)\;.
    \label{eq:factorized-matrix-element}
\end{align}

Equation~\eqref{eq:Bd-pipi-qcdf} should be viewed as a factorized reference expression. In mesonic EFT parametrization, the same hadronic information is reorganized into the Wilson coefficients $C_{1b}$, $C_{2b}$, $C_{4a}$, $C_{4c}$, and $C_{4d}$. To make the correspondence explicit, consider the chiral limit. In the chiral limit, two-body kinematics ($p_B\cdot p_{\pi^\pm}=m_B^2/2$, $p_{\pi^+}\!\cdot p_{\pi^-}=(m_B^2-2m_\pi^2)/2$) collapses eq.\,\eqref{eq:Bd-pipi-amplitude-expanded} to 
\begin{align}
	\mathcal{A}(\bar B^0_d\to\pi^+\pi^-)
	=\frac{4i}{3}\left(m_B^2-m_\pi^2\right)C_{4c}\,V_{d\bar u}V_{\bar b u}
	+\mathcal{O}\!\left(m_\pi^2\right)\;.
	\label{eq:chirallimitpipi}
\end{align}
Eliminating $V_{\bar b t}V_{d\bar t}$ in eq.\,\eqref{eq:Bd-pipi-qcdf} by CKM unitarity and matching the $V_{d\bar u}V_{\bar b u}$ structures gives
\begin{align}
	\frac{4i}{3}\,C_{4c}
	=\frac{G_F}{\sqrt{2}}\,f_\pi\,F_0^{B\to\pi}(m_\pi^2)
	\left(a_1+a_4^{\rm eff}\right)\;,
	\label{eq:C4cidentification}
\end{align}
while $C_{1b}$, $C_{2b}$, $C_{4a}$, and $C_{4d}$ enter this mode only at $\mathcal{O}(m_\pi^2/m_B^2)$, absorbing the power-suppressed and non-factorizable corrections. Keeping this correspondence in mind, quantities such as decay constants, transition form factors, effective coefficients $a_i$, and non-factorizable contributions are not introduced as independent building blocks at this stage; rather, their symmetry-allowed effects are absorbed into the EFT Wilson coefficients.
	
This reorganization has an important consequence.  Several $B\to P_1P_2$ amplitudes depend on the same finite set of Wilson coefficients.  Therefore, the EFT construction predicts correlations among different physical decay amplitudes.  These correlations are independent of the size of the Wilson coefficients and follow only from the flavor symmetry, the chosen spurion insertions, and the operator basis.  In the next subsection, we use this observation to derive amplitude relations among the charmless two-body decay modes.

%%%%%%%%%%%%%%%%%%%%%%%%%%%%%%%%%%%%%%%%%%%%%%%%%%%%%%%%%
\subsection{Amplitude sum rules}\label{sec:BPP_amp}
%%%%%%%%%%%%%%%%%%%%%%%%%%%%%%%%%%%%%%%%%%%%%%%%%%%%%%%%%

An important consequence of the flavor-covariant EFT construction is that amplitude relations follow directly from the operator basis. For a set of $N$ decay amplitudes receiving contributions from $r$ independent EFT operators, we write
\begin{align}
	\begin{pmatrix}
		{\cal A}_1 \\
		{\cal A}_2\\
		\vdots\\
		{\cal A}_N
	\end{pmatrix}
	&=
	M
	\begin{pmatrix}
		C_{1}\\
		C_{2}\\
		\vdots\\
		C_{r}
	\end{pmatrix}\;,
\end{align}
where $C_i$ are the EFT Wilson coefficients and $M$ is the amplitude matrix obtained by projecting the invariant operators onto physical meson states. If
\begin{align}
	{\rm rank}(M)=\rho < N\;, \Rightarrow \quad {\rm Null\; space:}\; N-\rho\;,
\end{align}
independent linear relations among the decay amplitudes, corresponding to the flavor-symmetry sum rules. We illustrate this procedure for the charmless two-body decays $B\to P_1P_2$, showing how the familiar amplitude relations arise directly from the EFT operator basis\footnote{In case of two-body decay, the amplitude, in the center-of-mass frame, can be parametrized in terms of a single momentum variable. Thus, the sum rules remain unaffected whether we express the individual amplitudes in terms of the momentum of the meson multiplet ($p_{\Sigma}$) or the meson ($p_{\pi/K/\eta_8}$) themselves. In Appendix~\ref{app:BPP}, we provide the amplitudes in terms of the meson momentum to be on par with the results in existing literature. However, this is not valid for the three-body decay, as the amplitudes have non-uniform dependence on the momenta. Our operator constitute of meson multiplets, and the derivatives acting on them. Thus, the amplitudes are naturally expressed in terms of the momenta of the meson-multiplets. This has been discussed in detail in Sec.~\ref{sec:comparison}.}. 
Since the number of physical decay modes exceeds the number of independent hadronic amplitudes, flavor symmetry implies linear relations among the decay amplitudes. 
%%%%%%%%%%%%%%%%%%%%%%%%%%%%%%%%%%%%%%%%%%%%%%%%%%
\begin{table}[h]
	\centering
	\renewcommand{\arraystretch}{1}
	\resizebox{.7\textwidth}{!}{$
		\left(\begin{array}{l|c|ccccc}
			& & \multicolumn{2}{c}{\alpha=c} & \multicolumn{3}{c}{\alpha=u} \\
			%\cline{3-7}
			{\rm Mode} & {\rm CKM} & O_{1b} & O_{2b} \,(\times\lambda_b/f_\pi) & O_{4a} & O_{4c} & O_{4d} \\ \hline
			{\cal A}(B^- \to \pi^-\pi^0) & V_{d\bar\alpha}V_{\bar b \alpha} & 0 & 0 & -\dfrac{2\sqrt2}{3} & -\dfrac{2\sqrt2}{3} & -\dfrac{2\sqrt2}{9} \\[4pt]
			{\cal A}(\bar B_d^0 \to \pi^+\pi^-) & V_{d\bar\alpha}V_{\bar b \alpha} & -8 & -\dfrac{20}{3} & -\dfrac{4}{3} & \dfrac{4}{3} & -\dfrac{28}{9} \\[4pt]
			{\cal A}(\bar B_d^0 \to \pi^0\pi^0) & V_{d\bar\alpha}V_{\bar b \alpha} & -4\sqrt2 & -\dfrac{10\sqrt2}{3} & 0 & \dfrac{4\sqrt2}{3} & -\dfrac{4\sqrt2}{3} \\[4pt]
			{\cal A}(\bar B_d^0 \to K^+K^-) & V_{d\bar\alpha}V_{\bar b \alpha} & 0 & -8 & -4 & 8 & -\dfrac{16}{3} \\[4pt]
			{\cal A}(\bar B_d^0 \to K^0\bar K^0) & V_{d\bar\alpha}V_{\bar b \alpha} & -8 & -\dfrac{20}{3} & 0 & \dfrac{16}{3} & 0 \\[4pt]
			{\cal A}(\bar B_s^0 \to K^+\pi^-) & V_{d\bar\alpha}V_{\bar b \alpha} & -8 & \dfrac{4}{3} & \dfrac{8}{3} & -\dfrac{20}{3} & \dfrac{20}{9} \\[4pt]
			{\cal A}(\bar B_s^0 \to K^0\pi^0) & V_{d\bar\alpha}V_{\bar b \alpha} & 4\sqrt2 & -\dfrac{2\sqrt2}{3} & -2\sqrt2 & \dfrac{8\sqrt2}{3} & -\dfrac{4\sqrt2}{3} \\[4pt]
			{\cal A}(B^- \to \bar K^0\pi^-) & V_{\bar b \alpha}V_{s\bar\alpha} & -8 & \dfrac{4}{3} & 0 & -\dfrac{20}{3} & 4 \\[4pt]
			{\cal A}(B^- \to K^-\pi^0) & V_{\bar b \alpha}V_{s\bar\alpha} & -4\sqrt2 & \dfrac{2\sqrt2}{3} & -\dfrac{2\sqrt2}{3} & -4\sqrt2 & \dfrac{16\sqrt2}{9} \\[4pt]
			{\cal A}(\bar B_d^0 \to K^-\pi^+) & V_{\bar b \alpha}V_{s\bar\alpha} & -8 & \dfrac{4}{3} & \dfrac{8}{3} & -\dfrac{20}{3} & \dfrac{20}{9} \\[4pt]
			{\cal A}(\bar B_d^0 \to \bar K^0\pi^0) & V_{\bar b \alpha}V_{s\bar\alpha} & 4\sqrt2 & -\dfrac{2\sqrt2}{3} & -2\sqrt2 & \dfrac{8\sqrt2}{3} & -\dfrac{4\sqrt2}{3} \\[4pt]
			{\cal A}(\bar B_s^0 \to K^0\bar K^0) & V_{\bar b \alpha}V_{s\bar\alpha} & -8 & -\dfrac{20}{3} & 0 & \dfrac{16}{3} & 0 \\[4pt]
			{\cal A}(\bar B_s^0 \to K^+K^-) & V_{\bar b \alpha}V_{s\bar\alpha} & -8 & -\dfrac{20}{3} & -\dfrac{4}{3} & \dfrac{4}{3} & -\dfrac{28}{9} \\[4pt]
			{\cal A}(B^- \to K^0K^-) & V_{d\bar\alpha}V_{\bar b \alpha} & -8 & \dfrac{4}{3} & 0 & -\dfrac{20}{3} & 4 \\[4pt]
			{\cal A}(\bar B_s^0 \to \pi^0\pi^0) & V_{\bar b \alpha}V_{s\bar\alpha} & 0 & -4\sqrt2 & -2\sqrt2 & 4\sqrt2 & -\dfrac{8\sqrt2}{3} \\[4pt]
			{\cal A}(\bar B_s^0 \to \pi^+\pi^-) & V_{\bar b \alpha}V_{s\bar\alpha} & 0 & -8 & -4 & 8 & -\dfrac{16}{3} \\[4pt]
			{\cal A}(B^- \to \pi^-\eta_8) & V_{d\bar\alpha}V_{\bar b \alpha} & -\dfrac{8\sqrt6}{3} & \dfrac{4\sqrt6}{9} & -\dfrac{2\sqrt6}{9} & -\dfrac{22\sqrt6}{9} & \dfrac{34\sqrt6}{27} \\[4pt]
			{\cal A}(B^- \to K^-\eta_8) & V_{\bar b \alpha}V_{s\bar\alpha} & \dfrac{4\sqrt6}{3} & -\dfrac{2\sqrt6}{9} & -\dfrac{2\sqrt6}{9} & \dfrac{8\sqrt6}{9} & -\dfrac{20\sqrt6}{27} \\[4pt]
			{\cal A}(\bar B_d^0 \to \pi^0\eta_8) & V_{d\bar\alpha}V_{\bar b \alpha} & \dfrac{8\sqrt3}{3} & -\dfrac{4\sqrt3}{9} & -\dfrac{4\sqrt3}{3} & \dfrac{8\sqrt3}{9} & -\dfrac{16\sqrt3}{9} \\[4pt]
			{\cal A}(\bar B_d^0 \to \eta_8\eta_8) & V_{d\bar\alpha}V_{\bar b \alpha} & -\dfrac{4\sqrt2}{3} & -\dfrac{34\sqrt2}{9} & -\dfrac{4\sqrt2}{3} & 4\sqrt2 & -\dfrac{4\sqrt2}{3} \\[4pt]
			{\cal A}(\bar B_d^0 \to \bar K^0\eta_8) & V_{\bar b \alpha}V_{s\bar\alpha} & \dfrac{4\sqrt6}{3} & -\dfrac{2\sqrt6}{9} & -\dfrac{2\sqrt6}{3} & \dfrac{8\sqrt6}{9} & -\dfrac{4\sqrt6}{9} \\[4pt]
			{\cal A}(\bar B_s^0 \to K^0\eta_8) & V_{d\bar\alpha}V_{\bar b \alpha} & \dfrac{4\sqrt6}{3} & -\dfrac{2\sqrt6}{9} & -\dfrac{2\sqrt6}{3} & \dfrac{8\sqrt6}{9} & -\dfrac{4\sqrt6}{9} \\[4pt]
			{\cal A}(\bar B_s^0 \to \pi^0\eta_8) & V_{\bar b \alpha}V_{s\bar\alpha} & 0 & 0 & 0 & -\dfrac{8\sqrt3}{9} & -\dfrac{8\sqrt3}{9} \\[4pt]
			{\cal A}(\bar B_s^0 \to \eta_8\eta_8) & V_{\bar b \alpha}V_{s\bar\alpha} & -\dfrac{16\sqrt2}{3} & -\dfrac{28\sqrt2}{9} & \dfrac{2\sqrt2}{3} & \dfrac{4\sqrt2}{3} & 0 
		\end{array}
		\right)$}
	\caption{Amplitudes for the charmless $B\to P_1P_2$ modes. The CKM column gives the	common factor for each row, with $\alpha=c$ for the operators $O_{1b},O_{2b}$ and $\alpha=u$ for $O_{4a},O_{4c},O_{4d}$.}
	\label{tab:BPPamps24}
\end{table}
%%%%%%%%%%%%%%%%%%%%%%%%%%%%%%%%%%%%%%%%%%%%%%%%%%
Conventionally, these sum rules are derived using topological amplitudes or $SU(3)_V$ reduced matrix elements. In our EFT framework, they instead follow directly from the flavor-invariant operator basis: the amplitude matrix automatically exhibits the corresponding null directions. For two-body decays, we work in the $B$-meson rest frame and use momentum conservation to eliminate redundant kinematic structures. After factoring out the common CKM and the momentum dependence, the amplitudes can be written in the matrix form $(\mathcal{A}|M)$, where $\mathcal{A}$ is the vector of physical decay amplitudes and $M$ encodes the coefficients of the independent EFT operator structures. We apply this to the full set of 24 charmless $B\to P_1P_2$ modes. The resulting amplitude matrix is given in Table~\ref{tab:BPPamps24}.

Row-reduction of the matrix $M$ gives Rank($M$)$ = 5$, implying $(24-5 = 19)$ relations among the 24 decay modes.  
There are eight relations among the twelve $b\to d$ modes and eight among the twelve $b\to s$ modes, in which all CKM factors cancel. Three relations connect the two sectors through a CKM ratio, $r_u\equiv V_{s\bar u}/V_{d\bar u}$ or  $r_c\equiv V_{s\bar c}/V_{d\bar c}$. We group the relations by the symmetry needed to derive them. Relations shown in blue have no explicit counterpart in the literature. They are either derivable from published $SU(3)$ tabulations but never written down, or, in two cases, specific to the chiral construction and absent from any $SU(3)_V$-based analysis.

\noindent The iso-spin relations are 
\begin{align}
    \sqrt2\,{\cal A}(B^-\!\to\pi^-\pi^0) &= {\cal A}(\bar B_d^0\!\to\pi^+\pi^-)-\sqrt2\,{\cal A}(\bar B_d^0\!\to\pi^0\pi^0)\,,\label{eq:iso1}\\
	{\cal A}(\bar B_s^0\!\to\pi^+\pi^-) &= \sqrt2\,{\cal A}(\bar B_s^0\!\to\pi^0\pi^0)\,,\label{eq:iso2}\\
	\sqrt2\,{\cal A}(B^-\!\to\bar K^0\pi^-)&-2{\cal A}(B^-\!\to K^-\pi^0)+\sqrt2\,{\cal A}(\bar B_d^0\!\to K^-\pi^+)+2{\cal A}(\bar B_d^0\!\to\bar K^0\pi^0)=0\,.\label{eq:iso3}
\end{align}
The U-spin relations, five in the $b\to d$ sector,
\begin{align}
	&{\cal A}(\bar B_d^0\!\to\pi^+\pi^-) = {\cal A}(\bar B_d^0\!\to K^+K^-)+{\cal A}(\bar B_s^0\!\to K^+\pi^-)\,,\label{eq:su1}\\
	&\sqrt3\,{\cal A}(B^-\!\to\pi^-\eta_8) = {\cal A}(B^-\!\to\pi^-\pi^0)+\sqrt2\,{\cal A}(B^-\!\to K^0K^-)\,,\label{eq:su7}\\
	&\textcolor{oceanblue}{{\cal A}(\bar B_s^0\!\to K^0\pi^0) = \sqrt3\,{\cal A}(\bar B_s^0\!\to K^0\eta_8)\,,}\label{eq:usd3}\\
	&\textcolor{oceanblue}{\sqrt3\,{\cal A}(\bar B_d^0\!\to\pi^0\pi^0)-\sqrt2\,{\cal A}(\bar B_d^0\!\to\pi^0\eta_8)-\sqrt3\,{\cal A}(\bar B_d^0\!\to\eta_8\eta_8)+4\,{\cal A}(\bar B_s^0\!\to K^0\eta_8)=0\,,}\label{eq:usd4}\\
	&\textcolor{oceanblue}{{\cal A}(\bar B_d^0\!\to K^0\bar K^0)-\sqrt2\,{\cal A}(\bar B_d^0\!\to\pi^0\pi^0)+\sqrt3\,{\cal A}(\bar B_d^0\!\to\pi^0\eta_8)-\sqrt6\,{\cal A}(\bar B_s^0\!\to K^0\eta_8)=0\,,}\label{eq:usd5}
\end{align}
and five in the $b\to s$ sector,
\begin{align}
	&{\cal A}(\bar B_d^0\!\to K^-\pi^+) = {\cal A}(\bar B_s^0\!\to K^+K^-)-{\cal A}(\bar B_s^0\!\to\pi^+\pi^-)\,,\label{eq:su10}\\
	&\sqrt3\,{\cal A}(B^-\!\to K^-\eta_8) = {\cal A}(B^-\!\to K^-\pi^0)-\sqrt2\,{\cal A}(B^-\!\to\bar K^0\pi^-)\,,\label{eq:su9}\\
	&{\cal A}(\bar B_d^0\!\to\bar K^0\pi^0) = \sqrt3\,{\cal A}(\bar B_d^0\!\to\bar K^0\eta_8)\,,\label{eq:su8}\\
	&\textcolor{oceanblue}{\sqrt3\,{\cal A}(\bar B_s^0\!\to\pi^0\pi^0)-\sqrt2\,{\cal A}(\bar B_s^0\!\to\pi^0\eta_8)-\sqrt3\,{\cal A}(\bar B_s^0\!\to\eta_8\eta_8)-4\,{\cal A}(\bar B_d^0\!\to\bar K^0\eta_8)=0\,,}\label{eq:su12}\\
	&\textcolor{oceanblue}{{\cal A}(\bar B_s^0\!\to K^0\bar K^0)-\sqrt2\,{\cal A}(\bar B_s^0\!\to\pi^0\pi^0)+\sqrt3\,{\cal A}(\bar B_s^0\!\to\pi^0\eta_8)+\sqrt6\,{\cal A}(\bar B_d^0\!\to\bar K^0\eta_8)=0\,.}\label{eq:uss5}
\end{align}
The remaining six relations are
\begin{align}
	&{\cal A}(B^-\!\to\pi^-\pi^0) = \tfrac{1}{\sqrt2}{\cal A}(\bar B_d^0\!\to\pi^+\pi^-)-\tfrac{1}{\sqrt2}{\cal A}(\bar B_d^0\!\to K^+K^-)+{\cal A}(\bar B_s^0\!\to K^0\pi^0)\,,\label{eq:su2}\\
	&\sqrt2\,{\cal A}(B^-\!\to K^-\pi^0)-{\cal A}(B^-\!\to\bar K^0\pi^-)
	= r_u\,\sqrt2\,{\cal A}(B^-\!\to\pi^-\pi^0)\,,\label{eq:us1}\\
	&\textcolor{oceanblue}{{\cal A}(\bar B_d^0\!\to\pi^+\pi^-)-{\cal A}(\bar B_d^0\!\to K^0\bar K^0)-\sqrt2\,{\cal A}(B^-\!\to\pi^-\pi^0)
	= \frac{\sqrt3}{r_u}\,{\cal A}(\bar B_s^0\!\to\pi^0\eta_8)\,,}\label{eq:us2}\\
	&\textcolor{oceanblue}{{\cal A}(\bar B_d^0\!\to K^+K^-)-{\cal A}(\bar B_d^0\!\to K^0\bar K^0)+{\cal A}(B^-\!\to K^0K^-)=3\sqrt2\,{\cal A}(B^-\!\to\pi^-\pi^0)\,,}\label{eq:su3}\\
	&\textcolor{oceanblue}{3{\cal A}(B^-\!\to\bar K^0\pi^-)-2\sqrt2\,{\cal A}(B^-\!\to K^-\pi^0)-{\cal A}(\bar B_d^0\!\to K^-\pi^+)+\sqrt3\,{\cal A}(\bar B_s^0\!\to\pi^0\eta_8)=0\,,}\label{eq:su11}\\
	&\textcolor{oceanblue}{r_c\big[3\sqrt2\,{\cal A}(\bar B_d^0\!\to\pi^0\pi^0)+{\cal A}(\bar B_d^0\!\to K^0\bar K^0)+2{\cal A}(B^-\!\to K^0K^-)\big]}
	\nonumber\\
	&\qquad\qquad\textcolor{oceanblue}{= 2{\cal A}(B^-\!\to\bar K^0\pi^-)+4{\cal A}(\bar B_s^0\!\to K^0\bar K^0)+3\sqrt3\,{\cal A}(\bar B_s^0\!\to\pi^0\eta_8)\,,}\label{eq:us3}
\end{align}
Ten of the nineteen relations are present in the literature, either directly or as trivial rearrangements of published relations~\cite{Gronau:1990ka,Nir:1991cu,Gronau:1991dq,Lipkin:1991st,Lipkin:1990us,Savage:1989ub,Gronau:1994bn,Gronau:1994rj,Deshpande:1995vc,Gronau:1995ng,Dighe:1995gq,Lipkin:1997ad,Feldmann:1998vh,Grossman:2003qp,Buras:2004ub,Raz:2005hu,Grinstein:1996us,Fleischer:2007hj,Grossman:2013lya}. The $\pi\pi$ triangle in eq.\,\eqref{eq:iso1}, the $K\pi$ quadrangle in eq.\,\eqref{eq:iso3}, and the $B_s$ mode relation in eq.\,\eqref{eq:iso2} are well known isospin relations~\cite{Gronau:1990ka,Nir:1991cu,Gronau:1991dq,Lipkin:1991st, Gronau:1994rj}. 

Among the U-spin relations, eq.\,\eqref{eq:su8} is the proportionality underlying the $K\pi^0$--$K\eta$--$K\eta'$ relations of Refs.~\cite{Lipkin:1990us,Lipkin:1997ad,Grossman:2003qp,Raz:2005hu}, eq.\,\eqref{eq:su9} is the $\{\pi^+K^0,\pi^0K^+,\eta_8K^+\}$ triangle of Refs.~\cite{Deshpande:1995vc,Gronau:1995ng,Dighe:1995gq}, and eqs.\,\eqref{eq:su1}, \eqref{eq:su7} and \eqref{eq:su10} are contained in the $SU(3)$ decompositions of Refs.~\cite{Savage:1989ub,Grinstein:1996us,Dighe:1995gq,Raz:2005hu,Buras:2004ub,Fleischer:2007hj}, as is eq.\,\eqref{eq:su2}. Eq.\,\eqref{eq:us1} is the relation between the $\Delta I=3/2$ $K\pi$ amplitude and $B^-\to\pi^-\pi^0$ with the tree ratio $r_u$~\cite{Gronau:1994bn}. Eqs.\,\eqref{eq:su1}, \eqref{eq:su10} and \eqref{eq:su2} are typically omitted from phenomenological analyses because they involve the exchange- and annihilation-populated modes $\bar B_d\to K^+K^-$ and $\bar B_s\to\pi^+\pi^-$, or the neutral kaon mode $\bar B_s\to K^0\pi^0$. They are nevertheless genuine null directions.

Among the six remaining relations, four of them, namely eqs.\,\eqref{eq:su2}, \eqref{eq:us1}, \eqref{eq:us2} and \eqref{eq:su3}, hold for $SU(3)$-symmetry, not for isospin or U-spin alone. The remaining three relations eqs.\,\eqref{eq:su3}, \eqref{eq:us3}, and eq.\,\eqref{eq:su11} are specific to the leading-order chiral operator, and they have no counterpart in $SU(3)$-based analyses. That these relations are absent from the literature is not an oversight but a structural consequence of how the conventional analyses are set up. We discuss further on this point in Sec.~\ref{sec:comparison}.

%%%%%%%%%%%%%%%%%%%%%%%%%%%%%%%%%%%%%%%%%%%%%%%%
\section{Application to $B\to D P$}
\label{sec:4}
%%%%%%%%%%%%%%%%%%%%%%%%%%%%%%%%%%%%%%%%%%%%%%%%
We now consider two-body decays that contain a charmed meson in the final state. Unlike the charmless modes discussed previously, the process $B\to DP$ involves one initial and one final heavy-light meson together with a light pseudoscalar. The relevant quark-level transitions are $b\to c\bar uq$ and $b\to u\bar cq$ ($q=d,s$), producing the final states $D$ and $\bar D$, respectively.
%%%%%%%%%%%%%%%%%%%%%%%%%%%%%%%%%%%%%%%%%%%%%%%%
\subsection{Effective operators for $B\to D P$}
%%%%%%%%%%%%%%%%%%%%%%%%%%%%%%%%%%%%%%%%%%%%%%%%%%%%%%
The operators contributing to $B\to DP$ are listed in Table~\ref{tab:BDoplist}. The operators $O_{5a}$, $O_{5b}$, and $O_{6a}$ with the spurion combination $\widehat Q_{qq}, \widehat Q_{QQ}^\dagger$, contribute to the $b\to c\bar uq$ class. Whereas $O_{7a}$, $O_{7b}$, and $O_{8a,\ldots,d}$, with the spurion combination $\{\hQ_{qQ},\hQ_{Qq}^\dagger\}$, generate the $b\to u\bar cq$ transitions. The spurion $\widehat Q$ assignments follow from the underlying quark level transition~\cite{Chakraborty:2026vdz}. For a charm meson in the final state, there should be only one $\widehat Q$ with a heavy-quark index corresponding to a $c$ quark etc.

%%%%%%%%%%%%%%%%%%%%%%%%%%%%%%%%%%%%%%%%%%%%%%%%%%
\begin{table}[h]
\centering
\renewcommand{\arraystretch}{1.25}
\resizebox{.95\textwidth}{!}{%
\begin{tabular}{p{0.2\textwidth} p{0.2\textwidth} p{0.12\textwidth} p{0.50\textwidth}}
\toprule
\textbf{Field content} & \textbf{Non-redundant operators} & \textbf{Operator} & \textbf{Covariant form} \\
\midrule
\multirow{2}{=}{ \(\Sigma^\dagger,\Sigma,\Phi^\dagger,\Phi\)\\
 \(\{\hQ_{QQ}^\dagger,\hQ_{qq}\}\)}
& \multirow{2}{*}{2}
& \(O_{5a}\) &
\((\Sigma^\dagger \, \hQ_{qq} T_L^A\, T_L^A \delmu \Sigma)
(\Phi^\dagger \,\hQ_{QQ}^\dagger \tau_L^I \, \tau_L^I \delmu \Phi)\)
\\[2ex]
&& \(O_{5b}\) &
\((\Sigma^\dagger \, \hQ_{qq} T_L^A\, T_L^A T_R^B\delmu \Sigma)
(\Phi^\dagger \,\hQ_{QQ}^\dagger \tau_L^I \, \tau_L^I T_R^B \delmu \Phi)\)
\\
\midrule 
\multirow{2}{=}{ \(\Sigma^\dagger_H,\Sigma,\Phi^\dagger,\Phi_c\)\\
 \(\{\hQ_{QQ}^\dagger,\hQ_{qq}\}\)}
& \multirow{1}{*}{1} & $O_{6a}$ &\((\Sigma^\dagger \langle \hQ_{qq}  T_L^A\rangle T_L^A \delmu \Phi_c)
(\Sigma_H^\dagger \,\hQ_{QQ}^\dagger \tau_L^I \, \tau_L^I \delmu \Phi)\)\\
&&&\\
\midrule 
\midrule
\multirow{2}{=}{ \(\Sigma^\dagger,\Sigma^\dagger, \Phi,\Phi\)\\
 \(\{\hQ_{qQ},\hQ_{Qq}^\dagger\}\)}
& \multirow{2}{*}{2} & $O_{7a}$ &  \((\Sigma^\dagger  T_L^A\hQ_{qQ} \tau_L^X \, T_R^B \delmu \Phi)
(\Sigma^\dagger T_L^A\hQ_{Qq}^\dagger \tau_L^X \,T_R^B \delmu \Phi)\) \\
&&$O_{7b}$ & \((\Sigma^\dagger  \hQ_{qQ} \delmu \Phi)
(\Sigma^\dagger \hQ_{Qq}^\dagger \delmu \Phi)\)\\
%&&$O_{7b}$ & \((\Sigma^\dagger  \hQ_{qQ} T_R^B \delmu \Phi)
%(\Sigma^\dagger \hQ_{Qq}^\dagger  T_R^B \delmu \Phi)\)\\
\midrule
\multirow{4}{=}{ \(\Sigma^\dagger,\Sigma_H, \Phi_c^\dagger,\Phi\)\\
 \(\{\hQ_{qQ},\hQ_{Qq}^\dagger\}\)}
& \multirow{4}{*}{4} & $O_{8a}$ &  \((\Phi_c^\dagger  \hQ_{qQ} \, \delmu \Phi)
(\Sigma^\dagger \hQ_{Qq}^\dagger  \delmu \Sigma_H)\) \\
&&$O_{8b}$  &  \((\Phi_c^\dagger  \hQ_{qQ} \tau_L^X\, \delmu \Phi)
(\Sigma^\dagger \hQ_{Qq}^\dagger \tau_L^X \delmu \Sigma_H)\) \\
%&&$O_{7b}$ & \((\Sigma^\dagger  \hQ_{qQ} T_R^B \delmu \Phi)
%(\Sigma^\dagger \hQ_{Qq}^\dagger  T_R^B \delmu \Phi)\)\\
&&$O_{8c}$  &  \((\Phi_c^\dagger  T_L^A \,\hQ_{qQ} \, \delmu \Phi)
(\Sigma^\dagger T_L^A\hQ_{Qq}^\dagger  \delmu \Sigma_H)\) \\
&&$O_{8d}$  &  \((\Phi_c^\dagger  T_L^A \,\hQ_{qQ} \tau_L^X\, \delmu \Phi)
(\Sigma^\dagger T_L^A\hQ_{Qq}^\dagger \tau_L^X \delmu \Sigma_H)\) \\
\bottomrule
\bottomrule
\end{tabular}}
\caption{\label{tab:BDoplist}Operator list in covariant form contributing to $B\to D P$ where $P$ is a light meson.}
\end{table}
%%%%%%%%%%%%%%%%%%%%%%%%%%%%%%%%%%%%%%%%%%%%%%%%%%

%%%%%%%%%%%%%%%%%%%%%%%%%%%%%%%%%%%%%%%%%%%%%%%%%%
\subsection{Amplitudes and sum rules for $B\to D P$}
%%%%%%%%%%%%%%%%%%%%%%%%%%%%%%%%%%%%%%%%%%%%%%%%%%%
We now project the effective operators onto the physical decay amplitudes for $B\to DP$ with $P\in \{\pi, K, \eta_8\}$ similar to Sec.~\ref{sec:BPP_amp}. For illustration, consider the color-allowed mode $\bar B_d^0\to D_d^+\pi^-$, induced by $b\to c\bar u d$: the heavy-flavor transition is selected by $(\hQ_{QQ})_{c\bar b}=V_{\bar bc}\equiv V_{cb}$ and the light-flavor transition by $(\hQ_{qq})_{d\bar u}=V_{d\bar u}\equiv V_{ud}^\ast$, so that only $O_{5a}$, $O_{5b}$, and $O_{6a}$ contribute. Expanding these operators in component meson fields and selecting the term proportional to $D_d^+\,\pi^-\,\bar B_d^0$ gives the corresponding physical decay amplitude.

As a result,
\begin{align}
	\Sigma=\frac{f_\pi}{2}\exp\left(\frac{2iM}{f_\pi}\right)
	=\frac{f_\pi}{2}+iM+\mathcal O(M^2)\;,
\end{align}
the light bilinear gives (with $\partial_\mu\to-ip_\mu$ for incoming and $+ip_\mu$ for outgoing fields)
\begin{align}
	\langle\pi^-(p_\pi)|\Sigma^\dagger\hQ_{qq}^\dagger
	\overleftrightarrow{\partial}_\mu\Sigma|0\rangle
	\longrightarrow -f_\pi V_{ud}^\ast\,p_{\pi\mu}\;,
\end{align}
the mesonic analogue of $\langle\pi^-|\bar d\gamma_\mu\gamma_5 u|0\rangle=if_\pi p_{\pi\mu}$.
Similarly, with $\bar B_d^0=\Phi_{2,2}$ and $D_d^+=(\Phi^\dagger)_{2,1}$, the heavy bilinear gives
\begin{align}
	\langle D_d^+(p_D)|\Phi^\dagger\hQ_{QQ}^\dagger
	\overleftrightarrow{\partial}^{\mu}\Phi|\bar B_d^0(p_B)\rangle
	=-iV_{cb}\,(p_B+p_D)^\mu\;.
	\label{eq:heavybilinear}
\end{align}
This structure fixes the form-factor content of the EFT vertex. Decomposing the QCD current as~\cite{Wirbel:1985ji, Bauer:1986bm}
\begin{align}
	\langle D_d^+|\bar c\gamma^\mu b|\bar B_d^0\rangle
	=F_+(q^2)(p_B+p_D)^\mu+F_-(q^2)\,q^\mu\;,
	\qquad q=p_B-p_D=p_\pi\;,
\end{align}
the two-derivative operators produce only the $(p_B+p_D)^\mu$ invariant of eq.\,\eqref{eq:heavybilinear}, and thus map onto $F_+$. The $F_-$ structure contracts with the pion momentum as $q\cdot p_\pi=q^2=m_\pi^2$
and is chirally suppressed. Equivalently, in terms of the scalar form factor $(m_B^2-m_D^2)F_0(q^2)=(m_B^2-m_D^2)F_+(q^2)+q^2F_-(q^2)$, the amplitude is governed by $F_0(m_\pi^2)=F_+(m_\pi^2)+\mathcal
O(m_\pi^2)$.

Assembling the operator contributions from Table~\ref{tab:BDPmode1}, all three operators generate the single kinematic invariant $(p_B+p_D)\cdot p_\pi$,
\begin{align}
	{\cal A}(\bar B_d^0\to D_d^+\pi^-)
	=i\,V_{d\bar u}V_{\bar bc}
	\left[8C_{5a}+\frac{8}{3}C_{5b}
	+4\frac{\lambda_b}{f_\pi}C_{6a}\right]
	(p_B+p_D)\cdot p_\pi\;,
	\label{eq:BDpimom}
\end{align}
in contrast to the charmless case eq.\,\eqref{eq:Bd-pipi-amplitude-expanded}, where three independent momentum structures appear. Momentum conservation gives $(p_B+p_D)\cdot p_\pi=m_B^2-m_D^2$ exactly, so that
\begin{align}
	{\cal A}(\bar B_d^0\to D_d^+\pi^-)
	=i\,V_{d\bar u}V_{\bar bc}\,(m_B^2-m_D^2)\,\widetilde C^{D\pi}\;,
	\qquad
	\widetilde C^{D\pi}=8C_{5a}+\frac{8}{3}C_{5b}
	+4\frac{\lambda_b}{f_\pi}C_{6a}\;.
\end{align}

For comparison, the conventional factorized expression is~\cite{Beneke:2000ry,Neubert:1997uc}
\begin{align}
	{\cal A}_{\rm fact}(\bar B_d^0\to D_d^+\pi^-)
	=i\frac{G_F}{\sqrt2}\,V_{ud}^\ast V_{cb}\,
	a_1 f_\pi\,(m_B^2-m_D^2)\,F_0^{B\to D}(m_\pi^2)\;,
\end{align}
where $a_1$ is the effective coefficient mentioned in Sec:~\ref{sec:BPP_amp}. Matching the two expressions gives
\begin{align}
	\widetilde C^{D\pi}
	\longleftrightarrow
	\frac{G_F}{\sqrt2}\,a_1 f_\pi\,F_0^{B\to D}(m_\pi^2)\;.
\end{align}
This provides a mapping between the conventional factorization approach and the present mesonic EFT. The coefficient $\widetilde C^{D\pi}$ which is a combinations of the Wilson coefficients of the effective operators of Table~\ref{tab:BDoplist}, is an effective low-energy coupling absorbing both the factorized contribution and possible non-factorizable effects.

The $B\to DP$ amplitudes separate into two independent classes. The transition $b\to c\bar uq$, producing a charmed meson $D$, is generated by $O_{5a}$, $O_{5b}$, and $O_{6a}$ through the spurion pair ${\hQ_{QQ},\hQ_{qq}^\dagger}$, while the CKM-suppressed transition $b\to u\bar cq$, producing an anti-charmed meson $\bar D$, is generated by $O_{7a}$, $O_{7b}$, and $O_{8a,\ldots,d}$ through ${\hQ_{qQ},\hQ_{Qq}^\dagger}$. Since the two classes involve distinct operators and final states, they are analyzed separately in Secs.~\ref{sec:BDPmode1} and \ref{sec:BDPmode2}, respectively.

\subsubsection{With $b\to c \bar u q$ transition} \label{sec:BDPmode1}
%%%%%%%%%%%%%%%%%%%%%%%%%%%%%%%%%%%%%%%%%%%%%%%%%
In Table~\ref{tab:BDPmode1}, we suppress the common two-body momentum dependence, and display only the relative operator coefficients. The resulting matrices should therefore be interpreted as the flavor and operator-coefficient part of the full amplitudes. With 14 modes and contributions from 3 independent operators and $\mathrm{rank}(M)=3$, there are 11 relations among the amplitudes. In this class every mode carries a single CKM factor, $V_{cb}V_{ud}^*$ for $b\to c\bar u d$ or $V_{cb}V_{us}^*$ for $b\to c\bar u s$, so that attaching the CKM factors leaves the rank unchanged.

%%%%%%%%%%%%%%%%%%%%%%%%%%%%%%%%%%%%%%%%%%%%%%%%%%
\begin{table}
\centering
\[
\resizebox{.5\textwidth}{!}{$
	\left(
	\begin{array}{c|c|ccc}
				{\rm Amplitude} & {\rm CKM} & O_{5a} & O_{5b} & O_{6a}\,(\times\lambda_b/f_\pi) \\
				\hline
				{\cal A}\!\left(B_u^- \to D_u\,\pi^-\right)      & V_{d\bar u}V_{\bar b c} & 8 & \dfrac{8}{3} & 0 \\[0.7em]
				{\cal A}\!\left(\bar B_d^0 \to D_d^+\,\pi^-\right)& V_{d\bar u}V_{\bar b c} & 8 & \dfrac{8}{3} & 4 \\[0.7em]
				{\cal A}\!\left(\bar B_d^0 \to D_u\,\pi^0\right)  & V_{d\bar u}V_{\bar b c} & 0 & 0 & 2\sqrt2 \\[0.7em]
				{\cal A}\!\left(\bar B_s^0 \to D_s^+\,\pi^-\right)& V_{d\bar u}V_{\bar b c} & 8 & -\dfrac{16}{3} & 0 \\[0.7em]
				{\cal A}\!\left(\bar B_s^0 \to D_u\,\pi^0\right)  & V_{\bar b c}V_{s\bar u} & 0 & 4\sqrt2 & 2\sqrt2 \\[0.7em]
				{\cal A}\!\left(\bar B_s^0 \to D_d^+\,\pi^-\right)& V_{\bar b c}V_{s\bar u} & 0 & 8 & 4 \\[0.7em]
				{\cal A}\!\left(B_u^- \to D_u\,K^-\right)         & V_{\bar b c}V_{s\bar u} & 8 & \dfrac{8}{3} & 0 \\[0.7em]
				{\cal A}\!\left(\bar B_d^0 \to D_d^+\,K^-\right)  & V_{\bar b c}V_{s\bar u} & 8 & -\dfrac{16}{3} & 0 \\[0.7em]
				{\cal A}\!\left(\bar B_d^0 \to D_u\,\bar K^0\right)& V_{\bar b c}V_{s\bar u} & 0 & 8 & 0 \\[0.7em]
				{\cal A}\!\left(\bar B_d^0 \to D_s^+\,K^-\right)  & V_{d\bar u}V_{\bar b c} & 0 & 8 & 4 \\[0.7em]
				{\cal A}\!\left(\bar B_s^0 \to D_s^+\,K^-\right)  & V_{\bar b c}V_{s\bar u} & 8 & \dfrac{8}{3} & 4 \\[0.7em]
				{\cal A}\!\left(\bar B_s^0 \to D_u\,K^0\right)    & V_{d\bar u}V_{\bar b c} & 0 & 8 & 0 \\[0.7em]
				{\cal A}\!\left(\bar B_d^0 \to D_u\,\eta_8\right) & V_{d\bar u}V_{\bar b c} & 0 & \dfrac{8\sqrt6}{3} & \dfrac{2\sqrt6}{3} \\[0.7em]
				{\cal A}\!\left(\bar B_s^0 \to D_u\,\eta_8\right) & V_{\bar b c}V_{s\bar u} & 0 & -\dfrac{4\sqrt6}{3} & \dfrac{2\sqrt6}{3} \\
			\end{array}
	\right)
	$}
\]
\caption{Amplitudes for $B\to D P$ modes mediated via $b\to c \bar u q$.}
\label{tab:BDPmode1}
\end{table}
%%%%%%%%%%%%%%%%%%%%%%%%%%%%%%%%%%%%%%%%%%%%%%%%%%

The iso-spin relations are
\begin{align}
	{\cal A}(B^-_u\to D_u\pi^-)+\sqrt2\,{\cal A}(\bar B^0_d\to D_u\pi^0)-{\cal A}(\bar B^0_d\to D^+_d\pi^-)&=0\,,\label{eq:DPiso1}\\
	{\cal A}(B^-_u\to D_u K^-)-{\cal A}(\bar B^0_d\to D_u\bar K^0)-{\cal A}(\bar B^0_d\to D^+_d K^-)&=0\,,\label{eq:DPiso2}\\
	{\cal A}(\bar B^0_s\to D_u\pi^0)-\frac{1}{\sqrt2}\,{\cal A}(\bar B^0_s\to D^+_d\pi^-)&=0\,.\label{eq:DPiso3}
\end{align}
The U-spin relations are
\begin{align}
	{\cal A}(B^-_u\to D_u K^-)&=r_u\,{\cal A}(B^-_u\to D_u\pi^-)\,,\label{eq:DPus1}\\
	{\cal A}(\bar B^0_s\to D^+_s K^-)&=r_u\,{\cal A}(\bar B^0_d\to D^+_d\pi^-)\,,\label{eq:DPus2}\\
	{\cal A}(\bar B^0_d\to D^+_d K^-)&=r_u\,{\cal A}(\bar B^0_s\to D^+_s\pi^-)\,,\label{eq:DPus3}
\end{align}
\begin{align}
	{\cal A}(\bar B^0_d\to D^+_s K^-)&={\cal A}(\bar B^0_d\to D^+_d\pi^-)-{\cal A}(\bar B^0_s\to D^+_s\pi^-)\,,\label{eq:DPus4}\\
	{\cal A}(\bar B^0_s\to D^+_s K^-)&={\cal A}(\bar B^0_s\to D^+_d\pi^-)+{\cal A}(\bar B^0_d\to D^+_d K^-)\,,\label{eq:DPus5}\\
	\textcolor{oceanblue}{{\cal A}(\bar B^0_s\to D_u\pi^0)}&\textcolor{oceanblue}{=\sqrt2\,{\cal A}(\bar B^0_d\to D_u\bar K^0)+\sqrt3\,{\cal A}(\bar B^0_s\to D_u\eta_8)\,.}\label{eq:DPus6}
\end{align}
The full $SU(3)$ projections are
\begin{align}
	{\cal A}(B^-_u\to D_u\pi^-)&={\cal A}(\bar B^0_s\to D^+_s\pi^-)+{\cal A}(\bar B^0_s\to D_u K^0)\,,\label{eq:DPsu1}\\
	\textcolor{oceanblue}{{\cal A}(\bar B^0_d\to D_u\eta_8)}&\textcolor{oceanblue}{=\frac{\sqrt6}{3}\,{\cal A}(B^-_u\to D_u\pi^-)+\frac{\sqrt3}{3}\,{\cal A}(\bar B^0_d\to D_u\pi^0)-\frac{\sqrt6}{3}\,{\cal A}(\bar B^0_s\to D^+_s\pi^-)\,.}\label{eq:DPsu2}
\end{align}
Eqs.~\eqref{eq:DPiso1}--\eqref{eq:DPiso3} are the isospin relations. The $D\pi$ and $DK$ triangles and the $\bar B^0_s\to D\pi$ relation are familiar from the $I=\tfrac12,\tfrac32$ analyses of $B\to D\pi$ decays~\cite{Neubert:2001sj,Chiang:2002tv,Colangelo:2005hh}. Eqs.~\eqref{eq:DPus1}--\eqref{eq:DPus6} are the U-spin relations. Eqs.~\eqref{eq:DPus1}--\eqref{eq:DPus3} connect the two transition sectors and carry $r_u$. They are the standard relations connecting the $DK$ modes to the $D\pi$ ones~\cite{Chiang:2002tv,Chiang:2007bd}. Eqs.~\eqref{eq:DPus4}--\eqref{eq:DPus6} stay within one sector and are free of CKM factors.  The first two of them (eqs.~\eqref{eq:DPus4}--\eqref{eq:DPus5}) are implicit in the topological tabulations~\cite{Colangelo:2005hh,Chiang:2007bd},though not quoted as sum rules. Eqs.~\eqref{eq:DPsu1}--\eqref{eq:DPsu2} are not individually isospin or U-spin relations rather their combinations~\footnote{With different choice of basis, it is possible to write all the eleven relations as isospin and U-spin relations only.}.

The two relations in blue, eqs.~\eqref{eq:DPus6} and \eqref{eq:DPsu2}, are those involving an $\eta_8$ in the final state. By the criterion of Sec.~\ref{sec:BPP_amp}, they are derivable from the published $SU(3)$ tabulations but have not been written down. When expressed in terms of the
mass eigenstates, both become relations among the measured $B\to D\eta^{(\prime)}$, $D\pi$ and $DK$ rates in which the singlet mixing enters as a correction. 
With the precision anticipated at Belle~II~\cite{Belle-II:2018jsg} for $\bar B^0_d\to D^0\eta^{(\prime)}$ and the $\bar B^0_s\to D^0\bar K^0$ modes at LHCb~\cite{LHCb:2016lky}, any violation of these relations measures $SU(3)$ breaking and singlet contamination.

Since $B\to DP$ involves a single light meson, the general $SU(3)$ decomposition contains exactly three reduced amplitudes. The three operators of Table~\ref{tab:BDoplist} map onto these reduced amplitudes. Unlike the charmless case of Sec.~\ref{sec:BPP_amp}, the leading-order chiral truncation adds no constraint here and all eleven relations are flavor-symmetry identities.

%%%%%%%%%%%%%%%%%%%%%%%%%%%%%%%%%%%%%%%%%%%%%%%%
\subsubsection{With $b\to u\bar c q$ transition}\label{sec:BDPmode2}
%%%%%%%%%%%%%%%%%%%%%%%%%%%%%%%%%%%%%%%%%%%%%%%%
%%%%%%%%%%%%%%%%%%%%%%%%%%%%%%%%%%%%%%%%%%%%%%%%%%
\begin{table}
	\centering
	\[
	\resizebox{.6\textwidth}{!}{$
		\left(\begin{array}{c|c|cc|cccc}
			& & & & \multicolumn{4}{c}{(\propto\,\lambda_b/f_\pi\, , \, \lambda_c/f_\pi)} \\
			{\rm Amplitude} & {\rm CKM} & O_{7a} & O_{7b} & O_{8a} & O_{8b} & O_{8c} & O_{8d} \\
			\hline
			{\cal A}\!\left(B^-_u \to \bar D_u\pi^-\right)          & V_{\bar b u}V_{d\bar c} & -\frac{40}{9} & 0 & 0 & 0 & 2 & -2 \\
			{\cal A}\!\left(B^-_u \to D^-_d\pi^0\right)             & V_{\bar b u}V_{d\bar c} & \frac{16\sqrt2}{9} & \frac{1}{\sqrt2} & 0 & 0 & -\sqrt2 & \sqrt2 \\
			{\cal A}\!\left(B^-_u \to D^-_s\pi^0\right)             & V_{\bar b u}V_{s\bar c} & -\frac{14\sqrt2}{9} & \frac{1}{\sqrt2} & 0 & 0 & 0 & 0 \\
			{\cal A}\!\left(\bar B^0_d \to D^-_d\pi^+\right)        & V_{\bar b u}V_{d\bar c} & \frac{56}{9} & 1 & 1 & -1 & -\frac{2}{3} & \frac{2}{3} \\
			{\cal A}\!\left(\bar B^0_d \to D^-_s\pi^+\right)        & V_{\bar b u}V_{s\bar c} & -\frac{28}{9} & 1 & 0 & 0 & 0 & 0 \\
			{\cal A}\!\left(\bar B^0_d \to \bar D_u\pi^0\right)     & V_{\bar b u}V_{d\bar c} & \frac{32\sqrt2}{9} & 0 & \frac{1}{\sqrt2} & -\frac{1}{\sqrt2} & -\frac{\sqrt2}{3} & \frac{\sqrt2}{3} \\
			{\cal A}\!\left(\bar B^0_s \to \bar D_u\pi^0\right)     & V_{\bar b u}V_{s\bar c} & \frac{14\sqrt2}{3} & 0 & \frac{1}{\sqrt2} & -\frac{1}{\sqrt2} & -\frac{\sqrt2}{3} & \frac{\sqrt2}{3} \\
			{\cal A}\!\left(\bar B^0_s \to D^-_d\pi^+\right)        & V_{\bar b u}V_{s\bar c} & \frac{28}{3} & 0 & 1 & -1 & -\frac{2}{3} & \frac{2}{3} \\
			{\cal A}\!\left(B^-_u \to \bar D_uK^-\right)            & V_{\bar b u}V_{s\bar c} & -\frac{40}{9} & 0 & 0 & 0 & 2 & -2 \\
			{\cal A}\!\left(B^-_u \to D^-_d\bar K^0\right)          & V_{\bar b u}V_{s\bar c} & -\frac{20}{3} & 0 & 0 & 0 & 2 & -2 \\
			{\cal A}\!\left(B^-_u \to D^-_sK^0\right)               & V_{\bar b u}V_{d\bar c} & -\frac{20}{3} & 0 & 0 & 0 & 2 & -2 \\
			{\cal A}\!\left(\bar B^0_d \to \bar D_u\bar K^0\right)  & V_{\bar b u}V_{s\bar c} & \frac{20}{9} & 0 & 0 & 0 & 0 & 0 \\
			{\cal A}\!\left(\bar B^0_d \to D^-_sK^+\right)          & V_{\bar b u}V_{d\bar c} & \frac{28}{3} & 0 & 1 & -1 & -\frac{2}{3} & \frac{2}{3} \\
			{\cal A}\!\left(\bar B^0_s \to D^-_d K^+\right)         & V_{\bar b u}V_{d\bar c} & -\frac{28}{9} & 1 & 0 & 0 & 0 & 0 \\
			{\cal A}\!\left(\bar B^0_s \to D^-_s K^+\right)         & V_{\bar b u}V_{s\bar c} & \frac{56}{9} & 1 & 1 & -1 & -\frac{2}{3} & \frac{2}{3} \\
			{\cal A}\!\left(\bar B^0_s \to \bar D_uK^0\right)       & V_{\bar b u}V_{d\bar c} & \frac{20}{9} & 0 & 0 & 0 & 0 & 0 \\
			{\cal A}\!\left(B^-_u \to D^-_d\eta_8\right)            & V_{\bar b u}V_{d\bar c} & -\frac{44\sqrt6}{27} & \frac{1}{\sqrt6} & 0 & 0 & \frac{\sqrt6}{3} & -\frac{\sqrt6}{3} \\
			{\cal A}\!\left(B^-_u \to D^-_s\eta_8\right)            & V_{\bar b u}V_{s\bar c} & \frac{46\sqrt6}{27} & \frac{1}{\sqrt6} & 0 & 0 & -\frac{2\sqrt6}{3} & \frac{2\sqrt6}{3} \\
			{\cal A}\!\left(\bar B^0_d \to \bar D_u\eta_8\right)    & V_{\bar b u}V_{d\bar c} & \frac{52\sqrt6}{27} & 0 & \frac{1}{\sqrt6} & -\frac{1}{\sqrt6} & -\frac{\sqrt6}{9} & \frac{\sqrt6}{9} \\
			{\cal A}\!\left(\bar B^0_s \to \bar D_u\eta_8\right)    & V_{\bar b u}V_{s\bar c} & \frac{22\sqrt6}{27} & 0 & \frac{1}{\sqrt6} & -\frac{1}{\sqrt6} & -\frac{\sqrt6}{9} & \frac{\sqrt6}{9} \\
		\end{array}
		\right)$}
	\]
	\caption{Amplitudes for $B\to \bar D P$ modes mediated via $b\to u\bar c q$.}
	\label{tab:BDPmode2}
\end{table}
%%%%%%%%%%%%%%%%%%%%%%%%%%%%%%%%%%%%%%%%%%%%%%%%%%

The second class corresponds to the CKM-suppressed transition generated by the spurion pair $\{\hQ_{qQ},\hQ_{Qq}^\dagger\}$. 
Contributions from the relevant operators are expressed as $({\cal A}|{\cal M})$ in Table~\ref{tab:BDPmode2}, with the momentum dependence kept implicit as before.
The columns $O_{8a}$ and $O_{8b}$ enter only through the difference $O_{8a}-O_{8b}$, and $O_{8c}$, $O_{8d}$ only through $O_{8c}-O_{8d}$, so that four independent operator structures contribute. 
With 20 modes, $\mathrm{rank}(M)=4$ and there are 16 relations among the amplitudes.

The iso-spin relations are
\begin{align}
	{\cal A}(B^-_u\to\bar D_u\pi^-)+\sqrt2\,{\cal A}(B^-_u\to D^-_d\pi^0)&={\cal A}(\bar B^0_d\to D^-_d\pi^+)-\sqrt2\,{\cal A}(\bar B^0_d\to\bar D_u\pi^0)\,,\label{eq:DbPiso1}\\
	{\cal A}(B^-_u\to\bar D_u K^-)&={\cal A}(B^-_u\to D^-_d\bar K^0)+{\cal A}(\bar B^0_d\to\bar D_u\bar K^0)\,,\label{eq:DbPiso2}\\
	\sqrt2\,{\cal A}(B^-_u\to D^-_s\pi^0)&={\cal A}(\bar B^0_d\to D^-_s\pi^+)\,,\label{eq:DbPiso3}\\
	{\cal A}(\bar B^0_s\to D^-_d\pi^+)&=\sqrt2\,{\cal A}(\bar B^0_s\to\bar D_u\pi^0)\,.\label{eq:DbPiso4}
\end{align}
The U-spin relations are
\begin{align}
	{\cal A}(B^-_u\to\bar D_u K^-)&=r_c\,{\cal A}(B^-_u\to\bar D_u\pi^-)\,,\label{eq:DbPus1}\\
	{\cal A}(\bar B^0_s\to D^-_s K^+)&=r_c\,{\cal A}(\bar B^0_d\to D^-_d\pi^+)\,,\label{eq:DbPus2}\\
	\sqrt2\,{\cal A}(B^-_u\to D^-_s\pi^0)-{\cal A}(B^-_u\to D^-_d\bar K^0)&=r_c\,\sqrt2\,{\cal A}(B^-_u\to D^-_d\pi^0)\,,\label{eq:DbPus3}\\
	\sqrt2\,{\cal A}(B^-_u\to D^-_s\pi^0)&=r_c\,\big[\sqrt2\,{\cal A}(B^-_u\to D^-_d\pi^0)+{\cal A}(B^-_u\to D^-_s K^0)\big]\,,\label{eq:DbPus4}\\
	\textcolor{oceanblue}{2\,{\cal A}(B^-_u\to D^-_s\pi^0)}&\textcolor{oceanblue}{=r_c\,\big[{\cal A}(B^-_u\to D^-_d\pi^0)+\sqrt3\,{\cal A}(B^-_u\to D^-_d\eta_8)\big]\,,}\label{eq:DbPus5}\\
	\textcolor{oceanblue}{{\cal A}(B^-_u\to D^-_s\pi^0)+\sqrt3\,{\cal A}(B^-_u\to D^-_s\eta_8)}&\textcolor{oceanblue}{=2\,r_c\,{\cal A}(B^-_u\to D^-_d\pi^0)\,.}\label{eq:DbPus6}
\end{align}
The full $SU(3)$ projections are
\begin{align}
	&\sqrt2\,{\cal A}(B^-_u\to D^-_s\pi^0)+\sqrt2\,{\cal A}(\bar B^0_s\to\bar D_u\pi^0)=r_c\,{\cal A}(\bar B^0_d\to D^-_d\pi^+)\,,\label{eq:DbPsu1}\\
	&\sqrt2\,{\cal A}(B^-_u\to D^-_s\pi^0)=r_c\,\big[{\cal A}(\bar B^0_d\to D^-_d\pi^+)-{\cal A}(\bar B^0_d\to D^-_s K^+)\big]\,,\label{eq:DbPsu2}\\
	&\sqrt2\,{\cal A}(B^-_u\to D^-_s\pi^0)=r_c\,{\cal A}(\bar B^0_s\to D^-_d K^+)\,,\label{eq:DbPsu3}\\
	&\sqrt2\,{\cal A}(B^-_u\to D^-_s\pi^0)=r_c\,\big[{\cal A}(B^-_u\to\bar D_u\pi^-)+\sqrt2\,{\cal A}(B^-_u\to D^-_d\pi^0)-{\cal A}(\bar B^0_s\to\bar D_u K^0)\big]\,,\label{eq:DbPsu4}\\
	&\textcolor{oceanblue}{2\sqrt2\,{\cal A}(B^-_u\to D^-_s\pi^0)}\textcolor{oceanblue}{=r_c\,\big[{\cal A}(B^-_u\to\bar D_u\pi^-)+\sqrt2\,{\cal A}(B^-_u\to D^-_d\pi^0)}\nonumber\\
	&\qquad\textcolor{oceanblue}{+{\cal A}(\bar B^0_d\to D^-_d\pi^+)-\sqrt6\,{\cal A}(\bar B^0_d\to\bar D_u\eta_8)\big]\,,}\label{eq:DbPsu5}\\
	&\textcolor{oceanblue}{{\cal A}(B^-_u\to D^-_s\pi^0)-\sqrt3\,{\cal A}(\bar B^0_s\to\bar D_u\eta_8)}\textcolor{oceanblue}{=r_c\,\big[\sqrt2\,{\cal A}(B^-_u\to\bar D_u\pi^-)+2\,{\cal A}(B^-_u\to D^-_d\pi^0)}\nonumber\\
	&\qquad\textcolor{oceanblue}{-\tfrac{1}{\sqrt2}\,{\cal A}(\bar B^0_d\to D^-_d\pi^+)\big]\,.}\label{eq:DbPsu6}
\end{align}
Eqs.~\eqref{eq:DbPiso1}--\eqref{eq:DbPiso4} are the isospin relations. They are, respectively, the $\bar D\pi$ quadrangle, the $\bar DK$ triangle and the two two-mode relations of Ref.~\cite{Gronau:1995hm,Fayyazuddin:2004ac}. Eqs.~\eqref{eq:DbPus1}--\eqref{eq:DbPus6} are the U-spin relations~\cite{Gronau:1995hm, Fleischer:2003yb}. All of them connect the two transition sectors and carry $r_c$. Eqs.~\eqref{eq:DbPsu1}--\eqref{eq:DbPsu6} are neither isospin nor U-spin relations by themselves and each follows only on combining the two. 

The four relations in blue, eqs.~\eqref{eq:DbPus5}, \eqref{eq:DbPus6}, \eqref{eq:DbPsu5} and \eqref{eq:DbPsu6}, are those involving an $\eta_8$ and are rarely quoted previously as the literature works with the physical $\eta,\eta'$ after octet--singlet mixing~\cite{Feldmann:1998vh}.

%%%%%%%%%%%%%%%%%%%%%%%%%%%%%%%%%%%%%%%%%%%
\section{Three body decays}\label{sec:3body}
%%%%%%%%%%%%%%%%%%%%%%%%%%%%%%%%%%%%%%%%%%%
 
We now extend the analysis to three-body $B$-meson decays. The same symmetry-based $D6$ operators contributing to two-body decays also generate local contact interactions for three-body final states. Here we consider only these contact contributions, excluding cascade processes mediated by intermediate resonances. This is complementary to analyses in Ref.~\cite{Bhattacharya:2014eca}, where flavor symmetry is applied directly to the full momentum-dependent amplitudes symmetrized with respect to momentum interchange, and in Ref.~\cite{Wang:2024tnx}, where isospin sum rules were derived in the presence of intermediate resonances. Our sum rules are instead leading-order statements about the non-resonant amplitudes, with resonant substructure entering through operators with additional derivatives. We study both charmless decays, $B\to P_1P_2P_3$, and modes with one charmed meson, $B\to DP_1P_2$ and $B\to \bar DP_1P_2$. Fewer operators contribute than in the two-body case, since operators involving $\Sigma_H$ do not generate the required external states. 

The amplitudes are obtained by expanding the effective operators to the corresponding four-point vertices. Owing to the two-derivative structure of the operators, they depend on scalar products of the external momenta. We keep the flavor structure manifest by expressing the kinematic dependence in terms of the momenta associated with the meson multiplets. When $n$ identical mesons appear in the final state, the corresponding contribution is multiplied by the normalization factor $\sqrt{n!}$.

%%%%%%%%%%%%%%%%%%%%%%%%%%%%%%%%%%%%%%%%%%%%%%%%%%
\begin{table}[h]
	\centering
	\renewcommand{\arraystretch}{1.45}
	\setlength{\tabcolsep}{3pt}
	\resizebox{\textwidth}{!}{%
		\begin{tabular}{p{0.32\textwidth} c c c c c c}
			\toprule
			\textbf{Amplitude}
			& \(\boldsymbol{O_{1a}}\)
			& \(\boldsymbol{O_{1b}}\)
			& \(\boldsymbol{O_{4a}}\)
			& \(\boldsymbol{O_{4b}}\)
			& \(\boldsymbol{O_{4c}}\)
			& \(\boldsymbol{O_{4d}}\) \\
			\midrule
			
			\({\cal A}\!\left(\bar B_d \to \pi^+\pi^-\pi^0\right)\)
			&
			\(2i\sqrt2\,V_{d\bar c}V_{\bar b c}\,X\)
			&
			\(\dfrac{2i\sqrt2}{3}V_{d\bar c}V_{\bar b c}\,X\)
			&
			\(-2i\sqrt2\,V_{d\bar u}V_{\bar b u}\,Y\)
			&
			\(-\dfrac{2i\sqrt2}{3}V_{d\bar u}V_{\bar b u}\,Y\)
			&
			\(-4\sqrt2\,V_{d\bar u}V_{\bar b u}\,Y\)
			&
			\(-\dfrac{8i\sqrt2}{9}V_{d\bar u}V_{\bar b u}\,Y\)
			\\
			
			\({\cal A}\!\left(\bar B_d \to \pi^0\pi^0\pi^0\right)\)
			&
			\(2i\sqrt3\,V_{d\bar c}V_{\bar b c}\,X\)
			&
			\(\dfrac{2i\sqrt3}{3}V_{d\bar c}V_{\bar b c}\,X\)
			&
			\(-2i\sqrt3\,V_{d\bar u}V_{\bar b u}\,Y\)
			&
			\(-\dfrac{2i\sqrt3}{3}V_{d\bar u}V_{\bar b u}\,Y\)
			&
			\(-4\sqrt3\,V_{d\bar u}V_{\bar b u}\,Y\)
			&
			\(-\dfrac{8i\sqrt3}{9}V_{d\bar u}V_{\bar b u}\,Y\)
			\\
			
			\({\cal A}\!\left(\bar B_d \to K^+K^-\pi^0\right)\)
			&
			\(2i\sqrt2\,V_{d\bar c}V_{\bar b c}\,X\)
			&
			\(-\dfrac{4i\sqrt2}{3}V_{d\bar c}V_{\bar b c}\,X\)
			&
			\(-2i\sqrt2\,V_{d\bar u}V_{\bar b u}\,Y\)
			&
			\(\dfrac{4i\sqrt2}{3}V_{d\bar u}V_{\bar b u}\,Y\)
			&
			\(-\dfrac{14\sqrt2}{3}V_{d\bar u}V_{\bar b u}\,Y\)
			&
			\(-\dfrac{26i\sqrt2}{9}V_{d\bar u}V_{\bar b u}\,Y\)
			\\
			
			\({\cal A}\!\left(\bar B_d \to K^0\bar K^0\pi^0\right)\)
			&
			\(2i\sqrt2\,V_{d\bar c}V_{\bar b c}\,X\)
			&
			\(\dfrac{8i\sqrt2}{3}V_{d\bar c}V_{\bar b c}\,X\)
			&
			\(0\)
			&
			\(-\dfrac{8i\sqrt2}{3}V_{d\bar u}V_{\bar b u}\,Y\)
			&
			\(-\dfrac{8\sqrt2}{3}V_{d\bar u}V_{\bar b u}\,Y\)
			&
			\(-\dfrac{8i\sqrt2}{9}V_{d\bar u}V_{\bar b u}\,Y\)
			\\
			
			\({\cal A}\!\left(\bar B_d \to K^+\bar K^0\pi^-\right)\)
			&
			\(0\)
			&
			\(-4i\,V_{d\bar c}V_{\bar b c}\,X\)
			&
			\(0\)
			&
			\(4i\,V_{d\bar u}V_{\bar b u}\,Y\)
			&
			\(-\dfrac{4}{3}V_{d\bar u}V_{\bar b u}\,Y\)
			&
			\(-4i\,V_{d\bar u}V_{\bar b u}\,Y\)
			\\
			
			\({\cal A}\!\left(\bar B_d \to K^-K^0\pi^+\right)\)
			&
			\(0\)
			&
			\(-4i\,V_{d\bar c}V_{\bar b c}\,X\)
			&
			\(-\dfrac{4i}{3}V_{d\bar u}V_{\bar b u}\,Y\)
			&
			\(4i\,V_{d\bar u}V_{\bar b u}\,Y\)
			&
			\(-\dfrac{16}{3}V_{d\bar u}V_{\bar b u}\,Y\)
			&
			\(\dfrac{8i}{9}V_{d\bar u}V_{\bar b u}\,Y\)
			\\
			
			\({\cal A}\!\left(\bar B_d \to K^-\pi^+\pi^0\right)\)
			&
			\(0\)
			&
			\(0\)
			&
			\(-\dfrac{2i\sqrt{2}}{3}V_{\bar b u}V_{s \bar u	}\,Y\)
			&
			\(0\)
			&
			\(-2\sqrt{2}V_{\bar b u}V_{s \bar u	}\,Y\)
			&
			\(\frac{22i\sqrt{2}}{9}V_{\bar b u}V_{s \bar u	}\,Y\)
			\\
			
			\({\cal A}\!\left(\bar B_d \to \bar K^0\pi^+\pi^-\right)\)
			&
			\(-4i\,V_{\bar b c}V_{s\bar c}\,X\)
			&
			\(-\dfrac{4i}{3}V_{\bar b c}V_{s\bar c}\,X\)
			&
			\(\dfrac{8i}{3}V_{\bar b u}V_{s\bar u}\,Y\)
			&
			\(\dfrac{4i}{3}V_{\bar b u}V_{s\bar u}\,Y\)
			&
			\(4V_{\bar b u}V_{s\bar u}\,Y\)
			&
			\(\dfrac{20i}{3}V_{\bar b u}V_{s\bar u}\,Y\)
			\\
			
			\({\cal A}\!\left(\bar B_d \to \bar K^0\pi^0\pi^0\right)\)
			&
			\(-2i\sqrt2\,V_{\bar b c}V_{s\bar c}\,X\)
			&
			\(-\dfrac{2i\sqrt2}{3}V_{\bar b c}V_{s\bar c}\,X\)
			&
			\(2i\sqrt2\,V_{\bar b u}V_{s\bar u}\,Y\)
			&
			\(\dfrac{2i\sqrt2}{3}V_{\bar b u}V_{s\bar u}\,Y\)
			&
			\(4\sqrt2\,V_{\bar b u}V_{s\bar u}\,Y\)
			&
			\(\dfrac{8i\sqrt2}{9}V_{\bar b u}V_{s\bar u}\,Y\)
			\\
			
			\({\cal A}\!\left(\bar B_d \to K^+K^-\bar K^0\right)\)
			&
			\(-4i\,V_{\bar b c}V_{s\bar c}\,X\)
			&
			\(-\dfrac{4i}{3}V_{\bar b c}V_{s\bar c}\,X\)
			&
			\(4i\,V_{\bar b u}V_{s\bar u}\,Y\)
			&
			\(\dfrac{4i}{3}V_{\bar b u}V_{s\bar u}\,Y\)
			&
			\(8V_{\bar b u}V_{s\bar u}\,Y\)
			&
			\(\dfrac{16i}{9}V_{\bar b u}V_{s\bar u}\,Y\)
			\\
			
			\({\cal A}\!\left(\bar B_d \to K^0\bar K^0\bar K^0\right)\)
			&
			\(-4i\sqrt2\,V_{\bar b c}V_{s\bar c}\,X\)
			&
			\(-\dfrac{4i\sqrt2}{3}V_{\bar b c}V_{s\bar c}\,X\)
			&
			\(0\)
			&
			\(\dfrac{4i\sqrt2}{3}V_{\bar b u}V_{s\bar u}\,Y\)
			&
			\(\dfrac{20\sqrt2}{3}V_{\bar b u}V_{s\bar u}\,Y\)
			&
			\(\dfrac{52i\sqrt2}{9}V_{\bar b u}V_{s\bar u}\,Y\)
			\\
			\hline
			
			\({\cal A}\!\left(B^- \to \pi^+\pi^-\pi^-\right)\)
			&
			\(-4i\sqrt2\,V_{d\bar c}V_{\bar b c}\,X\)
			&
			\(-\dfrac{4i\sqrt2}{3}V_{d\bar c}V_{\bar b c}\,X\)
			&
			\(4i\sqrt2\,V_{d\bar u}V_{\bar b u}\,Y\)
			&
			\(\dfrac{4i\sqrt2}{3}V_{d\bar u}V_{\bar b u}\,Y\)
			&
			\(8\sqrt2\,V_{d\bar u}V_{\bar b u}\,Y\)
			&
			\(\dfrac{16i\sqrt2}{9}V_{d\bar u}V_{\bar b u}\,Y\)
			\\
			
			\({\cal A}\!\left(B^- \to \pi^-\pi^0\pi^0\right)\)
			&
			\(-2i\sqrt2\,V_{d\bar c}V_{\bar b c}\,X\)
			&
			\(-\dfrac{2i\sqrt2}{3}V_{d\bar c}V_{\bar b c}\,X\)
			&
			\(2i\sqrt2\,V_{d\bar u}V_{\bar b u}\,Y\)
			&
			\(\dfrac{2i\sqrt2}{3}V_{d\bar u}V_{\bar b u}\,Y\)
			&
			\(4\sqrt2\,V_{d\bar u}V_{\bar b u}\,Y\)
			&
			\(\dfrac{8i\sqrt2}{9}V_{d\bar u}V_{\bar b u}\,Y\)
			\\
			
			\({\cal A}\!\left(B^- \to \pi^-K^+K^-\right)\)
			&
			\(-4i\,V_{d\bar c}V_{\bar b c}\,X\)
			&
			\(-\dfrac{4i}{3}V_{d\bar c}V_{\bar b c}\,X\)
			&
			\(4i\,V_{d\bar u}V_{\bar b u}\,Y\)
			&
			\(\dfrac{4i}{3}V_{d\bar u}V_{\bar b u}\,Y\)
			&
			\(8V_{d\bar u}V_{\bar b u}\,Y\)
			&
			\(\dfrac{16i}{9}V_{d\bar u}V_{\bar b u}\,Y\)
			\\
			
			\({\cal A}\!\left(B^- \to \pi^-\bar K^0K^0\right)\)
			&
			\(-4i\,V_{d\bar c}V_{\bar b c}\,X\)
			&
			\(-\dfrac{4i}{3}V_{d\bar c}V_{\bar b c}\,X\)
			&
			\(0\)
			&
			\(\dfrac{4i}{3}V_{d\bar u}V_{\bar b u}\,Y\)
			&
			\(\dfrac{20}{3}V_{d\bar u}V_{\bar b u}\,Y\)
			&
			\(\dfrac{52i}{9}V_{d\bar u}V_{\bar b u}\,Y\)
			\\
			
			\({\cal A}\!\left(B^- \to \pi^0K^-K^0\right)\)
			&
			\(0\)
			&
			\(0\)
			&
			\(-\dfrac{2i\sqrt2}{3}V_{d\bar u}V_{\bar b u}\,Y\)
			&
			\(0\)
			&
			\(-2\sqrt2\,V_{d\bar u}V_{\bar b u}\,Y\)
			&
			\(\dfrac{22i\sqrt2}{9}V_{d\bar u}V_{\bar b u}\,Y\)
			\\
			
			\({\cal A}\!\left(B^- \to \pi^+\pi^-K^-\right)\)
			&
			\(-4i\,V_{\bar b c}V_{s\bar c}\,X\)
			&
			\(-\dfrac{4i}{3}V_{\bar b c}V_{s\bar c}\,X\)
			&
			\(4i\,V_{\bar b u}V_{s\bar u}\,Y\)
			&
			\(\dfrac{4i}{3}V_{\bar b u}V_{s\bar u}\,Y\)
			&
			\(8V_{\bar b u}V_{s\bar u}\,Y\)
			&
			\(\dfrac{16i}{9}V_{\bar b u}V_{s\bar u}\,Y\)
			\\
			
			\({\cal A}\!\left(B^- \to \pi^-\pi^0\bar K^0\right)\)
			&
			\(0\)
			&
			\(0\)
			&
			\(-\dfrac{2i\sqrt2}{3}V_{\bar b u}V_{s\bar u}\,Y\)
			&
			\(0\)
			&
			\(-2\sqrt2\,V_{\bar b u}V_{s\bar u}\,Y\)
			&
			\(\dfrac{22i\sqrt2}{9}V_{\bar b u}V_{s\bar u}\,Y\)
			\\
			
			\({\cal A}\!\left(B^- \to \pi^0\pi^0K^-\right)\)
			&
			\(-2i\sqrt2\,V_{\bar b c}V_{s\bar c}\,X\)
			&
			\(-\dfrac{2i\sqrt2}{3}V_{\bar b c}V_{s\bar c}\,X\)
			&
			\(\dfrac{4i\sqrt2}{3}V_{\bar b u}V_{s\bar u}\,Y\)
			&
			\(\dfrac{2i\sqrt2}{3}V_{\bar b u}V_{s\bar u}\,Y\)
			&
			\(2\sqrt2\,V_{\bar b u}V_{s\bar u}\,Y\)
			&
			\(\dfrac{10i\sqrt2}{3}V_{\bar b u}V_{s\bar u}\,Y\)
			\\
			
			\({\cal A}\!\left(B^- \to K^+K^-K^-\right)\)
			&
			\(-4i\sqrt2\,V_{\bar b c}V_{s\bar c}\,X\)
			&
			\(-\dfrac{4i\sqrt2}{3}V_{\bar b c}V_{s\bar c}\,X\)
			&
			\(4i\sqrt2\,V_{\bar b u}V_{s\bar u}\,Y\)
			&
			\(\dfrac{4i\sqrt2}{3}V_{\bar b u}V_{s\bar u}\,Y\)
			&
			\(8\sqrt2\,V_{\bar b u}V_{s\bar u}\,Y\)
			&
			\(\dfrac{16i\sqrt2}{9}V_{\bar b u}V_{s\bar u}\,Y\)
			\\
			
			\({\cal A}\!\left(B^- \to K^-\bar K^0K^0\right)\)
			&
			\(-4i\,V_{\bar b c}V_{s\bar c}\,X\)
			&
			\(-\dfrac{4i}{3}V_{\bar b c}V_{s\bar c}\,X\)
			&
			\(0\)
			&
			\(\dfrac{4i}{3}V_{\bar b u}V_{s\bar u}\,Y\)
			&
			\(\dfrac{20}{3}V_{\bar b u}V_{s\bar u}\,Y\)
			&
			\(\dfrac{52i}{9}V_{\bar b u}V_{s\bar u}\,Y\)
			\\
			\hline
			
			\({\cal A}\!\left(\bar B_s \to \pi^+\pi^-\pi^0\right)\)
			& \(0\) & \(0\) & \(0\) & \(0\) & \(0\) & \(0\) \\
			
			\({\cal A}\!\left(\bar B_s \to \pi^+K^0K^-\right)\)
			&
			\(0\)
			&
			\(-4i\,V_{\bar b c}V_{s\bar c}\,X\)
			&
			\(0\)
			&
			\(4i\,V_{\bar b u}V_{s\bar u}\,Y\)
			&
			\(-\dfrac{4}{3}V_{\bar b u}V_{s\bar u}\,Y\)
			&
			\(-4i\,V_{\bar b u}V_{s\bar u}\,Y\)
			\\
			
			\({\cal A}\!\left(\bar B_s \to \pi^-K^+\bar K^0\right)\)
			&
			\(0\)
			&
			\(-4i\,V_{\bar b c}V_{s\bar c}\,X\)
			&
			\(-\dfrac{4i}{3}V_{\bar b u}V_{s\bar u}\,Y\)
			&
			\(4i\,V_{\bar b u}V_{s\bar u}\,Y\)
			&
			\(-\dfrac{16}{3}V_{\bar b u}V_{s\bar u}\,Y\)
			&
			\(\dfrac{8i}{9}V_{\bar b u}V_{s\bar u}\,Y\)
			\\
			
			\({\cal A}\!\left(\bar B_s \to \pi^0\pi^0\pi^0\right)\)
			& \(0\) & \(0\) & \(0\) & \(0\) & \(0\) & \(0\) \\
			
			\({\cal A}\!\left(\bar B_s \to \pi^0K^+K^-\right)\)
			&
			\(0\)
			&
			\(-2i\sqrt2\,V_{\bar b c}V_{s\bar c}\,X\)
			&
			\(-\dfrac{2i\sqrt2}{3}V_{\bar b u}V_{s\bar u}\,Y\)
			&
			\(2i\sqrt2\,V_{\bar b u}V_{s\bar u}\,Y\)
			&
			\(-\dfrac{8\sqrt2}{3}V_{\bar b u}V_{s\bar u}\,Y\)
			&
			\(\dfrac{4i\sqrt2}{9}V_{\bar b u}V_{s\bar u}\,Y\)
			\\
			
			\({\cal A}\!\left(\bar B_s \to \pi^0K^0\bar K^0\right)\)
			&
			\(0\)
			&
			\(2i\sqrt2\,V_{\bar b c}V_{s\bar c}\,X\)
			&
			\(0\)
			&
			\(-2i\sqrt2\,V_{\bar b u}V_{s\bar u}\,Y\)
			&
			\(\dfrac{2\sqrt2}{3}V_{\bar b u}V_{s\bar u}\,Y\)
			&
			\(2i\sqrt2\,V_{\bar b u}V_{s\bar u}\,Y\)
			\\
			
			\({\cal A}\!\left(\bar B_s \to \pi^+\pi^-K^0\right)\)
			&
			\(-4i\,V_{d\bar c}V_{\bar b c}\,X\)
			&
			\(-\dfrac{4i}{3}V_{d\bar c}V_{\bar b c}\,X\)
			&
			\(4i\,V_{d\bar u}V_{\bar b u}\,Y\)
			&
			\(\dfrac{4i}{3}V_{d\bar u}V_{\bar b u}\,Y\)
			&
			\(8V_{d\bar u}V_{\bar b u}\,Y\)
			&
			\(\dfrac{16i}{9}V_{d\bar u}V_{\bar b u}\,Y\)
			\\
			
			\({\cal A}\!\left(\bar B_s \to \pi^-\pi^0K^+\right)\)
			& \(0\) & \(0\) & \(0\) & \(0\) & \(0\) & \(0\) \\
			
			\({\cal A}\!\left(\bar B_s \to \pi^0\pi^0K^0\right)\)
			&
			\(-2i\sqrt2\,V_{d\bar c}V_{\bar b c}\,X\)
			&
			\(-\dfrac{2i\sqrt2}{3}V_{d\bar c}V_{\bar b c}\,X\)
			&
			\(2i\sqrt2\,V_{d\bar u}V_{\bar b u}\,Y\)
			&
			\(\dfrac{2i\sqrt2}{3}V_{d\bar u}V_{\bar b u}\,Y\)
			&
			\(4\sqrt2\,V_{d\bar u}V_{\bar b u}\,Y\)
			&
			\(\dfrac{8i\sqrt2}{9}V_{d\bar u}V_{\bar b u}\,Y\)
			\\
			
			\({\cal A}\!\left(\bar B_s \to K^+K^-K^0\right)\)
			&
			\(-4i\,V_{d\bar c}V_{\bar b c}\,X\)
			&
			\(-\dfrac{4i}{3}V_{d\bar c}V_{\bar b c}\,X\)
			&
			\(\dfrac{8i}{3}V_{d\bar u}V_{\bar b u}\,Y\)
			&
			\(\dfrac{4i}{3}V_{d\bar u}V_{\bar b u}\,Y\)
			&
			\(4V_{d\bar u}V_{\bar b u}\,Y\)
			&
			\(\dfrac{20i}{3}V_{d\bar u}V_{\bar b u}\,Y\)
			\\
			
			\({\cal A}\!\left(\bar B_s \to K^0K^0\bar K^0\right)\)
			&
			\(-4i\sqrt2\,V_{d\bar c}V_{\bar b c}\,X\)
			&
			\(-\dfrac{4i\sqrt2}{3}V_{d\bar c}V_{\bar b c}\,X\)
			&
			\(0\)
			&
			\(\dfrac{4i\sqrt2}{3}V_{d\bar u}V_{\bar b u}\,Y\)
			&
			\(\dfrac{20\sqrt2}{3}V_{d\bar u}V_{\bar b u}\,Y\)
			&
			\(\dfrac{52i\sqrt2}{9}V_{d\bar u}V_{\bar b u}\,Y\)
			\\
			\hline
	\end{tabular}}
	\caption{\label{tab:BPPPamps}$B\to P_1P_2P_3$ amplitudes with
		$
		X \equiv (p_2-p_3)\!\cdot\!(p_1+p_B),
		\quad
		Y \equiv (p_1-p_2)\!\cdot\!(p_3+p_B).
		$}
\end{table}
%%%%%%%%%%%%%%%%%%%%%%%%%%%%%%%%%%%%%%%%%%%%%%%%%%
%%%%%%%%%%%%%%%%%%%%%%%%%%%%%%%%%%%%%%%%%%%
\subsection{$B\to P_1 P_2 P_3$}
%%%%%%%%%%%%%%%%%%%%%%%%%%%%%%%%%%%%%%%%%%%

We first consider the charmless three-body decays $B\to P_1P_2P_3$, with the complete set of modes listed in Table~\ref{tab:BPPPamps}. In $D6$, only six independent operators contribute, namely $O_{1a}$, $O_{1b}$, and $O_{4a\text{–}d}$, since operators involving $\Sigma_H$ do not generate the required external states. Due to the two-derivative structure of the operators, the amplitudes depend on the two independent kinematic combinations
\begin{align}
X &\equiv (p_2-p_3)\cdot(p_1+p_B)\;, &
Y &\equiv (p_1-p_2)\cdot(p_3+p_B)\;,
\end{align}
where $p_{1,2,3}$ are the light-meson momenta. The resulting amplitudes are summarized in Table~\ref{tab:BPPPamps}, which defines a $(32\times6)$ amplitude matrix for the 32 possible $B\to P_1 P_2 P_3$ modes. The reduction of the row yields $\mathrm{Rank}(M)=6$, implying $32-6=26$ independent amplitude relations.

The iso-spin relations are
\begin{align}
	&{\cal A}(B^- \to \pi^-\pi^0\bar K^0)-{\cal A}(\bar B_d \to K^-\pi^+\pi^0)=0\,,\label{eq:PPPiso1}\\[0.5em]
	&{\cal A}(B^- \to \pi^-\pi^0\bar K^0)-\frac{1}{\sqrt{2}}{\cal A}(\bar B_d \to \bar K^0\pi^+\pi^-)+{\cal A}(\bar B_d \to \bar K^0\pi^0\pi^0)=0\,,\label{eq:PPPiso2}\\[0.5em]
	&{\cal A}(\bar B_d \to K^-\pi^+\pi^0)+\frac{1}{\sqrt{2}}{\cal A}(B^- \to \pi^+\pi^-K^-)-{\cal A}(B^- \to \pi^0\pi^0K^-)=0\,,\label{eq:PPPiso3}\\[0.5em]
	&{\cal A}(B^- \to K^+K^-K^-)-\sqrt{2}\,{\cal A}(B^- \to K^-\bar K^0K^0)-\sqrt{2}\,{\cal A}(\bar B_d \to K^+K^-\bar K^0)+{\cal A}(\bar B_d \to K^0\bar K^0\bar K^0)=0\,,\label{eq:PPPiso4}\\[0.5em]
	&{\cal A}(\bar B_s \to \pi^0K^+K^-)-{\cal A}(\bar B_s \to \pi^0K^0\bar K^0)-\frac{1}{\sqrt2}{\cal A}(\bar B_s \to \pi^+K^0K^-)-\frac{1}{\sqrt2}{\cal A}(\bar B_s \to \pi^-K^+\bar K^0)=0\,,\label{eq:PPPiso5}\\[0.5em]
	&{\cal A}(\bar B_s \to \pi^0\pi^0\pi^0)-\frac{3}{\sqrt{6}}{\cal A}(\bar B_s \to \pi^+\pi^-\pi^0)=0\,,\label{eq:PPPiso6}\\[0.5em]
	&{\cal A}(\bar B_d \to \pi^0\pi^0\pi^0)-\frac{3}{\sqrt{6}}{\cal A}(\bar B_d \to \pi^+\pi^-\pi^0)=0\,,\label{eq:PPPiso7}\\[0.5em]
	&{\cal A}(B^- \to \pi^-\pi^0\pi^0)-\frac{1}{2}{\cal A}(B^- \to \pi^+\pi^-\pi^-)=0\,,\label{eq:PPPiso8}\\[0.5em]
	&{\cal A}(\bar B_s \to \pi^0\pi^0K^0)-\frac{1}{\sqrt{2}}{\cal A}(\bar B_s \to \pi^+\pi^-K^0)+{\cal A}(\bar B_s \to \pi^-\pi^0K^+)=0\,,\label{eq:PPPiso9}\\[0.5em]
	&{\cal A}(\bar B_d \to K^+K^-\pi^0)-\frac{1}{\sqrt2}{\cal A}(\bar B_d \to K^+\bar K^0\pi^-)+\frac{1}{\sqrt2}{\cal A}(B^- \to \pi^-K^+K^-)-{\cal A}(\bar B_d \to K^0\bar K^0\pi^0)\nonumber\\
	&\quad-\frac{1}{\sqrt2}{\cal A}(\bar B_d \to K^-K^0\pi^+)-\frac{1}{\sqrt2}{\cal A}(B^- \to \pi^-\bar K^0K^0)+{\cal A}(B^- \to \pi^0K^-K^0)=0\,.\label{eq:PPPiso10}
\end{align}
The ten isospin relations coincide with those of Ref.~\cite{He:2014xha}\footnote{In \cite{He:2014xha}, the relations are at the Wilson coefficient level, i.e., before multiplying with the $\sqrt{n!}$ for identical $n$ meson in the mode.} and, up to the sign convention for $\bar u$, with the isospin relations of Ref.~\cite{Bhattacharya:2014eca} for the fully symmetric final state.
The further relations among the $b\to s$ modes are
\begin{align}
	&\textcolor{oceanblue}{{\cal A}(\bar B_s \to \pi^+\pi^-\pi^0)=0\,,}\label{eq:Bs3pi}\\[0.5em]
	&{\cal A}(B^- \to K^+K^-K^-)-\sqrt2\,{\cal A}(B^- \to \pi^+\pi^-K^-)=0\,,\label{eq:PPPs2}\\[0.5em]
	&\textcolor{oceanblue}{{\cal A}(\bar B_d \to \bar K^0\pi^+\pi^-)-\sqrt2\,{\cal A}(B^- \to \pi^0\pi^0K^-)=0\,,}\label{eq:PPPs3}\\[0.5em]
	&\textcolor{oceanblue}{\sqrt2\,{\cal A}(\bar B_d \to \bar K^0\pi^0\pi^0)-{\cal A}(\bar B_d \to K^+K^-\bar K^0)=0\,,}\label{eq:PPPs4}\\[0.5em]
	&\textcolor{oceanblue}{{\cal A}(\bar B_s \to \pi^+K^0K^-)+\sqrt2\,{\cal A}(\bar B_s \to \pi^0K^0\bar K^0)=0\,,}\label{eq:PPPs5}\\[0.5em]
	&\textcolor{oceanblue}{\sqrt2\,{\cal A}(\bar B_d \to K^-\pi^+\pi^0)+{\cal A}(\bar B_s \to \pi^+K^0K^-)-{\cal A}(\bar B_s \to \pi^-K^+\bar K^0)=0\,,}\label{eq:PPPs6}
\end{align}
and among the $b\to d$ modes
\begin{align}
	&\textcolor{oceanblue}{{\cal A}(\bar B_d \to K^+\bar K^0\pi^-)+\sqrt{2}\,{\cal A}(\bar B_d \to \pi^+\pi^-\pi^0)-\sqrt{2}\,{\cal A}(\bar B_d \to K^+K^-\pi^0)=0\,,}\label{eq:PPPd1}\\[0.5em]
	&\textcolor{oceanblue}{{\cal A}(B^- \to \pi^+\pi^-\pi^-)+2\,{\cal A}(\bar B_d \to \pi^+\pi^-\pi^0)=0\,,}\label{eq:PPPd2}\\[0.5em]
	&\textcolor{oceanblue}{{\cal A}(B^- \to \pi^-K^+K^-)+\sqrt{2}\,{\cal A}(\bar B_d \to \pi^+\pi^-\pi^0)=0\,,}\label{eq:PPPd3}\\[0.5em]
	&\textcolor{oceanblue}{{\cal A}(\bar B_s \to \pi^-\pi^0K^+)=0\,,}\label{eq:BPPPsr28}\\[0.5em]
	&\textcolor{oceanblue}{\sqrt2\,{\cal A}(\bar B_d \to \pi^+\pi^-\pi^0)+{\cal A}(\bar B_s \to \pi^+\pi^-K^0)=0\,,}\label{eq:PPPd5}\\[0.5em]
	&\textcolor{oceanblue}{\sqrt2\,{\cal A}(B^- \to \pi^-\bar K^0K^0)-{\cal A}(\bar B_s \to K^0K^0\bar K^0)=0\,,}\label{eq:PPPd6}\\[0.5em]
	&\textcolor{oceanblue}{\sqrt2\,{\cal A}(\bar B_d \to \pi^+\pi^-\pi^0)-\sqrt2\,{\cal A}(B^- \to \pi^0K^-K^0)+{\cal A}(\bar B_s \to K^+K^-K^0)=0\,,}\label{eq:PPPd7}\\[0.5em]
	&\textcolor{oceanblue}{\sqrt2\,{\cal A}(\bar B_d \to K^+K^-\pi^0)-{\cal A}(\bar B_d \to K^-K^0\pi^+)+{\cal A}(\bar B_s \to K^+K^-K^0)=0\,.}\label{eq:PPPd8}
\end{align}
The two relations connecting the sectors are
\begin{align}
	&\textcolor{oceanblue}{{\cal A}(\bar B_d \to K^-\pi^+\pi^0)=r_u\Big[{\cal A}(\bar B_d \to \pi^+\pi^-\pi^0)-{\cal A}(\bar B_d \to K^+K^-\pi^0)+\frac{1}{\sqrt2}\,{\cal A}(\bar B_d \to K^-K^0\pi^+)\Big]\,,}\label{eq:PPPx1}\\[0.5em]
	&\textcolor{oceanblue}{{\cal A}(\bar B_d \to K^0\bar K^0\bar K^0)-\sqrt2\,{\cal A}(\bar B_d \to \bar K^0\pi^+\pi^-)}\nonumber\\
	&\qquad\textcolor{oceanblue}{=r_u\Big[2\,{\cal A}(\bar B_d \to \pi^+\pi^-\pi^0)-2\,{\cal A}(\bar B_d \to K^0\bar K^0\pi^0)-\sqrt2\,{\cal A}(\bar B_d \to K^-K^0\pi^+)\Big]\,.}\label{eq:PPPx2}
\end{align}
The combination eq.~\eqref{eq:PPPd2}$\,-\sqrt2\,$eq.~\eqref{eq:PPPd3} are the two $SU(3)$ equalities of Ref.~\cite{Bhattacharya:2014eca}. 
%As in Sec.~\ref{sec:BPP_amp}, the U-spin pairs are used there only through the CP relation $A_s{\cal B}_s/(A_d{\cal B}_d)=-1$, not as amplitude relations. 
The relations in blue have not been written down before~\footnote{For the CKM-stripped amplitudes, i.e.\ separately for the
$V_{\bar bc}V_{q\bar c}$ and $V_{\bar bu}V_{q\bar u}$ components, our basis reproduces the eight U-spin pairs of Ref.~\cite{Bhattacharya:2014eca}, such as
${\cal A}(\bar B_d\to K^0\bar K^0\bar K^0)={\cal A}(\bar B_s\to K^0K^0\bar K^0)$
and ${\cal A}(B^-\to K^+K^-K^-)={\cal A}(B^-\to\pi^+\pi^-\pi^-)$, up to the sign convention for $\bar u$. They do not survive as relations among physical amplitudes, since the two components scale differently between the $b\to d$
and $b\to s$ sectors.}. 
Within each sector, $SU(3)$ gives nine relations~\cite{Bhattacharya:2014eca} against our twelve, so three directions per sector, including the zeros eqs.~\eqref{eq:Bs3pi} and \eqref{eq:BPPPsr28}, are specific to the $D6$ operator basis and receive corrections from operators with more than four meson fields.

These relations share common phenomenological implications. Since each contact amplitude is linear combinations of the invariants $X$ and $Y$, the sum rules hold locally in phase space once the momenta of flavor-related mesons are identified. They can therefore be tested against the non-resonant components extracted in amplitude analyses of related modes, and used to transfer information from the well-measured $B_{u,d}\to K\pi\pi$ and $KK\bar K$ Dalitz plots~\cite{LHCb:2013ptu} to the corresponding, largely unmeasured $\bar B_s$ channels.

%%%%%%%%%%%%%%%%%%%%%%%%%%%%%%%%%%%%%%%%%%%%%%%%
\subsection{$B \to D P_1 P_2$}
%%%%%%%%%%%%%%%%%%%%%%%%%%%%%%%%%%%%%%%%

We next consider three-body decays with one charmed meson in the final state. The relevant modes are of the form $B\to D P_1P_2$ and $B\to \bar D P_1P_2$, where $P_1$ and $P_2$ denote light pseudoscalar mesons. As in the charmless case, only those operators that generate the required four-point vertices contribute to these amplitudes. The operators involving $\Sigma_H$ do not contribute to the modes considered here, and the remaining relevant operators are $O_{5a}$, $O_{5b}$, $O_{7a}$, and $O_{7b}$. We organize the decay modes according to whether the final state contains a $D$ or a $\bar D$ meson, since the two classes are generated by different underlying flavor structures.

The contributions from $O_{5a}$ and $O_{5b}$ are proportional to a single kinematic structure,
\begin{align}
	Z \equiv (p_1-p_2)\cdot(p_B+p_D)\;.\label{eq:Z}
\end{align}
On the other hand, the operators $O_{7a}$ and $O_{7b}$ generate four possible momentum structures, which we define as
\begin{align}
	P &\equiv p_1\cdot(11p_2+5p_B-6p_D)
	+ p_2\cdot(6p_B-5p_D)
	-11p_B\cdot p_D\;, \\
	Q &\equiv p_1\cdot(13p_2+3p_B-10p_D)
	+ p_2\cdot(10p_B-3p_D)
	-13p_B\cdot p_D\;, \\
	R &\equiv p_1\cdot(p_2-p_B-2p_D)
	-p_B\cdot p_D
	+p_2\cdot(2p_B+p_D)\;, \\
	S &\equiv (p_2+p_B)\cdot(p_1-p_D)\;.\label{eq:S}
\end{align}
Here, $p_1$ and $p_2$ denote the momenta associated with the two light-meson multiplet fields, while $p_D$ denotes the momentum of the charmed meson. With these definitions, the operator contributions to the $B\to DP_1P_2$ and $B\to \bar D P_1P_2$ amplitudes are summarized in Tables~\ref{tab:BDPP_Bd}--\ref{tab:BDPP_Bs}.

%%%%%%%%%%%%%%%%%%%%%%%%%%%%%%%%%%%%%%%%%%%%%%%%%%
\begin{table}[h]
	\centering
	\renewcommand{\arraystretch}{1.1}
	\setlength{\tabcolsep}{3pt}
	\resizebox{.95\textwidth}{!}{%
		\begin{tabular}{
				p{0.27\textwidth} c c
				@{\hspace{1.5em}} p{0.05\textwidth}@{\hspace{1.5em}}
				p{0.27\textwidth} c c
			}
			\hline
			\multicolumn{3}{c}{\(b\to c\bar u q\)}
			&
			& \multicolumn{3}{c}{\(b\to u\bar c q\)} \\
			\cline{1-3}\cline{5-7}
			\textbf{Mode}
			& \(\boldsymbol{O_{5a}}\)
			& \(\boldsymbol{O_{5b}}\)
			&
			& \textbf{Amplitudes}
			& \(\boldsymbol{O_{7a}}\)
			& \(\boldsymbol{O_{7b}}\) \\
			\hline
			
			\({\cal A}\!\left(\bar B_d \to D_u\,\pi^+\pi^- \right)\)
			& \(0\)
			& \(0\)
			&
			& \({\cal A}\!\left(\bar B_d \to \bar D_u\,\pi^+\pi^- \right)\)
			& \(\dfrac{8}{9}V_{d\bar c}V_{\bar b u}\,P\)
			& \(-V_{d\bar c}V_{\bar b u}\,S\) \\
			
			\({\cal A}\!\left(\bar B_d \to D_u\,\pi^+K^- \right)\)
			& \(0\)
			& \(0\)
			&
			& \({\cal A}\!\left(\bar B_d \to \bar D_u\,\pi^+K^- \right)\)
			& \(\dfrac{8}{9}V_{\bar b u}V_{s\bar c}\,P\)
			& \(-V_{\bar b u}V_{s\bar c}\,S\) \\
			
			\({\cal A}\!\left(\bar B_d \to D_u\,\pi^0\pi^0 \right)\)
			& \(0\)
			& \(-4\sqrt{2}\,V_{d\bar u}V_{\bar b c}\,Z\)
			&
			& \({\cal A}\!\left(\bar B_d \to \bar D_u\,\pi^0\pi^0 \right)\)
			& \(\dfrac{4\sqrt{2}}{9}V_{d\bar c}V_{\bar b u}\,Q\)
			& \(0\) \\
			
			\({\cal A}\!\left(\bar B_d \to D_u\,\pi^0\bar K^0 \right)\)
			& \(0\)
			& \(4\sqrt{2}\,V_{\bar b c}V_{s\bar u}\,Z\)
			&
			& \({\cal A}\!\left(\bar B_d \to \bar D_u\,\pi^0\bar K^0 \right)\)
			& \(-\dfrac{4\sqrt2}{9}V_{\bar b u}V_{s\bar c}\,Q\)
			& \(0\) \\
			
			\({\cal A}\!\left(\bar B_d \to D_u\,K^+K^- \right)\)
			& \(0\)
			& \(0\)
			&
			& \({\cal A}\!\left(\bar B_d \to \bar D_u\,K^+K^- \right)\)
			& \(0\)
			& \(0\) \\
			
			\({\cal A}\!\left(\bar B_d \to D_u\,K^0\bar K^0 \right)\)
			& \(0\)
			& \(0\)
			&
			& \({\cal A}\!\left(\bar B_d \to \bar D_u\,K^0\bar K^0 \right)\)
			& \(0\)
			& \(0\) \\
			
			\({\cal A}\!\left(\bar B_d \to D_d\,\pi^0\pi^- \right)\)
			& \(0\)
			& \(-4\sqrt{2}\,V_{d\bar u}V_{\bar b c}\,Z\)
			&
			& \({\cal A}\!\left(\bar B_d \to D_d^-\,\pi^+\pi^0 \right)\)
			& \(\dfrac{8\sqrt2}{9}V_{d\bar c}V_{\bar b u}\,R\)
			& \(\dfrac{1}{\sqrt2}V_{d\bar c}V_{\bar b u}\,S\) \\
			
			\({\cal A}\!\left(\bar B_d \to D_d\,\pi^0K^- \right)\)
			& \(2\sqrt{2}\,V_{\bar b c}V_{s\bar u}\,Z\)
			& \(-\dfrac{4\sqrt{2}}{3}\,V_{\bar b c}V_{s\bar u}\,Z\)
			&
			& \({\cal A}\!\left(\bar B_d \to D_d^-\,\pi^+\bar K^0 \right)\)
			& \(-\dfrac{16}{9}V_{\bar b u}V_{s\bar c}\,R\)
			& \(-V_{\bar b u}V_{s\bar c}\,S\) \\
			
			\({\cal A}\!\left(\bar B_d \to D_d\,\pi^-\bar K^0 \right)\)
			& \(4\,V_{\bar b c}V_{s\bar u}\,Z\)
			& \(\dfrac{16}{3}\,V_{\bar b c}V_{s\bar u}\,Z\)
			&
			& \({\cal A}\!\left(\bar B_d \to D_d^-\,K^+\bar K^0 \right)\)
			& \(0\)
			& \(0\) \\
			
			\({\cal A}\!\left(\bar B_d \to D_d\,K^0K^- \right)\)
			& \(4\,V_{d\bar u}V_{\bar b c}\,Z\)
			& \(-\dfrac{8}{3}\,V_{d\bar u}V_{\bar b c}\,Z\)
			&
			& \({\cal A}\!\left(\bar B_d \to D_s^-\,\pi^+\pi^0 \right)\)
			& \(0\)
			& \(0\) \\
			
			\({\cal A}\!\left(\bar B_d \to D_s\,\pi^0K^- \right)\)
			& \(0\)
			& \(-4\sqrt{2}\,V_{d\bar u}V_{\bar b c}\,Z\)
			&
			& \({\cal A}\!\left(\bar B_d \to D_s^-\,\pi^+K^0 \right)\)
			& \(\dfrac{8}{9}V_{d\bar c}V_{\bar b u}\,P\)
			& \(-V_{d\bar c}V_{\bar b u}\,S\) \\
			
			\({\cal A}\!\left(\bar B_d \to D_s\,\pi^-\bar K^0 \right)\)
			& \(0\)
			& \(0\)
			&
			& \({\cal A}\!\left(\bar B_d \to D_s^-\,\pi^0K^+ \right)\)
			& \(\dfrac{4\sqrt2}{9}V_{d\bar c}V_{\bar b u}\,Q\)
			& \(0\) \\
			
			\({\cal A}\!\left(\bar B_d \to D_s\,K^-\bar K^0 \right)\)
			& \(0\)
			& \(8\,V_{\bar b c}V_{s\bar u}\,Z\)
			&
			& \({\cal A}\!\left(\bar B_d \to D_s^-\,K^+\bar K^0 \right)\)
			& \(-\dfrac{8}{9}V_{\bar b u}V_{s\bar c}\,Q\)
			& \(0\) \\
			
			\bottomrule
		\end{tabular}
	}
    \caption{$B_d \to D P_1 P_2$ amplitudes. Here $P$, $Q$, $R$, $S$ and $Z$ contain the momentum dependencies as defined in eqs.\,(\ref{eq:Z}) to (\ref{eq:S}).}
    \label{tab:BDPP_Bd}
\end{table}
%%%%%%%%%%%%%%%%%%%%%%%%%%%%%%%%%%%%%%%%%%%%%%%%%%

%%%%%%%%%%%%%%%%%%%%%%%%%%%%%%%%%%%%%%%%%%%%%%%%%%
\begin{table}[h]
	\centering
	\renewcommand{\arraystretch}{1.1}
	\setlength{\tabcolsep}{3pt}
	\resizebox{.95\textwidth}{!}{%
		\begin{tabular}{
				p{0.27\textwidth} c c
				@{\hspace{1.5em}} c @{\hspace{1.5em}}
				p{0.27\textwidth} c c
			}
			\toprule
			\multicolumn{3}{c}{\(b\to c\bar u q\)}
			&
			& \multicolumn{3}{c}{\(b\to u\bar c q\)} \\
			\cmidrule(lr){1-3}\cmidrule(lr){5-7}
			\textbf{Mode}
			& \(\boldsymbol{O_{5a}}\)
			& \(\boldsymbol{O_{5b}}\)
			&
			& \textbf{Amplitudes}
			& \(\boldsymbol{O_{7a}}\)
			& \(\boldsymbol{O_{7b}}\) \\
			\midrule
			
			\({\cal A}\!\left(B^- \to D_u\,\pi^0\pi^-\right)\)
			& \(0\)
			& \(4\sqrt{2}\,V_{d\bar u}V_{\bar b c}\,Z\)
			&
			& \({\cal A}\!\left(B^- \to \bar D_u\,\pi^0\pi^- \right)\)
			& \(-\dfrac{8\sqrt2}{9}V_{d\bar c}V_{\bar b u}\,R\)
			& \(-\dfrac{1}{\sqrt2}V_{d\bar c}V_{\bar b u}\,S\) \\
			
			\({\cal A}\!\left(B^- \to D_u\,\pi^0K^- \right)\)
			& \(2\sqrt{2}\,V_{\bar b c}V_{s\bar u}\,Z\)
			& \(\dfrac{8\sqrt{2}}{3}\,V_{\bar b c}V_{s\bar u}\,Z\)
			&
			& \({\cal A}\!\left(B^- \to \bar D_u\,\pi^0K^- \right)\)
			& \(-\dfrac{8\sqrt2}{9}V_{\bar b u}V_{s\bar c}\,R\)
			& \(-\dfrac{1}{\sqrt2}V_{\bar b u}V_{s\bar c}\,S\) \\
			
			\({\cal A}\!\left(B^- \to D_u\,\pi^-\bar K^0 \right)\)
			& \(4\,V_{\bar b c}V_{s\bar u}\,Z\)
			& \(-\dfrac{8}{3}\,V_{\bar b c}V_{s\bar u}\,Z\)
			&
			& \({\cal A}\!\left(B^- \to \bar D_u\,\pi^-\bar K^0 \right)\)
			& \(0\)
			& \(0\) \\
			
			\({\cal A}\!\left(B^- \to D_u\,K^0K^- \right)\)
			& \(4\,V_{d\bar u}V_{\bar b c}\,Z\)
			& \(-\dfrac{8}{3}\,V_{d\bar u}V_{\bar b c}\,Z\)
			&
			& \({\cal A}\!\left(B^- \to \bar D_u\,K^0K^- \right)\)
			& \(0\)
			& \(0\) \\
			
			\({\cal A}\!\left(B^- \to D_d\,\pi^-\pi^- \right)\)
			& \(0\)
			& \(8\sqrt{2}\,V_{d\bar u}V_{\bar b c}\,Z\)
			&
			& \({\cal A}\!\left(B^- \to D_d^-\,\pi^+\pi^- \right)\)
			& \(-\dfrac{8}{9}V_{d\bar c}V_{\bar b u}\,Q\)
			& \(0\) \\
			
			\({\cal A}\!\left(B^- \to D_d\,\pi^-K^- \right)\)
			& \(0\)
			& \(8\,V_{\bar b c}V_{s\bar u}\,Z\)
			&
			& \({\cal A}\!\left(B^- \to D_d^-\,\pi^+K^- \right)\)
			& \(-\dfrac{8}{9}V_{\bar b u}V_{s\bar c}\,Q\)
			& \(0\) \\
			
			\({\cal A}\!\left(B^- \to D_s\,\pi^-K^- \right)\)
			& \(0\)
			& \(8\,V_{d\bar u}V_{\bar b c}\,Z\)
			&
			& \({\cal A}\!\left(B^- \to D_d^-\,\pi^0\pi^0 \right)\)
			& \(-\dfrac{4\sqrt{2}}{9}V_{d\bar c}V_{\bar b u}\,P\)
			& \(\dfrac{\sqrt{2}}{2}V_{d\bar c}V_{\bar b u}\,S\) \\
			
			\({\cal A}\!\left(B^- \to D_s\,K^-K^- \right)\)
			& \(0\)
			& \(8\sqrt{2}\,V_{\bar b c}V_{s\bar u}\,Z\)
			&
			& \({\cal A}\!\left(B^- \to D_d^-\,\pi^0\bar K^0 \right)\)
			& \(\dfrac{4\sqrt2}{9}V_{\bar b u}V_{s\bar c}\,P\)
			& \(-\dfrac{1}{\sqrt2}V_{\bar b u}V_{s\bar c}\,S\) \\
			
			-- 
			& -- 
			& -- 
			&
			& \({\cal A}\!\left(B^- \to D_d^-\,K^+K^- \right)\)
			& \(0\)
			& \(0\) \\
			
			-- 
			& -- 
			& -- 
			&
			& \({\cal A}\!\left(B^- \to D_d^-\,K^0\bar K^0 \right)\)
			& \(0\)
			& \(0\) \\
			
			-- 
			& -- 
			& -- 
			&
			& \({\cal A}\!\left(B^- \to D_s^-\,\pi^+\pi^- \right)\)
			& \(0\)
			& \(0\) \\
			
			-- 
			& -- 
			& -- 
			&
			& \({\cal A}\!\left(B^- \to D_s^-\,\pi^0\pi^0 \right)\)
			& \(0\)
			& \(0\) \\
			
			-- 
			& -- 
			& -- 
			&
			& \({\cal A}\!\left(B^- \to D_s^-\,\pi^0K^0 \right)\)
			& \(\dfrac{4\sqrt2}{9}V_{d\bar c}V_{\bar b u}\,P\)
			& \(-\dfrac{1}{\sqrt2}V_{d\bar c}V_{\bar b u}\,S\) \\
			
			-- 
			& -- 
			& -- 
			&
			& \({\cal A}\!\left(B^- \to D_s^-\,\pi^-K^+ \right)\)
			& \(-\dfrac{8}{9}V_{d\bar c}V_{\bar b u}\,Q\)
			& \(0\) \\
			
			-- 
			& -- 
			& -- 
			&
			& \({\cal A}\!\left(B^- \to D_s^-\,K^+K^- \right)\)
			& \(-\dfrac{8}{9}V_{\bar b u}V_{s\bar c}\,Q\)
			& \(0\) \\
			
			-- 
			& -- 
			& -- 
			&
			& \({\cal A}\!\left(B^- \to D_s^-\,K^0\bar K^0 \right)\)
			& \(0\)
			& \(0\) \\
			
			\bottomrule
		\end{tabular}
	}
    \caption{$B_u \to D P_1 P_2$ modes.}
    \label{tab:BDPP_Bu}
\end{table}
%%%%%%%%%%%%%%%%%%%%%%%%%%%%%%%%%%%%%%%%%%%%%%%%%%

%%%%%%%%%%%%%%%%%%%%%%%%%%%%%%%%%%%%%%%%%%%%%%%%%%
\begin{table}[h]
	\centering
	\renewcommand{\arraystretch}{1.15}
	\setlength{\tabcolsep}{3pt}
	\resizebox{\textwidth}{!}{%
		\begin{tabular}{
				p{0.27\textwidth} c c
				@{\hspace{1.5em}} c @{\hspace{1.5em}}
				p{0.27\textwidth} c c
			}
			\toprule
			\multicolumn{3}{c}{\(b\to c\bar u q\)}
			&
			& \multicolumn{3}{c}{\(b\to u\bar c q\)} \\
			\cmidrule(lr){1-3}\cmidrule(lr){5-7}
			\textbf{Amplitudes}
			& \(\boldsymbol{O_{5a}}\)
			& \(\boldsymbol{O_{5b}}\)
			&
			& \textbf{Amplitudes}
			& \(\boldsymbol{O_{7a}}\)
			& \(\boldsymbol{O_{7b}}\) \\
			\midrule
			
			\({\cal A}\!\left(\bar B_s \to D_u\,\pi^+\pi^- \right)\)
			& \(0\)
			& \(0\)
			&
			& \({\cal A}\!\left(\bar B_s \to \bar D_u\,\pi^+\pi^- \right)\)
			& \(0\)
			& \(0\) \\
			
			\({\cal A}\!\left(\bar B_s \to D_u\,\pi^0\pi^0 \right)\)
			& \(0\)
			& \(0\)
			&
			& \({\cal A}\!\left(\bar B_s \to \bar D_u\,\pi^0\pi^0 \right)\)
			& \(0\)
			& \(0\) \\
			
			\({\cal A}\!\left(\bar B_s \to D_u\,\pi^0K^0 \right)\)
			& \(0\)
			& \(4\sqrt{2}\,V_{d\bar u}V_{\bar b c}\,Z\)
			&
			& \({\cal A}\!\left(\bar B_s \to \bar D_u\,\pi^0K^0 \right)\)
			& \(-\dfrac{4\sqrt2}{9}V_{d\bar c}V_{\bar b u}\,Q\)
			& \(0\) \\
			
			\({\cal A}\!\left(\bar B_s \to D_u\,\pi^-K^+ \right)\)
			& \(0\)
			& \(0\)
			&
			& \({\cal A}\!\left(\bar B_s \to \bar D_u\,\pi^-K^+ \right)\)
			& \(\dfrac{8}{9}V_{d\bar c}V_{\bar b u}\,P\)
			& \(-V_{d\bar c}V_{\bar b u}\,S\) \\
			
			\({\cal A}\!\left(\bar B_s \to D_u\,K^+K^- \right)\)
			& \(0\)
			& \(0\)
			&
			& \({\cal A}\!\left(\bar B_s \to \bar D_u\,K^+K^- \right)\)
			& \(\dfrac{8}{9}V_{\bar b u}V_{s\bar c}\,P\)
			& \(-V_{\bar b u}V_{s\bar c}\,S\) \\
			
			\({\cal A}\!\left(\bar B_s \to D_u\,K^0\bar K^0 \right)\)
			& \(0\)
			& \(0\)
			&
			& \({\cal A}\!\left(\bar B_s \to \bar D_u\,K^0\bar K^0 \right)\)
			& \(0\)
			& \(0\) \\
			
			\({\cal A}\!\left(\bar B_s \to D_d\,\pi^0\pi^- \right)\)
			& \(0\)
			& \(0\)
			&
			& \({\cal A}\!\left(\bar B_s \to D_d^-\,\pi^+\pi^0 \right)\)
			& \(0\)
			& \(0\) \\
			
			\({\cal A}\!\left(\bar B_s \to D_d\,\pi^-K^0 \right)\)
			& \(0\)
			& \(8\,V_{d\bar u}V_{\bar b c}\,Z\)
			&
			& \({\cal A}\!\left(\bar B_s \to D_d^-\,\pi^+K^0 \right)\)
			& \(-\dfrac{8}{9}V_{d\bar c}V_{\bar b u}\,Q\)
			& \(0\) \\
			
			\({\cal A}\!\left(\bar B_s \to D_d\,K^0K^- \right)\)
			& \(0\)
			& \(0\)
			&
			& \({\cal A}\!\left(\bar B_s \to D_d^-\,\pi^0K^+ \right)\)
			& \(-\dfrac{4\sqrt2}{9}V_{d\bar c}V_{\bar b u}\,P\)
			& \(\dfrac{1}{\sqrt2}V_{d\bar c}V_{\bar b u}\,S\) \\
			
			\({\cal A}\!\left(\bar B_s \to D_s\,\pi^0\pi^- \right)\)
			& \(0\)
			& \(0\)
			&
			& \({\cal A}\!\left(\bar B_s \to D_d^-\,K^+\bar K^0 \right)\)
			& \(\dfrac{8}{9}V_{\bar b u}V_{s\bar c}\,P\)
			& \(-V_{\bar b u}V_{s\bar c}\,S\) \\
			
			\({\cal A}\!\left(\bar B_s \to D_s\,\pi^0K^- \right)\)
			& \(2\sqrt{2}\,V_{\bar b c}V_{s\bar u}\,Z\)
			& \(-\dfrac{4\sqrt{2}}{3}\,V_{\bar b c}V_{s\bar u}\,Z\)
			&
			& \({\cal A}\!\left(\bar B_s \to D_s^-\,\pi^+K^0 \right)\)
			& \(0\)
			& \(0\) \\
			
			\({\cal A}\!\left(\bar B_s \to D_s\,\pi^-\bar K^0 \right)\)
			& \(4\,V_{\bar b c}V_{s\bar u}\,Z\)
			& \(-\dfrac{8}{3}\,V_{\bar b c}V_{s\bar u}\,Z\)
			&
			& \({\cal A}\!\left(\bar B_s \to D_s^-\,\pi^0K^+ \right)\)
			& \(0\)
			& \(0\) \\
			
			\({\cal A}\!\left(\bar B_s \to D_s\,K^0K^- \right)\)
			& \(4\,V_{d\bar u}V_{\bar b c}\,Z\)
			& \(\dfrac{16}{3}\,V_{d\bar u}V_{\bar b c}\,Z\)
			&
			& \({\cal A}\!\left(\bar B_s \to D_s^-\,K^+K^0 \right)\)
			& \(-\dfrac{16}{9}V_{d\bar c}V_{\bar b u}\,R\)
			& \(-V_{d\bar c}V_{\bar b u}\,S\) \\
			
			\bottomrule
		\end{tabular}
	}
    \caption{$B_s \to D P_1 P_2$ modes.}
    \label{tab:BDPP_Bs}
\end{table}
%%%%%%%%%%%%%%%%%%%%%%%%%%%%%%%%%%%%%%%%%%%%%%%%%%

We combine the amplitude information from these tables into a single matrix and perform a row reduction to extract the corresponding null directions. This gives the sum rules among the three-body decay amplitudes with one charmed meson in the final state. For clarity, we present the resulting relations separately for modes containing a $D$ meson and those containing a $\bar D$ meson. For the modes involving $B\to DP_1P_2$ example relations are
\begin{align}
	&{\cal A}(\bar B_d \to D_u\,\pi^0\bar K^0)=-r_u\,{\cal A}(\bar B_d \to D_u\,\pi^0\pi^0)\,,\label{eq:DPP1}\\
	&{\cal A}(\bar B_d \to D_u\,\pi^0\bar K^0)=-r_u\,{\cal A}(\bar B_d \to D_d\,\pi^0\pi^-)\,,\\
	&{\cal A}(\bar B_d \to D_s\,\pi^0K^-)-{\cal A}(\bar B_d \to D_d\,\pi^0\pi^-)=0\,,\\
	&{\cal A}(\bar B_d \to D_s\,K^-\bar K^0)=-\sqrt{2}\,r_u\,{\cal A}(\bar B_d \to D_u\,\pi^0\pi^0)\,,\\
	&{\cal A}(B^- \to D_u\,\pi^0\pi^-)+{\cal A}(\bar B_d \to D_d\,\pi^0\pi^-)=0\,,\\
	&\sqrt{2}\,{\cal A}(\bar B_d \to D_s\,K^-\bar K^0)=r_u\,{\cal A}(B^- \to D_d\,\pi^-\pi^-)\,,\\
	&{\cal A}(B^- \to D_u\,\pi^-\bar K^0)=r_u\,{\cal A}(\bar B_d \to D_d\,K^0K^-)\,,\\
	&{\cal A}(\bar B_s \to D_s\,\pi^0K^-)-{\cal A}(\bar B_d \to D_d\,\pi^0K^-)=0\,,\\
	&{\cal A}(\bar B_d \to D_d\,\pi^-\bar K^0)=r_u\,{\cal A}(\bar B_s \to D_s\,K^0K^-)\,.\label{eq:DPP9}
\end{align}
Example relations among modes with $\bar D$ in the final state are
\begin{align}
	&{\cal A}(\bar B_d \to \bar D_u\,\pi^+K^-)=r_c\,{\cal A}(\bar B_d \to \bar D_u\,\pi^+\pi^-)\,,\label{eq:DbPP1}\\
	&{\cal A}(\bar B_d \to D_s^-\,\pi^+K^0)-{\cal A}(\bar B_d \to \bar D_u\,\pi^+\pi^-)=0\,,\\
	&{\cal A}(\bar B_d \to \bar D_u\,\pi^0\bar K^0)=-r_c\,{\cal A}(\bar B_d \to \bar D_u\,\pi^0\pi^0)\,,\\
	&{\cal A}(\bar B_d \to D_s^-\,\pi^0K^+)-{\cal A}(\bar B_d \to \bar D_u\,\pi^0\pi^0)=0\,,\\
	&{\cal A}(\bar B_d \to D_s^-\,K^+\bar K^0)=-\sqrt2\,r_c\,{\cal A}(\bar B_d \to \bar D_u\,\pi^0\pi^0)\,,\\
	&{\cal A}(\bar B_d \to D_d^-\,\pi^+\bar K^0)=-\sqrt2\,r_c\,{\cal A}(\bar B_d \to D_d^-\,\pi^+\pi^0)\,,\\
	&{\cal A}(B^- \to \bar D_u\,\pi^0\pi^-)+{\cal A}(\bar B_d \to D_d^-\,\pi^+\pi^0)=0\,,\\
	&{\cal A}(B^- \to D_d^-\,\pi^0\pi^0)+\frac{1}{\sqrt{2}}{\cal A}(\bar B_d \to \bar D_u\,\pi^+\pi^-)=0\,,\\
	&{\cal A}(\bar B_s \to D_s^-\,K^+K^0)+\sqrt2\,{\cal A}(\bar B_d \to D_d^-\,\pi^+\pi^0)=0\,.\label{eq:DbPP9}
\end{align}
The relations above illustrate that the symmetry-based operator construction provides a direct way to derive three-body sum rules without referring explicitly to quark-level topologies. The same procedure can be extended systematically to decay modes with larger final-state multiplicities by constructing the relevant operators order by order in the EFT expansion. The relations derived here are leading-order predictions at the $D6$ level. Operators involving more  ($>4$) meson multiplets can modify these sum rules, but their effects are expected to be suppressed by additional powers of the EFT expansion scale. Therefore, the $D6$ relations capture the leading symmetry-controlled structure of the three-body decay amplitudes. {\it While flavor-$SU(3)$ amplitude relations are well established for two-body modes and for charmless three-body decays, to our knowledge the analogous relations for charmed three-body contact modes $B\to DP_1P_2$ have not been tabulated previously.}

%%%%%%%%%%%%%%%%%%%%%%%%%%%%%%%%%%%%%%%%%%%%%%%%%%%%%%%%%%%%%
\section{Unfolding and SM roots} \label{sec:unfolding}
%%%%%%%%%%%%%%%%%%%%%%%%%%%%%%%%%%%%%%%%%%%%%%%%%%

The Hilbert series construction of Secs.~\ref{sec:HS-OP}--\ref{sec:3body} is purely symmetry based, identifying all independent flavor-invariant contact operators without reference to their microscopic origin. In this section, we connect this operator basis to the UV theory (e.g., SM) by decomposing each contact operator into a pair of flavor bilinears linked by an intermediate field transforming in a definite representation of the flavor group. This factorized form makes the underlying weak topology explicit, distinguishing operators generated by tree-level weak interactions from those of penguin origin, and thereby elucidates the expected hierarchy of Wilson coefficients. It also provides a nontrivial consistency check.

%%%%%%%%%%%%%%%%%%%%%%%%%%%%%%%%%%%%%%%%%%%%%%%%%%
\begin{table}[h]
	\centering
	\renewcommand{\arraystretch}{1.00}
	\setlength{\tabcolsep}{3pt}
	\resizebox{0.9\textwidth}{!}{%
		\begin{tabular}{cccc}
			\toprule
			Op. & Invariant operators& Op. & Traced form basis~\cite{Chakraborty:2026vdz} \\
			\midrule
			
			\multicolumn{4}{c}{
				Category I: 
				\(\Sigma^\dagger,\Sigma,\Sigma^\dagger,\Phi;\ 
				\hQ_{qQ},\hQ_{QQ}^\dagger;\ 
				D^2\)
			} \\
			\midrule
			
			\(O_{1a}\) &
			\((\Sigma^\dagger T_L^A\hQ_{qQ}\tau_L^I \delmu \Phi)
			(\Sigma^\dagger \,\hQ_{QQ}^\dagger \tau_L^I \, T_L^A \delmu\Sigma)\)
			&
			{\purp \({\cal O}_{1a}^{\rm tr}\)} &
			{\purp \(\langle \Sigma^\dagger \hQ_{qQ}\hQ_{QQ}^\dagger \delmu \Phi \rangle\,
			\langle \Sigma^\dagger \delmu \Sigma \rangle\)}
			\\[2ex]
			
			\(O_{1b}\) &
			\((\Sigma^\dagger T_L^A T_R^B\hQ_{qQ}\tau_L^I \delmu \Phi)
			(\Sigma^\dagger \,\hQ_{QQ}^\dagger \tau_L^I \, T_L^A T_R^B \delmu\Sigma)\)
			&
			{\blue \({\cal O}_{1b}^{\rm tr}\)} &
			{\blue \(\langle \Sigma^\dagger \hQ_{qQ}\hQ_{QQ}^\dagger \delmu \Phi \,\Sigma^\dagger \delmu \Sigma \rangle\)}
			\\[2ex]
			
			\midrule
			\multicolumn{4}{c}{
				Category II: 
				\(\Sigma^\dagger,\Sigma,\Phi_c^\dagger,\Sigma_H;\ 
				\hQ_{qQ},\hQ_{QQ}^\dagger;\ 
				D^2\)
			} \\
			\midrule
			
			\(O_{2a}\) &
			 \((\Sigma^\dagger\hQ_{qQ}\tau_L^I \delmu \Sigma_H)
			(\Phi_c^\dagger \,\hQ_{QQ}^\dagger \tau_L^I \, \delmu \Sigma)\)
			&
			{\purp \({\cal O}_{2a}^{\rm tr}\)} &
			{\purp \(\langle \Sigma^\dagger \delmu \Sigma \rangle\,
			\langle \Phi_c^\dagger \hQ_{qQ}\hQ_{QQ}^\dagger \delmu \Sigma_H \rangle\)}
			\\[2ex]
			
			\(O_{2b}\) &
			\((\Sigma^\dagger T_L^A\hQ_{qQ}\tau_L^I \delmu \Sigma_H)
			(\Phi_c^\dagger \,\hQ_{QQ}^\dagger \tau_L^I \, T_L^A \delmu \Sigma)\)
			&
			{\blue \({\cal O}_{2b}^{\rm tr}\)} &
			{\blue \(\langle \Sigma^\dagger \hQ_{qQ}\hQ_{QQ}^\dagger \delmu \Sigma_H \,\Phi_c^\dagger \delmu \Sigma \rangle\)}
			\\[2ex]
			
			\midrule
			\multicolumn{4}{c}{
				Category III: 
				\(\Sigma_H^\dagger,\Phi_c,\Sigma^\dagger,\Sigma;\ 
				\hQ_{Qq},\hQ_{qq}^\dagger;\ 
				D^2\)
			} \\
			\midrule
			
			\(O_{3a}\) &
			\((\Sigma^\dagger \,\hQ_{qq}^\dagger T_L^A\, T_L^A \delmu\Sigma)\,
			(\Sigma_H^\dagger\hQ_{Qq}\delmu \Phi_c)\)
			&
			{\magenta \({\cal O}_{3a}^{\rm tr}\)} &
			{\magenta \(\langle \Sigma^\dagger\hQ_{qq}^\dagger \delmu\Sigma \rangle\,
			\langle \Sigma_H^\dagger\hQ_{Qq}\delmu \Phi_c\rangle\)}
			\\[2ex]
			
			\(O_{3b}\) &
			\((\Sigma^\dagger \,\hQ_{qq}^\dagger T_L^A\, \delmu\Sigma)\,
			(\Sigma_H^\dagger\hQ_{Qq} T_L^A \delmu \Phi_c)\)
			&
			{\purp \({\cal O}_{3b}^{\rm tr}\)} &
			{\purp \(\langle \Sigma^\dagger \delmu\Sigma \rangle\,
			\langle \Sigma_H^\dagger\hQ_{Qq}\hQ_{qq}^\dagger \delmu \Phi_c\rangle\)}
			\\[2ex]
			
			\(O_{3c}\) &
			\(f^{ABC}(\Sigma^\dagger \,\hQ_{qq}^\dagger T_L^A\, T_L^B \delmu\Sigma)\,
			(\Sigma_H^\dagger\hQ_{Qq} T_L^C \delmu \Phi_c)\)
			&
			{\blue \({\cal O}_{3c}^{\rm tr}\)} &
			{\blue \(\langle \Sigma^\dagger \delmu \Phi_c \Sigma_H^\dagger\hQ_{Qq}\hQ_{qq}^\dagger \delmu \Sigma\rangle\)}
			\\[2ex]
			
			\(O_{3d}\) &
			\(d^{ABC}(\Sigma^\dagger \,\hQ_{qq}^\dagger T_L^A\, T_L^B \delmu\Sigma)\,
			(\Sigma_H^\dagger\hQ_{Qq} T_L^C \delmu \Phi_c)\)
			&
			{\Green \({\cal O}_{3d}^{\rm tr}\)} &
			{\Green \(\langle \Sigma^\dagger\hQ_{qq}^\dagger \delmu \Phi_c\, \Sigma_H^\dagger\hQ_{Qq} \delmu \Sigma\rangle\)}
			\\[2ex]
			
			\midrule
			\multicolumn{4}{c}{
				Category IV: 
				\(\Sigma^\dagger,\Sigma,\Phi^\dagger,\Sigma;\ 
				\hQ_{qq}^\dagger,\hQ_{Qq};\ 
				D^2\)
			} \\
			\midrule
			
			\(O_{4a}\) &
			\((\Sigma^\dagger T_R^D \,\hQ_{qq}^\dagger T_L^A \, T_L^A \delmu\Sigma)
			(\Phi^\dagger T_R^D\hQ_{Qq} \delmu\Sigma)\)
			&
			{\magenta \({\cal O}_{4a}^{\rm tr}\)} &
			{\magenta \(\langle \Sigma^\dagger \hQ_{qq}^\dagger \delmu \Sigma \rangle\,
			\langle \Phi^\dagger \hQ_{Qq} \delmu \Sigma \rangle\)}
			\\[2ex]
			
			\(O_{4b}\) &
			\((\Sigma^\dagger T_R^D \,\hQ_{qq}^\dagger T_L^A \, \delmu\Sigma)
			(\Phi^\dagger T_R^D\hQ_{Qq} T_L^A \delmu\Sigma)\)
			&
			{\Green \({\cal O}_{4b}^{\rm tr}\)} &
			{\Green\(\langle \Sigma^\dagger \hQ_{qq}^\dagger \delmu \Sigma\,
			\Phi^\dagger \hQ_{Qq} \delmu \Sigma \rangle\)}
			\\[2ex]
			
			\(O_{4c}\) &
			\(f^{ABC}(\Sigma^\dagger T_R^D \,\hQ_{qq}^\dagger T_L^A \, T_L^B \delmu\Sigma)
			(\Phi^\dagger T_R^D\hQ_{Qq} T_L^C \delmu\Sigma)\)
			&
			{\blue \({\cal O}_{4c}^{\rm tr}\)} &
			{\blue\(\langle \Sigma^\dagger \delmu \Sigma\,
			\Phi^\dagger \hQ_{Qq} \hQ_{qq}^\dagger \delmu \Sigma \rangle\)}
			\\[2ex]
			
			\(O_{4d}\) &
			\(d^{ABC}(\Sigma^\dagger T_R^D \,\hQ_{qq}^\dagger T_L^A \, T_L^B \delmu\Sigma)
			(\Phi^\dagger T_R^D\hQ_{Qq} T_L^C \delmu\Sigma)\)
			&
			{\purp \({\cal O}_{4d}^{\rm tr}\)} &
			{\purp\(\langle \Sigma^\dagger \delmu \Sigma \rangle\,
			\langle \Phi^\dagger \hQ_{Qq} \hQ_{qq}^\dagger \delmu \Sigma \rangle\)}
			\\
			
			\bottomrule
		\end{tabular}%
	}
	\caption{\label{tab:traced} List of operators for $B\to P_1P_2$ in two formats: (i) Covariant and  (ii) Current-current traced.}
\end{table}
%%%%%%%%%%%%%%%%%%%%%%%%%%%%%%%%%%%%%%%%%%%%%%%%%%

Using $SU(3)$ trace identities and Fierz rearrangements, every four-field, two-derivative invariant can be expressed as a product of two mesonic bilinears with appropriate weak-spurion insertions. The resulting current-factorized (traced) basis is listed in Table~\ref{tab:traced}, alongside the covariant operators of Table~\ref{tab:BPP_oplist}. This two-current structure was the starting point of the earlier phenomenological construction in Ref.~\cite{Chakraborty:2026vdz}, where the mesonic operators were built directly as products of hadronized currents; the present analysis derives the same structure from the covariant basis, establishing the equivalence of the two formulations. Accordingly, the invariant counting of Sec.~\ref{sec:HS-OP} translates directly into
the traced format.

The two-current structure mirrors the SM weak Hamiltonian, where two quark currents connected by a $W$ boson—or, after penguin contraction, by a gluon—are mapped onto two mesonic vertices after hadronization. In this language, double-trace operators, $\langle\cdots\rangle\langle\cdots\rangle$, correspond to two independent color-singlet currents, while single-trace operators, $\langle\cdots\cdots\rangle$, encode connected color flow mediated by an intermediate color-octet exchange~\cite{Chakraborty:2026vdz}. The SM topology of a given operator is then fixed by three data:
\begin{itemize}
\item The connected vs.\ factorized trace structure, separating color-allowed from color-suppressed/exchange contractions.
\item The mediator representation [$(1,1,1,1)$ vs.\ $(\cdots,8,\cdots)$], separating electroweak-mediated SM from new physics (NP) roots.
\item Whether a spurion pair closes back into a single vertex: two $\widehat Q$'s at one vertex force the CKM charge to circulate, i.e.\ a quark loop, marking a penguin rather than a tree topology.
\end{itemize}
Together with the assignment of the external mesons to the $\Sigma$ fields, this decomposition uniquely identifies the corresponding Standard Model weak topology, distinguishing the tree ($T$) and color-suppressed ($C$), exchange ($E$) and annihilation ($A$), as well as penguin ($P$) and penguin-annihilation ($PA$) contributions.
%%%%%%%%%%%%%%%%%%%%%%%%%%%%%%%%%%%%%%%%%%%%%%%%%%
\begin{table}[h]
	\centering
	\renewcommand{\arraystretch}{1.2}
	\resizebox{.9\textwidth}{!}{%
	\begin{tabular}{p{0.20\textwidth} |  p{0.18\textwidth} || p{0.20\textwidth}  |p{0.21\textwidth}  ||p{0.18\textwidth} }
		\hline
		Left vertex & Mediator Rep.& Right vertex & Mediator Rep.& Comment \\
         & under $ \mathcal{G}_{3322}$ & & under $\mathcal{G}_{3322}$ &\\
		\hline

		\adjustbox{valign=c}{%
			\begin{tikzpicture}[line cap=round, line join=round, scale=0.7]
				
				% Box
				\draw[thick] (0,-0.8) rectangle (2,0.8);
				
				% Text inside the box
				\node at (1,0.25) {\small $Q_{qQ}$};
				\node at (1,-0.25) {\small $Q_{QQ}^\dagger$};
				
				% Left incoming lines from the middle of top and bottom edges
				\draw[thick] (-1,1.2) -- (1,0.8)
				node[midway, above left=6pt] {$\Sigma^\dagger$};
				
				\draw[thick, dashed] (-1,-1.2) -- (1,-0.8)
				node[midway, below left=6pt] {$\Phi$};
				
				% Right outgoing double red line from center of right edge
				\draw[red, double distance=3pt, line width=0.8pt] (2,0) -- (3.5,0);
				
			\end{tikzpicture}
		}
		&
		\adjustbox{valign=c}{%
			\begin{minipage}{0.22\textwidth}
				{\purp $(1, 1, 1, 1)$}, \\{\blue $(8, 8, 1, 1)$}, \\ {\blue $(8, 1, 1, 1)$}, \\ {\blue $(1, 8, 1, 1)$}
			\end{minipage}
		}
		&
		\adjustbox{valign=c}{%
			\begin{tikzpicture}[line cap=round, line join=round, scale=.8]
				\coordinate (V) at (0,0);
				
				\draw[thick] (-2,1) -- (V)
				node[midway, above left = 10pt] {$\Sigma^\dagger$};
				
				\draw[thick] (-2,-1) -- (V)
				node[midway, below left=10pt] {$\Sigma$};
				
				\draw[red, double distance=3pt, line width=0.8pt] (V) -- (2,0);
				
			\end{tikzpicture}
		}
		&
		\adjustbox{valign=c}{%
			\begin{minipage}{0.22\textwidth}
				{\purp $(1, 1, 1, 1)$}, \\ {\blue $(8, 8, 1, 1)$}, \\{\blue $(8, 1, 1, 1)$}, \\{\blue $(1, 8, 1, 1)$}
			\end{minipage}
		}
		&
		\adjustbox{valign=c}{%
			\begin{minipage}{0.18\textwidth}
				{\purp ${\cal O}_{1a}^{tr}$} $(P_{SM}, PA_{SM})$,\\ \\ {\blue ${\cal O}_{1b}^{tr}$}
			\end{minipage}
		}
		\\
		\hline
        %%%%%%%%%%%%%%%%%%%%%%%%%%%%%%%%%%%%%%%%%%%
        %%%%%%%%%%%%%%%%%%%%%%%%%%%%%%%%%%%%%%%%%%%
        
        \adjustbox{valign=c}{%
			\begin{tikzpicture}[line cap=round, line join=round, scale=0.7]
				
				% Box
				\draw[thick] (0,-0.8) rectangle (2,0.8);
				
				% Text inside the box
				\node at (1,0.25) {\small $Q_{qQ}$};
				\node at (1,-0.25) {\small $Q_{QQ}^\dagger$};
				
				% Left incoming lines from the middle of top and bottom edges
				\draw[thick, dashed] (-1,1.2) -- (1,0.8)
				node[midway, above left=6pt] {$\Phi_c^\dagger$};
				
				\draw[thick, dash dot] (-1,-1.2) -- (1,-0.8)
				node[midway, below left=6pt] {$\Sigma_H$};
				
				% Right outgoing double red line from center of right edge
				\draw[red, double distance=3pt, line width=0.8pt] (2,0) -- (3.5,0);
				
			\end{tikzpicture}
		}
		&
		\adjustbox{valign=c}{%
			\begin{minipage}{0.22\textwidth}
				{\purp $(1, 1, 1, 1)$}, \\{\blue $(8, 8, 1, 1)$}, \\ {\blue $(8, 1, 1, 1)$}, \\ {\blue $(1, 8, 1, 1)$}
			\end{minipage}
		}
		&
		\adjustbox{valign=c}{%
			\begin{tikzpicture}[line cap=round, line join=round, scale=.8]
				\coordinate (V) at (0,0);
				
				\draw[thick] (-2,1) -- (V)
				node[midway, above left = 10pt] {$\Sigma^\dagger$};
				
				\draw[thick] (-2,-1) -- (V)
				node[midway, below left=10pt] {$\Sigma$};
				
				\draw[red, double distance=3pt, line width=0.8pt] (V) -- (1.8,0);
				
			\end{tikzpicture}
		}
		&
		\adjustbox{valign=c}{%
			\begin{minipage}{0.22\textwidth}
				{\purp $(1, 1, 1, 1)$}, \\{\blue $(8, 8, 1, 1)$}, \\ {\blue $(8, 1, 1, 1)$}, \\ {\blue $(1, 8, 1, 1)$}
			\end{minipage}
		}
		&
		\adjustbox{valign=c}{%
			\begin{minipage}{0.18\textwidth}
				 {\purp ${\cal O}_{2a}^{tr}$} $(PA_{SM})$,\\ \\ {\blue ${\cal O}_{2b}^{tr}$}
			\end{minipage}
		}
		\\
		\hline
        %%%%%%%%%%%%%%%%%%%%%%%%%%%%%%%%%%%%%%%%%%
        %%%%%%%%%%%%%%%%%%%%%%%%%%%%%%%%%%%%%%%%%%

        \adjustbox{valign=c}{%
			\begin{tikzpicture}[line cap=round, line join=round, scale=.8]
				\coordinate (V) at (0,0);
				
				\draw[thick, dash dot] (-2,1) -- (V)
				node[midway, above left = 10pt] {$\Sigma_H^\dagger$};
				
				\draw[thick, dashed] (-2,-1) -- (V)
				node[midway, below left=10pt] {$\Phi_c$};
				
				\draw[red, double distance=3pt, line width=0.8pt] (V) -- (1.8,0);
				
				\node[draw, fill=black, regular polygon, inner sep=2pt] at (V) {}
				node[left=10pt] {$\hQ_{Qq}$};
			\end{tikzpicture}
		}
		&
		\adjustbox{valign=c}{%
			\begin{minipage}{0.22\textwidth}
				{\magenta $(1, 1, 1, 1)$}, \\{\Green $(8, 8, 1, 1)$}, \\{\Green $(8, 1, 1, 1)$}, \\{\Green $(1, 8, 1, 1)$}
			\end{minipage}
		}
		& 
		\adjustbox{valign=c}{%
			\begin{tikzpicture}[line cap=round, line join=round, scale=.8]
				\coordinate (V) at (0,0);
				
				\draw[thick] (-2,1) -- (V)
				node[midway, above left = 10pt] {$\Sigma^\dagger$};
				
				\draw[thick] (-2,-1) -- (V)
				node[midway, below left=10pt] {$\Sigma$};
				
				\draw[red, double distance=3pt, line width=0.8pt] (V) -- (2,0);
				
				\node[draw, fill=black, star, inner sep=2pt] at (V) {}
				node[left=10pt] {$\hQ_{qq}^\dagger$};
			\end{tikzpicture}
		}
		&
		\adjustbox{valign=c}{%
			\begin{minipage}{0.22\textwidth}
			{\magenta $(1, 1, 1, 1)$}, \\{\Green $(8, 8, 1, 1)$}, \\{\Green $(8, 1, 1, 1)$}, \\{\Green $(1, 8, 1, 1)$}
			\end{minipage}
		}
		&
		\adjustbox{valign=c}{%
			\begin{minipage}{0.18\textwidth}
				{\magenta ${\cal O}_{3a}^{tr}$} $(A_{SM})$, \\ \\ {\Green ${\cal  O}_{3d}^{tr}$}
			\end{minipage}
		}
		
		\\
		
		\hdashline

        	\adjustbox{valign=c}{%
			\begin{tikzpicture}[line cap=round, line join=round, scale=0.7]
				
				% Box
				\draw[thick] (0,-0.8) rectangle (2,0.8);
				
				% Text inside the box
				\node at (1,0.25) {\small $Q_{Qq}$};
				\node at (1,-0.25) {\small $Q_{qq}^\dagger$};
				
				% Left incoming lines from the middle of top and bottom edges
				\draw[thick, dash dot] (-1,1.2) -- (1,0.8)
				node[midway, above left=6pt] {$\Sigma^\dagger$};
				
				\draw[thick, dashed] (-1,-1.2) -- (1,-0.8)
				node[midway, below left=6pt] {$\Phi_c$};
				
				% Right outgoing double red line from center of right edge
				\draw[red, double distance=3pt, line width=0.8pt] (2,0) -- (3.5,0);
				
			\end{tikzpicture}
		}
		&
		\adjustbox{valign=c}{%
			\begin{minipage}{0.22\textwidth}
				{\purp $(1, 1, 1, 1)$}, \\{\blue $(8, 8, 1, 1)$}, \\{\blue $(8, 1, 1, 1)$}, \\{\blue $(1, 8, 1, 1)$}
			\end{minipage}
		}
		&
		\adjustbox{valign=c}{%
			\begin{tikzpicture}[line cap=round, line join=round, scale=.8]
				\coordinate (V) at (0,0);
				
				\draw[thick] (-2,1) -- (V)
				node[midway, above left = 10pt] {$\Sigma^\dagger$};
				
				\draw[thick] (-2,-1) -- (V)
				node[midway, below left=10pt] {$\Sigma$};
				
				\draw[red, double distance=3pt, line width=0.8pt] (V) -- (2,0);
				
			\end{tikzpicture}
		}
		&
		\adjustbox{valign=c}{%
			\begin{minipage}{0.22\textwidth}
				{\purp $(1, 1, 1, 1)$}, \\{\blue $(8, 8, 1, 1)$}, \\{\blue $(8, 1, 1, 1)$}, \\{\blue $(1, 8, 1, 1)$}
			\end{minipage}
		}
		&
		\adjustbox{valign=c}{%
			\begin{minipage}{0.18\textwidth}
				{\purp ${\cal O}_{3b}^{tr}$} $(P_{SM}, PA_{SM})$, \\ {\blue ${\cal O}_{3c}^{tr}$} 
			\end{minipage}
		}
		\\
		\hline
        %%%%%%%%%%%%%%%%%%%%%%%%%%%%%%%%%%%%%
        %%%%%%%%%%%%%%%%%%%%%%%%%%%%%%%%%%%%%

		\adjustbox{valign=c}{%
			\begin{tikzpicture}[line cap=round, line join=round, scale=.8]
				\coordinate (V) at (0,0);
				
				\draw[thick, dashed] (-2,1) -- (V)
				node[midway, above left = 10pt] {$\Phi^\dagger$};
				
				\draw[thick] (-2,-1) -- (V)
				node[midway, below left=10pt] {$\Sigma$};
				
				\draw[red, double distance=3pt, line width=0.8pt] (V) -- (1.8,0);
				
				%\filldraw[black] (V) circle (4.2pt)
				\node[draw, fill=black, regular polygon, inner sep=2pt] at (V) {}
				node[left=10pt] {$\hQ_{Qq}$};

			\end{tikzpicture}
		}
		&
		\adjustbox{valign=c}{%
			\begin{minipage}{0.22\textwidth}
				{\magenta $(1, 1, 1, 1)$}, \\{\Green $(8, 8, 1, 1)$}, \\{\Green $(8, 1, 1, 1)$}, \\{\Green $(1, 8, 1, 1)$}
			\end{minipage}
		}
		& 
		\adjustbox{valign=c}{%
			\begin{tikzpicture}[line cap=round, line join=round, scale=.8]
				\coordinate (V) at (0,0);
				
				\draw[thick] (-2,1) -- (V)
				node[midway, above left = 10pt] {$\Sigma^\dagger$};
				
				\draw[thick] (-2,-1) -- (V)
				node[midway, below left=10pt] {$\Sigma$};
				
				\draw[red, double distance=3pt, line width=0.8pt] (V) -- (2,0);
				
				\node[draw, fill=black, star, inner sep=2pt] at (V) {}
				node[left=10pt] {$\hQ_{qq}^\dagger$};
			\end{tikzpicture}
		}
		&
		\adjustbox{valign=c}{%
			\begin{minipage}{0.22\textwidth}
				{\magenta $(1, 1, 1, 1)$}, \\{\Green $(8, 8, 1, 1)$}, \\{\Green $(8, 1, 1, 1)$}, \\{\Green $(1, 8, 1, 1)$}
			\end{minipage}
		}
		&
		\adjustbox{valign=c}{%
			\begin{minipage}{0.18\textwidth}
				{\magenta ${\cal O}_{4a}^{tr}$} $(T_{SM}, A_{SM})$, \\ \\{\Green ${\cal O}_{4b}^{tr}$}.
			\end{minipage}
		}
		
		\\
		
		\hdashline

		\adjustbox{valign=c}{%
			\begin{tikzpicture}[line cap=round, line join=round, scale=0.7]
				
				% Box
				\draw[thick] (0,-0.8) rectangle (2,0.8);
				
				% Text inside the box
				\node at (1,0.25) {\small $Q_{Qq}$};
				\node at (1,-0.25) {\small $Q_{qq}^\dagger$};
				
				% Left incoming lines from the middle of top and bottom edges
				\draw[thick, dashed] (-1,1.2) -- (1,0.8)
				node[midway, above left=6pt] {$\Phi^\dagger$};
				
				\draw[thick] (-1,-1.2) -- (1,-0.8)
				node[midway, below left=6pt] {$\Sigma$};
				
				% Right outgoing double red line from center of right edge
				\draw[red, double distance=3pt, line width=0.8pt] (2,0) -- (3.5,0);
				
			\end{tikzpicture}
		}
		&
		\adjustbox{valign=c}{%
			\begin{minipage}{0.22\textwidth}
				{\purp $(1, 1, 1, 1)$}, \\{\blue $(8, 8, 1, 1)$}, \\{\blue $(8, 1, 1, 1)$}, \\{\blue $(1, 8, 1, 1)$}
			\end{minipage}
		}
		&
		\adjustbox{valign=c}{%
			\begin{tikzpicture}[line cap=round, line join=round, scale=.8]
				\coordinate (V) at (0,0);
				
				\draw[thick] (-2,1) -- (V)
				node[midway, above left = 10pt] {$\Sigma^\dagger$};
				
				\draw[thick] (-2,-1) -- (V)
				node[midway, below left=10pt] {$\Sigma$};
				
				\draw[red, double distance=3pt, line width=0.8pt] (V) -- (2,0);
				
				%				\node[] at (V) {}
				%				node[left=10pt] {$\hQ_{QQ}^\dagger$};
			\end{tikzpicture}
		}
		&
		\adjustbox{valign=c}{%
			\begin{minipage}{0.22\textwidth}
				{\purp $(1, 1, 1, 1)$}, \\{\blue $(8, 8, 1, 1)$}, \\{\blue $(8, 1, 1, 1)$}, \\{\blue $(1, 8, 1, 1)$}
			\end{minipage}
		}
		&
		\adjustbox{valign=c}{%
			\begin{minipage}{0.20\textwidth}
				{\purp ${\cal O}_{4d}^{tr}$}($P_{SM}$, $PA_{SM}$), \\ \\ {\blue ${\cal O}_{4c}^{tr}$}
			\end{minipage}
		}
		\\
		\hline		
	\end{tabular}}
    \caption{\label{tab:unfoldingop4} UV roots of effective operators I: possible heavy fields as tree-mediators. It is worth mentioning that our HS construction rely on the flavor symmetry where the SM gauge symmetry is not explicitly employed. Thus, the virtual gluon (that are flavor singlet) exchange is not explicitly captured, but suitably masked within the WCs of these operators.}
\end{table}
%%%%%%%%%%%%%%%%%%%%%%%%%%%%%%%%%%%%%%%%%%%%%%%%%%

Each row of Table~\ref{tab:unfoldingop4} splits one contact operator into two flavor bilinear currents---a left and a right vertex---joined by a mediator carrying the representation shown. The diagrams accompanying each mediator charge in Table~\ref{tab:unfoldingop4} show the corresponding quark flow; the symbols on the internal and external lines follow the notation as below.

%%%%%%%%%%%%%%%%%%%%%%%%%%%%%%%%%%%%%%%%%%%%%%%%%%
\begin{figure}[H]
\tikz[baseline=-0.5ex]{\filldraw[black] (0,0) circle (4.2pt);}, \tikz[baseline=-0.5ex]{\node[draw, fill=black, diamond, inner sep=2.2pt] at (0,0) {};}, \tikz[baseline=-0.5ex]{\node[draw, fill=black, regular polygon, inner sep=2.2pt] at (0,0) {};}, \tikz[baseline=-0.5ex]{\node[draw, fill=black, star, inner sep=2.2pt] at (0,0) {};}
	\hspace{0.4em} denote  spurion insertions, \(\hQ_{qQ}\), \(\hQ_{QQ}\), \(\hQ_{Qq}\) and \(\hQ_{qq}\), respectively.\\
\tikz[baseline=-0.5ex]{\draw[thick] (0,0) -- (1.0,0);}
	\hspace{0.4em} denotes a \(\Sigma\)-type light meson field.\\
\tikz[baseline=-0.5ex]{\draw[thick, dashed] (0,0) -- (1.0,0);}
	\hspace{0.4em} denotes a \(\Phi\) or \(\Phi_c\)-type heavy-light meson field.\\
\tikz[baseline=-0.5ex]{\draw[thick, dash dot] (0,0) -- (1.0,0);}
	\hspace{0.4em} denotes a \(\Sigma_H\)-type heavy-heavy meson field.\\
\tikz[baseline=-0.5ex]{\draw[red, double distance=2pt, line width=0.6pt] (0,0) -- (1.0,0);}
	\hspace{0.4em} denotes a mediator line carrying the relevant representation.
\end{figure}
%%%%%%%%%%%%%%%%%%%%%%%%%%%%%%%%%%%%%%%%%%%%%%%%%%

The two vertices represent the mesonic counterparts of the quark currents in the underlying weak transition, while the intermediate field carries the corresponding flavor quantum numbers. Each row is interpreted by assigning the external mesons to the $\Sigma$ fields at  two vertices, thereby identifying the associated quark-flow topology. The corresponding SM realizations are summarized in the ``Comment'' column.

As an illustration, consider the singlet-mediator contribution $(1,1,1,1)$ associated with $O^{\rm tr}_{1a,1b}$. To obtain a non-vanishing contribution from the right-hand trace, $\langle\Sigma^\dagger\overset{\leftrightarrow}{\partial}\mu\Sigma\rangle$, both $\Sigma$ fields must correspond to external pseudoscalar mesons, so neither can be replaced by its background value. Consequently, the $\Sigma^\dagger$ field in the left-hand trace is fixed to the identity. The two final-state mesons are therefore emitted from the same current, while the second current annihilates the initial $B$ meson, yielding the penguin-annihilation ($PA$) topology in the SM.

%%%%%%%%%%%%%%%%%%%%%%%%%%%%%%%%%%%%%%%%%%%%%%%%
\subsection*{Unfolding of effective operators for $B\to D P$}
%%%%%%%%%%%%%%%%%%%%%%%%%%%%%%%%%%%%%%%%%%%%%%%%
For $B\to DP$ we do not carry out the full unfolding of every operator; in Table~\ref{tab:tracedBDP}, categories~V and~VI list only a few representative roots (e.g.\ $O^{\rm tr}_{5a,5b}$ and $O^{\rm tr}_{6a}$) to illustrate how the heavy--light vertex enters. The heavy line now acts as a near-spectator, so the surviving contractions are fewer than in the charmless case, and the examples shown suffice to fix the pattern.

%%%%%%%%%%%%%%%%%%%%%%%%%%%%%%%%%%%%%%%%%%%%%%%%%%
\begin{table}[h]
	\centering
	\renewcommand{\arraystretch}{1.0}
	\setlength{\tabcolsep}{3pt}
	\resizebox{0.9\textwidth}{!}{%
		\begin{tabular}{cccc}
			\toprule
			Op. & Covariant form & Op. & Traced form basis~\cite{Chakraborty:2026vdz} \\
			\midrule
			
			\multicolumn{4}{c}{
				Category V:
				\( \Sigma^\dagger,\Sigma,\Phi^\dagger,\Phi;\ 
				\{\hQ_{QQ},\hQ_{qq}^\dagger\};\ 
				D^2\)
			} \\
			\midrule
			
			\(O_{5a}\) &
			\((\Sigma^\dagger \langle \hQ_{qq} \cdot T_L^A\rangle T_L^A \delmu \Sigma)
			(\Phi^\dagger \,\hQ_{QQ}^\dagger \tau_L^I \, \tau_L^I \delmu \Phi)\)
			&
			{\magenta \({\cal O}_{5a}^{\rm tr}\)} &
			{\magenta
				\(\langle \Sigma^\dagger \hQ_{qq}^\dagger \delmu \Sigma\rangle\,
				\langle \Phi^\dagger \hQ_{QQ}^\dagger \delmu \Phi\rangle\)}
			\\[2ex]
			
			\(O_{5b}\) &
			\((\Sigma^\dagger \langle \hQ_{qq} \cdot T_L^A\rangle T_L^A T_R^B\delmu \Sigma)
			(\Phi^\dagger \,\hQ_{QQ}^\dagger \tau_L^I \, \tau_L^I T_R^B \delmu \Phi)\)
			&
			{\Green \({\cal O}_{5b}^{\rm tr}\)} &
			{\Green
				\(\langle \Sigma^\dagger \hQ_{qq}^\dagger \delmu \Sigma\,
				\Phi^\dagger \hQ_{QQ}^\dagger \delmu \Phi\rangle\)}
			\\[2ex]
			
			\midrule
			\multicolumn{4}{c}{
				Category VI:
				\( \Sigma_H^\dagger,\Sigma,\Phi^\dagger,\Phi_c;\ 
				\{\hQ_{QQ},\hQ_{qq}^\dagger\};\ 
				D^2\)
			} \\
			\midrule
			
			\(O_{6a}\) &
			\((\Sigma^\dagger \langle \hQ_{qq} \cdot T_L^A\rangle T_L^A \delmu \Phi_c)
			(\Sigma_H^\dagger \,\hQ_{QQ}^\dagger \tau_L^I \, \tau_L^I \delmu \Phi)\)
			&
			{\red \({\cal O}_{6a}^{\rm tr}\)} &
			{\red
				\(\langle \Sigma^\dagger \hQ_{qq}^\dagger \delmu \Phi_c \, \Sigma_H^\dagger \hQ_{QQ}^\dagger \delmu \Phi\rangle\)}
			\\[2ex]
			
			\bottomrule
		\end{tabular}%
	}
	\caption{\label{tab:tracedBDP} List of operators for $B\to DP$ in two formats: (i) Covariant and  (ii) Current-current traced. }
\end{table}
%%%%%%%%%%%%%%%%%%%%%%%%%%%%%%%%%%%%%%%%%%%%%%%%%%

%%%%%%%%%%%%%%%%%%%%%%%%%%%%%%%%%%%%%%%%%%%%%%%%%%
\begin{table}[h]
	\centering
	\renewcommand{\arraystretch}{1.4}
	\resizebox{\textwidth}{!}{%
		\begin{tabular}{p{0.20\textwidth} | p{0.21\textwidth} || p{0.20\textwidth} | p{0.21\textwidth}  |p{0.18\textwidth} }
			\hline
			\multicolumn{5}{c}{Category V: $\Sigma^\dagger$, $\Sigma$,$\Phi^\dagger$, $\Phi$, $\hQ_{qq}$, $\hQ_{QQ}^\dagger$}\\
			\hline
			Left vertex & Mediator Rep. & Right vertex & Mediator Rep. & Comment \\
               & under $ \mathcal{G}_{3322}$ & & under $\mathcal{G}_{3322}$ &\\
			\hline
			\adjustbox{valign=c}{%
				\begin{tikzpicture}[line cap=round, line join=round, scale=.8]
					\coordinate (V) at (0,0);
					
					\draw[thick] (-2,1) -- (V)
					node[midway, above left = 10pt] {$\Sigma^\dagger$};
					
					\draw[thick] (-2,-1) -- (V)
					node[midway, below left=10pt] {$\Sigma$};
					
					\draw[red, double distance=3pt, line width=0.8pt] (V) -- (1.8,0);
					
					%\filldraw[black] (V) circle (4.2pt)
					\node[draw, fill=black,star, inner sep=2pt] at (V) {}
					node[left=10pt] {$\hQ_{qq}$};

				\end{tikzpicture}
			}
			&
			\adjustbox{valign=c}{%
				\begin{minipage}{0.22\textwidth}
					{\magenta $(1, 1, 1, 1)$}, \\{\Green $(1, 8, 1, 1)$}
				\end{minipage}
			}
			& 
			\adjustbox{valign=c}{%
				\begin{tikzpicture}[line cap=round, line join=round, scale=.8]
					\coordinate (V) at (0,0);
					
					\draw[thick, dashed] (-2,1) -- (V)
					node[midway, above left = 10pt] {$\Phi^\dagger$};
					
					\draw[thick, dashed] (-2,-1) -- (V)
					node[midway, below left=10pt] {$\Phi$};
					
					\draw[red, double distance=3pt, line width=0.8pt] (V) -- (2,0);
					
					\node[draw, fill=black, diamond, inner sep=2pt] at (V) {}
					node[left=10pt] {$\hQ_{QQ}^\dagger$};
				\end{tikzpicture}
			}
			&
			\adjustbox{valign=c}{%
				\begin{minipage}{0.22\textwidth}
					{\magenta $(1, 1, 1, 1)$}, \\{\Green $(1, 8, 1, 1)$}
				\end{minipage}
			}
			&
			\adjustbox{valign=c}{%
				\begin{minipage}{0.18\textwidth}
					{\magenta${\cal O}_{5a}^{tr}$} $(T_{SM})$,\\ \\ {\Green ${\cal O}_{5b}^{tr}$}
				\end{minipage}
			}
			
			\\
			
			\hline
			\multicolumn{5}{c}{Category VI: $\Sigma^\dagger$, $\Phi_c$,$\Sigma_H^\dagger$, $\Phi$, $\hQ_{qq}$, $\hQ_{QQ}^\dagger$}\\
			\hline
			\adjustbox{valign=c}{%
				\begin{tikzpicture}[line cap=round, line join=round, scale=.8]
					\coordinate (V) at (0,0);
					
					\draw[thick] (-2,1) -- (V)
					node[midway, above left = 10pt] {$\Sigma^\dagger$};
					
					\draw[thick, dashed] (-2,-1) -- (V)
					node[midway, below left=10pt] {$\Phi_c$};
					
					\draw[red, double distance=3pt, line width=0.8pt] (V) -- (1.8,0);
					
					\node[draw, fill=black, star, inner sep=2pt] at (V) {}
					node[left=10pt] {$\hQ_{qq}$};
				\end{tikzpicture}
			}
			&
			\adjustbox{valign=c}{%
				\begin{minipage}{0.22\textwidth}
					{\red $(1, 3,1,2)$}
				\end{minipage}
			}
			& 
			\adjustbox{valign=c}{%
				\begin{tikzpicture}[line cap=round, line join=round, scale=.8]
					\coordinate (V) at (0,0);
					
					\draw[thick, dash dot] (-2,1) -- (V)
					node[midway, above left = 10pt] {$\Sigma_H^\dagger$};
					
					\draw[thick, dashed] (-2,-1) -- (V)
					node[midway, below left=10pt] {$\Phi$};
					
					\draw[red, double distance=3pt, line width=0.8pt] (V) -- (2,0);
					
					\node[draw, fill=black, diamond, inner sep=2pt] at (V) {}
					node[left=10pt] {$\hQ_{QQ}^\dagger$};
				\end{tikzpicture}
			}
			&
			\adjustbox{valign=c}{%
				\begin{minipage}{0.22\textwidth}
					{\red $(1, \bar 3,1,2)$}
				\end{minipage}
			}
			&
			\adjustbox{valign=c}{%
				\begin{minipage}{0.18\textwidth}
					{\red ${\cal O}_{6a}^{tr}$}
				\end{minipage}
			}\\
			\hline
	\end{tabular}}
    \caption{UV roots of effective operators II: possible heavy fields as tree-mediators. }
\end{table}
%%%%%%%%%%%%%%%%%%%%%%%%%%%%%%%%%%%%%%%%%%%%%%%%

%%%%%%%%%%%%%%%%%%%%%%%%%%%%%%%%%%%%%%%%%%%%%%%%%%
\section{Comparison}
\label{sec:comparison}
%%%%%%%%%%%%%%%%%%%%%%%%%%%%%%%%%%%%%%%%%%%%%%%%
%---------------------
Physical $B\to P_1P_2$ and $B\to P_1P_2P_3$ amplitudes satisfy flavor relations because the number of decay channels exceeds the number of independent hadronic parameters. These relations are commonly derived using the topological-amplitude approach~\cite{Chau:1982da, Chau:1987tk}, in which quark-flow contributions are classified as
\begin{equation}
T,\; C,\; P,\; E,\; A,\; PA\;,
\label{eq:topologies}
\end{equation}
corresponding to the color-allowed tree, color-suppressed tree, QCD penguin, exchange, annihilation, and penguin-annihilation topologies. Although six topological amplitudes are introduced, only five independent linear combinations contribute to the physical decay amplitudes~\cite{Gronau:1994rj}.

Alternatively, one may adopt a symmetry-driven approach~\cite{Zeppenfeld:1980ex,Savage:1989ub,Grinstein:1996us}, in which the weak Hamiltonian is decomposed into irreducible representations of $SU(3)_V$, the final-state mesons are organized into irreducible multiplets, and the Wigner--Eckart theorem is used to express each decay amplitude as a product of a reduced matrix element and an $SU(3)$ Clebsch--Gordan coefficient. Thus, in both the topological and symmetry-driven approaches, one first constructs a basis of reduced amplitudes and subsequently derives the physical decay amplitudes through the appropriate topological or Clebsch--Gordan decomposition.

In the present framework, these two steps are unified. The independent flavor-invariant operators constitute the reduced amplitudes, as they form the complete basis of flavor-covariant structures at a given order in the EFT expansion. Physical decay amplitudes are obtained directly by projecting these operators onto the desired external states through a straightforward expansion of the meson multiplets. Consequently, the amplitude relations emerge as the null space of the resulting projection matrix, eliminating the need for a separate reduced-amplitude basis or an explicit mode-by-mode evaluation of $SU(3)$ Clebsch--Gordan coefficients.

For the charmless two-body modes, the twenty-four amplitudes in Table~\ref{tab:BPPamps24} are governed by five independent operator structures. Once the CKM factors are attached, the resulting matrix of physical amplitudes has $\mathrm{rank}(M_{\rm CKM})=5$, resulting in the nineteen sum rules in eqs.~\eqref{eq:iso1}--\eqref{eq:uss5}. The familiar $B\to\pi\pi$ isospin triangles, the $K\pi$ quadrangle and the $\Delta I=3/2$ $K\pi$--$\pi\pi$ relation with its CKM ratio emerge as particular null vectors, while the remaining, less familiar identities follow automatically. Among the latter, eleven relations involve an $\eta_8$ in the final state. Because $\eta_8$ is an isosinglet and is not a $U$-spin eigenstate, they have no isospin or U-spin counterpart and require the full octet structure. Conventional analyses involving $\eta^{(\prime)}$ work directly with the physical mass eigenstates after octet--singlet mixing~\cite{Dighe:1995gq}, so relations for the pure octet $\eta_8$ are rarely written out. Several of them are nevertheless contained, as relations between CKM components, in the systematic $SU(3)$ catalogue of Ref.~\cite{Raz:2005hu}. The four involving $\bar B_s$ decays, eqs.~\eqref{eq:us2}, \eqref{eq:us3}, \eqref{eq:su12} and \eqref{eq:uss5}, are derivable from published tabulations but have not, to our knowledge, appeared explicitly. The full-$SU(3)$ relations mixing $\pi^0$ and kaon channels, as well as those probing $\pi^0$--$\eta_8$ mixing, are likewise reproduced within the same construction. Unlike conventional analyses, which typically present only selected relations, the null-space construction yields the complete set of leading-order flavor sum rules.

Comparing the counting with the SM topology analysis unveils one structural difference between the two frameworks. With CKM factors retained, the six topologies of eq.~\eqref{eq:topologies} together with their $\lambda_u$ and $\lambda_c$ penguin components span seven independent directions over the twenty-four modes, whereas the five operator structures span only five, so that the topological parametrization admits seventeen physical sum rules against our nineteen. The seventeen sum rules shared between the two frameworks, including every relation involving an $\eta_8$, 
%other than eq.~\eqref{eq:su11}, 
provide a nontrivial consistency check of the unfolding procedure described in Sec.~\ref{sec:unfolding}. 

The two relations eqs.~\eqref{eq:su3},
%\eqref{eq:su11} 
and \eqref{eq:us3} are the imprint of the chiral operator basis beyond the topological parametrization. 
Evaluated on the topological amplitudes, eqs.~\eqref{eq:su3} and eq.~\eqref{eq:us3} impose the following conditions 
\begin{align}
E + A &= 3\,(T + C)\,,
\label{eq:constraint1}\\
3\,C - 3\,E - 2\,A &= 6\,P_u + 4\,PA_u\,,
\label{eq:constraint2}
\end{align}
where $P_u$ and $PA_u$ denote the $\lambda_u$-components of the penguin amplitudes. The origin of these two constraints is the different symmetry input of the two frameworks. The topological decomposition, like the $SU(3)$ reduced-matrix-element analysis, applies the Wigner--Eckart theorem under $SU(3)_V$ to the external states. Every reduced amplitude allowed by the diagonal flavor group is an independent parameter. The two-octet final state is restricted to the symmetric product $\mathbf 1\oplus\mathbf 8_S\oplus\mathbf{27}$ and leaving five tree-class structures. In the present construction the operators are built from fields transforming under the full $SU(3)_L\times SU(3)_R$ and at leading order locking the 27-plet and the 15-sourced octet reduced amplitudes in a fixed ratio. As a result, the constraint in eq.\,\eqref{eq:constraint1} requires
$E+A$ to be as large as $3(T+C)$. In heavy-quark power counting $E$ and $A$ are $\mathcal O(\Lambda_{\rm QCD}/m_b)$ while $T$ and $C$ are leading power, so eq.~\eqref{eq:constraint1} is expected to be violated at $\mathcal O(1)$ for the hard final-state mesons of $B$ decays. The three chiral-specific
relations should therefore be read not as robust predictions but as the
falsifiable imprint of the leading-order chiral truncation.

The advantages of the present framework become even more evident for three-body decays. The thirty-two charmless amplitudes in Table~\ref{tab:BPPPamps} are generated by six operators with $\mathrm{rank}(M)=6$, yielding twenty-six independent sum rules. Unlike the conventional $SU(3)_V$ analysis~\cite{Bhattacharya:2014eca}, which requires the decomposition of the fully symmetric product $(\mathbf{8}\otimes\mathbf{8}\otimes\mathbf{8})_{FS}$ before reduced amplitudes can be defined, the present approach works directly with local flavor-invariant operators. Bose symmetry among identical pseudoscalars is therefore incorporated automatically through the field expansion, with the only additional ingredient being the standard normalization factor $\sqrt{n!}$ for $n$ identical mesons. The two constructions also differ in their domain of validity across the Dalitz plot. Sum rules derived by applying the symmetry directly to the physical states, such as the isospin relations of ref.~\cite{Wang:2024tnx}, constrain the full momentum-dependent amplitudes and remain valid in the presence of intermediate resonances and cascade contributions, $B\to RP\to P_1P_2P_3$. In the current analysis, the relations derived in Sec.~\ref{sec:3body} are instead momentum-independent statements about the leading contact amplitudes.

Moreover, the relations obtained in the present framework are more restrictive than those implied by exact $SU(3)$ symmetry alone, since the $D6$ chiral operator basis is smaller than the most general $SU(3)_V$ reduced-amplitude space. Consequently, additional flavor relations emerge. In particular, relations involving $\eta_8$ or multiple strange mesons are leading-order predictions of the mesonic EFT and receive corrections from operators beyond $D6$, mass-spurion insertions, additional trace structures, and octet--singlet mixing. By contrast, the familiar isospin and $U$-spin relations remain valid beyond leading order. Our conventions reproduce the ten $B\to PPP$ relations of Ref.~\cite{He:2014xha}, while a subset agrees with Ref.~\cite{Bhattacharya:2014eca}, up to the $-\bar u$ convention adopted there.

These differences reflect the underlying symmetry organization. Conventional $SU(3)_V$ analyses rely solely on the transformation properties of the weak Hamiltonian and do not incorporate chiral power counting, so a given reduced amplitude generally receives contributions from multiple chiral orders. In contrast, the present framework is formulated within the chiral symmetry-breaking pattern $SU(3)_L\times SU(3)_R\to SU(3)_V$, with $\Sigma\to L\Sigma R^\dagger$, and introduces electroweak interactions through spurions with definite chiral transformation properties. The resulting operator basis is therefore more restrictive at fixed order, leading to additional flavor relations.

The unfolding procedure of Sec.~\ref{sec:unfolding} provides the inverse perspective to the conventional topological approach. Rather than taking quark-flow diagrams as the starting point, it begins with the effective operators whose underlying weak topologies are revealed by their current-factorized form. Decomposing a four-field operator into two mesonic bilinears connected by an intermediate mediator exposes the corresponding Standard Model realization. The mediator representation distinguishes electroweak singlet exchange $(\mathbf{1},\mathbf{1},\mathbf{1},\mathbf{1})$, from nontrivial flavor representations, while the manner in which the spurion pair contracts identifies tree and penguin structures. This classification also isolates operators with no Standard Model tree-level origin. In particular, a mediator transforming as $(\mathbf{1},\mathbf{8},\mathbf{1},\mathbf{1})$ admits no electroweak realization and, therefore, signals a genuinely new-physics tree contribution, a feature not apparent in the conventional topology-based classification.

%%%%%%%%%%%%%%%%%%%%%%%%%%%%%%%%%
\section{Conclusions and outlook }
\label{sec:conclusion}
%%%%%%%%%%%%%%%%%%%%%%%%%%%%%%%%%

In this paper, we compute the non-redundant operator basis for pseudoscalar mesons. We adopt the methodology of calculating $D6$ operators using Hilbert series. We encapsulate the pseudo-scalar mesons within meson multiplets $(\Sigma,\Sigma_H,\Phi, \Phi_c)$ under global flavor symmetry $SU(3)_L \times SU(3)_R \times SU(2)_{L_H} \times SU(2)_{R_H}$. We also introduce the CKM-elements as spurions which transform under this symmetry to capture the footprints of the SM weak interactions. In this work, we focus on the computation of operators that consist of four meson multiplets and two derivatives. These are local composite operators and, on expansion, lead to interactions among four or fewer meson fields at the leading order. Thus, the Wilson coefficients of these operators, suitably blended by numerical factors originate from generators of the groups, can be identified as the amplitude associated with two- and three-body decays of relatively heavier mesons. 

Relying on this property, we list all the relevant and complete sets of amplitudes for all such decays. As an example, we demonstrate the two-body decay channels (with and without charm-quarks) of the $B$ meson, and calculate an extensive set of sum rules. We also analyse the three-body decays of $B$-meson (with and without charm-quarks), and note down the corresponding sum rules. We show that these sum rules are generic and do not depend on the choice of on-shell momenta of the mesons. We highlight our agreement in sum rules with existing ones and also comment on the {\it new findings}. Although we explicitly compute the $B$-decay related sum rules, but we list the complete list of operators that would allow the readers to deduce all other sum rules involving these set of meson fields. In that sense, our result is complete and exhaustive.

We also propose a method for unfolding these local operators to trace their origin. We note that some operators can be generated by integrating out flavor singlet heavy fields, thus can emerge from the tree-level SM process. We also identify a few operators that require heavy flavor non-singlet tree-propagators, and thus cannot originate from the SM tree-level process. In this context, we also pave the way for possible BSM origins of such operators.

We would like to stress that our methodology is universal and is equally applicable for any connected compact flavor symmetry. We can easily accommodate the vector mesons and axion fields within this framework and calculate modified amplitudes and sum rules. This proposal can also be extended in thermal background, see Ref.~\cite{Chakrabortty:2026swu}. We leave these directions for our future research. 

%%%%%%%%%%%%%%%%%%%%%%%%%%%%%%%%%
\subsection*{Acknowledgments}
%%%%%%%%%%%%%%%%%%%%%%%%%%%%%%%%%
%
%
JC acknowledges the hospitality of HRI, Allahabad, India, where part of the research was carried out, and also the support from Project SERB/PHY/2023799.  S.C. would also like to thank the IIT-Kanpur initiation grant (PHY/2022220) and the Science and Engineering Research
Board, Government of India (Grant No. SRG/2023/001162) for financial support. S.C. also acknowledges the hospitality of the Mainz Institute for Theoretical Physics (MITP), Germany, during the workshop ``ALPs Across Scales", where part of this work was carried out. S.C. also thanks Namit Mahajan and Tuhin S. Roy for helpful discussions. S.K. acknowledges support from the Institute Postdoctoral Fellowship, IIT Kanpur. S.K. also thanks the organizers and participants of the SATPP EFT school at ICTS-TIFR, Bengaluru for useful discussions relevant to this work. The authors are grateful to Margarita Gavrilova for providing useful feedback.

\clearpage
\appendix
%%%%%%%%%%%%%%%%%%%%%%%%%%%%%%%%%%%%%%%%%%%%%%%%%%
\section{Complete list of operators}\label{app:oplist}
%%%%%%%%%%%%%%%%%%%%%%%%%%%%%%%%%%%%%%%%%%%%%%%%%%%%%%%%
We employ the HS method, discussed in Sec.~\ref{sec:HS-OP},  to compute the complete non-redundant operators consisting of four meson fields and two derivatives. 
Here, we provide the full list of operators.

%%%%%%%%%%%%%%%%%%%%%%%%%%
%\hQ_qQ \hQ_qQ^\dagger
%%%%%%%%%%%%%%%%%%%%%%%%%%
\begin{table}[H]
	\centering
	\renewcommand{\arraystretch}{.9}
	\setlength{\tabcolsep}{3pt}
	\tiny
	
	\begin{minipage}[t]{0.49\textwidth}
		\vspace{0pt}
		\centering
\begin{tabular}{>{\black}p{0.94\linewidth}}
			\toprule
			
			\centering
			\(\hQ_{qQ}\hQ_{qQ}^{\dagger}D^2\)
			\tabularnewline[1.2ex]
			
			\hline
			
			\centering
			\(8\;\Phi\Sigma\Phi^\dagger\Sigma^\dagger\)
			\tabularnewline
			
			\hdashline
			
			\(
			(\Sigma^\dagger \hQ_{qQ}\hQ_{qQ}^{\dagger}\delmu\Sigma)
			(\Phi^\dagger \tau_L^I\delmu\Phi)
			\)
			\\[1.2ex]
			
			\(
			(\Sigma^\dagger T_L^A\hQ_{qQ}\hQ_{qQ}^{\dagger}
			T_L^A\delmu\Sigma)
			(\Phi^\dagger\delmu\Phi)
			\)
			\\[1.2ex]
			
			\(
			(\Sigma^\dagger \hQ_{qQ}\tau_L^I\hQ_{qQ}^{\dagger}
			\delmu\Sigma)
			(\Phi^\dagger \tau_L^I\delmu\Phi)
			\)
			\\[1.2ex]
			
			\(
			(\Sigma^\dagger T_L^A\hQ_{qQ}\tau_L^I\hQ_{qQ}^{\dagger}
			T_L^A\delmu\Sigma)
			(\Phi^\dagger \tau_L^I\delmu\Phi)
			\)
			\\[1.2ex]
			
			\(
			(\Sigma^\dagger \hQ_{qQ}\hQ_{qQ}^{\dagger}
			T_R^B\delmu\Sigma)
			(\Phi^\dagger \tau_L^I T_R^B\delmu\Phi)
			\)
			\\[1.2ex]
			
			\(
			(\Sigma^\dagger T_L^A\hQ_{qQ}\hQ_{qQ}^{\dagger}
			T_L^A T_R^B\delmu\Sigma)
			(\Phi^\dagger T_R^B\delmu\Phi)
			\)
			\\
\bottomrule
		\end{tabular}
	\end{minipage}
	\hfill
	\begin{minipage}[t]{0.49\textwidth}
		\vspace{0pt}
		\centering
		\begin{tabular}{>{\black}p{0.94\linewidth}}
			\toprule
			
			\centering
			\(\hQ_{Qq}\hQ_{Qq}^{\dagger}D^2\)
			\tabularnewline [1.2ex]
			
			\hline
			
			\centering
			\(8\;\Phi\Sigma\Phi^\dagger\Sigma^\dagger\)
			\tabularnewline
			
			\hdashline
			
			\(
			(\Sigma^\dagger \hQ_{Qq}\hQ_{Qq}^{\dagger}\delmu\Sigma)
			(\Phi^\dagger \tau_L^I\delmu\Phi)
			\)
			\\[1.2ex]
			
			\(
			(\Sigma^\dagger T_L^A\hQ_{Qq}\hQ_{Qq}^{\dagger}
			T_L^A\delmu\Sigma)
			(\Phi^\dagger\delmu\Phi)
			\)
			\\[1.2ex]
			
			\(
			(\Sigma^\dagger \hQ_{Qq}\tau_L^I\hQ_{Qq}^{\dagger}
			\delmu\Sigma)
			(\Phi^\dagger \tau_L^I\delmu\Phi)
			\)
			\\[1.2ex]
			
			\(
			(\Sigma^\dagger T_L^A\hQ_{Qq}\tau_L^I\hQ_{Qq}^{\dagger}
			T_L^A\delmu\Sigma)
			(\Phi^\dagger \tau_L^I\delmu\Phi)
			\)
			\\[1.2ex]
			
			\(
			(\Sigma^\dagger \hQ_{Qq}\hQ_{Qq}^{\dagger}
			T_R^B\delmu\Sigma)
			(\Phi^\dagger \tau_L^I T_R^B\delmu\Phi)
			\)
			\\[1.2ex]
			
			\(
			(\Sigma^\dagger T_L^A\hQ_{Qq}\hQ_{Qq}^{\dagger}
			T_L^A T_R^B\delmu\Sigma)
			(\Phi^\dagger T_R^B\delmu\Phi)
			\)
			\\
\bottomrule
		\end{tabular}
	\end{minipage}
	
	\caption{Explicit operator forms for the \(\widehat Q_{qQ}\widehat Q_{qQ}^{\dagger}D^2\) and \(\widehat Q_{Qq}\widehat Q_{Qq}^{\dagger}D^2\) classes: first \(23\) operators.}
\end{table}

\begin{table}[H]
	\centering
	\renewcommand{\arraystretch}{0.9}
	\setlength{\tabcolsep}{3pt}
	\tiny
	
	\begin{minipage}[t]{0.49\textwidth}
		\vspace{0pt}
		\centering
		\begin{tabular}{>{\black}p{0.94\linewidth}}
			\toprule
			
			\centering
			\(\hQ_{qQ}\hQ_{qQ}^{\dagger}D^2\)
			\tabularnewline[1.2ex]
			
			\hline

            \centering
			\(2\;(\Sigma^\dagger)^2\Sigma^2\)
			\tabularnewline
			
			\hdashline
			
            			\(
			f_{ABC}(\Sigma^\dagger T_L^A\hQ_{qQ}
			\hQ_{qQ}^{\dagger}T_L^B\delmu\Sigma)
			(\Sigma^\dagger T_L^C\delmu\Sigma)
			\)
			\\[1.2ex]
			
			\(
			(\Sigma^\dagger \hQ_{qQ}
			\hQ_{qQ}^{\dagger}\delmu\Sigma)
			(\Sigma^\dagger \delmu\Sigma)
			\)
			\\[1.5ex]
			
			\hdashline
			
			\centering
			\(8\;\Phi_c\Sigma_H\Phi_c^\dagger\Sigma_H^\dagger\)
			\tabularnewline
			
			\hdashline
			
			\(
			(\Sigma_H^\dagger \hQ_{qQ}^{\dagger}\hQ_{qQ}\delmu\Sigma_H)
			(\Phi_c^\dagger \tau_L^I\delmu\Phi_c)
			\)
			\\[1.2ex]
			
			\(
			(\Sigma_H^\dagger T_L^A\hQ_{qQ}^{\dagger}\hQ_{qQ}
			T_L^A\delmu\Sigma_H)
			(\Phi_c^\dagger\delmu\Phi_c)
			\)
			\\[1.2ex]
			
			\(
			(\Sigma_H^\dagger \hQ_{qQ}^{\dagger}\tau_L^I
			\hQ_{qQ}\delmu\Sigma_H)
			(\Phi_c^\dagger \tau_L^I\delmu\Phi_c)
			\)
			\\
\bottomrule
		\end{tabular}
	\end{minipage}
	\hfill
	\begin{minipage}[t]{0.49\textwidth}
		\vspace{0pt}
		\centering
		\begin{tabular}{>{\black}p{0.94\linewidth}}
			\toprule
			
			\centering
			\(\hQ_{Qq}\hQ_{Qq}^{\dagger}D^2\)
			\tabularnewline[1.2ex]
			
			\hline
			
			\centering
			\(2\;(\Sigma^\dagger)^2\Sigma^2\)
			\tabularnewline
			
			\hdashline
			
			\(
			f_{ABC}(\Sigma^\dagger T_L^A\hQ_{Qq}
			\hQ_{Qq}^{\dagger}T_L^B\delmu\Sigma)
			(\Sigma^\dagger T_L^C\delmu\Sigma)
			\)
			\\[1.2ex]
			
			\(
			(\Sigma^\dagger \hQ_{Qq}
			\hQ_{Qq}^{\dagger}\delmu\Sigma)
			(\Sigma^\dagger \delmu\Sigma)
			\)
			\\[1.5ex]
			
			\hdashline
			
			\centering
			\(8\;\Phi_c\Sigma_H\Phi_c^\dagger\Sigma_H^\dagger\)
			\tabularnewline
			
			\hdashline
			
			\(
			(\Sigma_H^\dagger \hQ_{Qq}^{\dagger}\hQ_{Qq}\delmu\Sigma_H)
			(\Phi_c^\dagger \tau_L^I\delmu\Phi_c)
			\)
			\\[1.2ex]
			
			\(
			(\Sigma_H^\dagger T_L^A\hQ_{Qq}^{\dagger}\hQ_{Qq}
			T_L^A\delmu\Sigma_H)
			(\Phi_c^\dagger\delmu\Phi_c)
			\)
			\\[1.2ex]
			
			\(
			(\Sigma_H^\dagger \hQ_{Qq}^{\dagger}\tau_L^I
			\hQ_{Qq}\delmu\Sigma_H)
			(\Phi_c^\dagger \tau_L^I\delmu\Phi_c)
			\)
			\\
\bottomrule
		\end{tabular}
	\end{minipage}
	
	\caption{Explicit operator forms for the \(\widehat Q_{qQ}\widehat Q_{qQ}^{\dagger}D^2\) and \(\widehat Q_{Qq}\widehat Q_{Qq}^{\dagger}D^2\) classes: remaining \(22\) operators.}
\end{table}

\begin{table}[H]
	\centering
	\renewcommand{\arraystretch}{1.00}
	\setlength{\tabcolsep}{2pt}
	\tiny
	
	\begin{minipage}[t]{0.49\textwidth}
		\vspace{0pt}
		\centering
		\begin{tabular}{>{\black}p{0.94\linewidth}}
			\toprule
			
			\centering
			\(\hQ_{QQ}\hQ_{QQ}^{\dagger}D^2\)
			\tabularnewline[1.2ex]
			
			\hline
			
			\centering
			\(4\;\Sigma_H^\dagger\Sigma_H\Phi^\dagger\Phi\)
			\tabularnewline
			
			\hdashline
			
			\(
			(\Sigma_H^\dagger
			\, \hQ_{QQ}\tau_L^I\,
			\delmu\Phi)
			(\Phi^\dagger
			\, \hQ_{QQ}^{\dagger}\tau_L^I\,
			\delmu\Sigma_H)
			\)
			\\[1.2ex]
			
			\(
			(\Sigma_H^\dagger \tau_L^J
			\, \hQ_{QQ}\tau_L^I\,
			\delmu\Phi)
			(\Phi^\dagger \tau_L^J
			\, \hQ_{QQ}^{\dagger}\tau_L^I\,
			\delmu\Sigma_H)
			\)
			\\[1.2ex]
			
			\(
			(\Sigma_H^\dagger \tau_L^I
			\, \hQ_{QQ}\tau_L^I\,
			\delmu\Phi)
			(\Phi^\dagger \tau_L^J
			\langle \hQ_{QQ}^{\dagger}\tau_L^J\rangle
			\delmu\Sigma_H)
			\)
			\\[1.2ex]
			
			\(
			\epsilon_{IJK}\epsilon_{LMK}
			(\Sigma_H^\dagger \tau_L^J
			\, \hQ_{QQ}\tau_L^I\,
			\delmu\Phi)
			(\Phi^\dagger \tau_L^L
			\, \hQ_{QQ}^{\dagger}\tau_L^M\,
			\delmu\Sigma_H)
			\)
			\\[1.5ex]
			
			\hdashline
			
			\centering
			\(4\;(\Sigma_H^\dagger)^2\Sigma_H^2\)
			\tabularnewline
			
			\hdashline
			
			\(
			(\Sigma_H^\dagger
			\, \hQ_{QQ}\tau_L^I\,
			\delmu\Sigma_H)
			(\Sigma_H^\dagger
			\, \hQ_{QQ}^{\dagger}\tau_L^I\,
			\delmu\Sigma_H)
			\)
			\\
\bottomrule
		\end{tabular}
	\end{minipage}
	\hfill
	\begin{minipage}[t]{0.49\textwidth}
		\vspace{0pt}
		\centering
		\begin{tabular}{>{\black}p{0.94\linewidth}}
			\toprule
			
			\centering
			\(\hQ_{QQ}\hQ_{QQ}^{\dagger}D^2\)
			\tabularnewline[1.2ex]
			
			\hline
			
			\centering
			\(4\;(\Phi^\dagger)^2\Phi^2\)
			\tabularnewline
			
			\hdashline
			
			\(
			(\Phi^\dagger
			\, \hQ_{QQ}\tau_L^I\,
			\delmu\Phi)
			(\Phi^\dagger
			\, \hQ_{QQ}^{\dagger}\tau_L^I\,
			\delmu\Phi)
			\)
			\\[1.2ex]
			
			\(
			(\Phi^\dagger \tau_L^J T_R^A
			\, \hQ_{QQ}\tau_L^I\,
			\delmu\Phi)
			(\Phi^\dagger \tau_L^J T_R^A
			\, \hQ_{QQ}^{\dagger}\tau_L^I\,
			\delmu\Phi)
			\)
			\\[1.2ex]
			
			\(
			(\Phi^\dagger T_R^A
			\, \hQ_{QQ}\tau_L^I\,
			\delmu\Phi)
			(\Phi^\dagger T_R^A
			\, \hQ_{QQ}^{\dagger}\tau_L^I\,
			\delmu\Phi)
			\)
			\\[1.2ex]
			
			\(
			\epsilon_{IJK}\epsilon_{LMK}
			(\Phi^\dagger \tau_L^J T_R^A
			\, \hQ_{QQ}\tau_L^I\,
			\delmu\Phi)
			(\Phi^\dagger \tau_L^M T_R^A
			\, \hQ_{QQ}^{\dagger}\tau_L^L\,
			\delmu\Phi)
			\)
			\\[1.5ex]
			
			\hdashline
			
			\centering
			\(4\;\Phi^\dagger\Phi\Sigma^\dagger\Sigma\)
			\tabularnewline
			
			\hdashline
			
			\(
			(\Phi^\dagger
			\, \hQ_{QQ}\tau_L^I\,
			\delmu\Phi)
			(\Sigma^\dagger
			\, \hQ_{QQ}^{\dagger}\tau_L^I\,
			\delmu\Sigma)
			\)
			\\
\bottomrule
		\end{tabular}
	\end{minipage}
	
	\caption{Explicit operator forms for the \(\widehat Q_{QQ}\widehat Q_{QQ}^{\dagger}D^2\) class.}
\end{table}

\begin{table}[H]
	\centering
	\renewcommand{\arraystretch}{1.2}
	\setlength{\tabcolsep}{2pt}
	\tiny
	
	\begin{minipage}[t]{0.49\textwidth}
		\vspace{0pt}
		\centering
		\begin{tabular}{>{\black}p{0.94\linewidth}}
			\toprule
			
			\centering
			\(\hQ_{qq}\hQ_{qq}^{\dagger}D^2\)
			\tabularnewline[1.2ex]
			
			\hline
			
			\centering
			\(4\;(\Sigma^\dagger)^2\Sigma^2\)
			\tabularnewline
			
			\hdashline
			
			\(
			(\Sigma^\dagger
			\, \hQ_{qq}T_L^A\,
			T_L^A\delmu\Sigma)
			(\Sigma^\dagger
			\, \hQ_{qq}^{\dagger}T_L^B\,
			T_L^B\delmu\Sigma)
			\)
			\\[1.2ex]
			
			\(
			f_{ABC}
			(\Sigma^\dagger
			\, \hQ_{qq}T_L^A\,
			T_L^B\delmu\Sigma)
			(\Sigma^\dagger
			\, \hQ_{qq}^{\dagger}T_L^C\,
			\delmu\Sigma)
			\)
			\\[1.2ex]
			
			\(
			d_{ABC}
			(\Sigma^\dagger
			\, \hQ_{qq}T_L^A\,
			T_L^B\delmu\Sigma)
			(\Sigma^\dagger
			\, \hQ_{qq}^{\dagger}T_L^C\,
			\delmu\Sigma)
			\)
			\\[1.2ex]
			
			\(
			(\Sigma^\dagger
			\, \hQ_{qq}T_L^A\,
			\delmu\Sigma)
			(\Sigma^\dagger
			\, \hQ_{qq}^{\dagger}T_L^A\,
			\delmu\Sigma)
			\)
			\\[1.5ex]
			
			\hdashline
			
			\centering
			\(6\;\Sigma\Sigma^\dagger\Phi\Phi^\dagger\)
			\tabularnewline
			
			\hdashline
			
			\(
			(\Sigma^\dagger
			\, \hQ_{qq}T_L^A\,
			\delmu\Sigma)
			(\Phi^\dagger
			\, \hQ_{qq}^{\dagger}T_L^A\,
			\delmu\Phi)
			\)
			\\
\bottomrule
		\end{tabular}
	\end{minipage}
	\hfill
	\begin{minipage}[t]{0.49\textwidth}
		\vspace{0pt}
		\centering
		\begin{tabular}{>{\black}p{0.94\linewidth}}
			\toprule
			
			\centering
			\(\hQ_{qq}\hQ_{qq}^{\dagger}D^2\)
			\tabularnewline[1.2ex]
			
			\hline
			
			\centering
			\(13\;\Sigma\Sigma^\dagger\Phi_c\Phi_c^\dagger\)
			\tabularnewline
			
			\hdashline
			
			\(
			(\Sigma^\dagger
			\, \hQ_{qq}T_L^A\,
			T_L^A\delmu\Phi_c)
			(\Phi_c^\dagger
			\, \hQ_{qq}^{\dagger}T_L^B\,
			T_L^B\delmu\Sigma)
			\)
			\\[1.2ex]
			
			\(
			(\Sigma^\dagger
			\, \hQ_{qq}T_L^A\,
			\delmu\Phi_c)
			(\Phi_c^\dagger
			\, \hQ_{qq}^{\dagger}T_L^A\,
			\delmu\Sigma)
			\)
			\\[1.2ex]
			
			\(
			d_{ABC}
			(\Sigma^\dagger
			\, \hQ_{qq}T_L^A\,
			\delmu\Phi_c)
			(\Phi_c^\dagger
			\, \hQ_{qq}^{\dagger}T_L^B\,
			T_L^C\delmu\Sigma)
			\)
			\\[1.2ex]
			
			\(
			f_{ABC}
			(\Sigma^\dagger
			\, \hQ_{qq}T_L^A\,
			\delmu\Phi_c)
			(\Phi_c^\dagger
			\, \hQ_{qq}^{\dagger}T_L^B\,
			T_L^C\delmu\Sigma)
			\)
			\\[1.2ex]
			
			\(
			d_{ABC}
			(\Sigma^\dagger
			\langle \hQ_{qq}T_L^B\rangle
			T_L^C\delmu\Phi_c)
			(\Phi_c^\dagger
			\, \hQ_{qq}^{\dagger}T_L^A\,
			\delmu\Sigma)
			\)
			\\[1.2ex]
			
			\(
			d_{ABC}d_{ADE}
			(\Sigma^\dagger
			\langle \hQ_{qq}T_L^B\rangle
			T_L^C\delmu\Phi_c)
			(\Phi_c^\dagger
			\langle \hQ_{qq}^{\dagger}T_L^D\rangle
			T_L^E\delmu\Sigma)
			\)
			\\
\bottomrule
		\end{tabular}
	\end{minipage}
	
	\caption{Explicit operator forms for the \(\widehat Q_{qq}\widehat Q_{qq}^{\dagger}D^2\) class.}
\end{table}

\begin{table}[H]
	\centering
	\renewcommand{\arraystretch}{0.90}
	\setlength{\tabcolsep}{2pt}
	\tiny
	
	\begin{minipage}[t]{0.49\textwidth}
		\vspace{0pt}
		\centering
		\begin{tabular}{p{0.94\linewidth}}
			\toprule
			
			{\black\centering
			\(\{\hQ_{qQ},\hQ_{QQ}^{\dagger}\}\)
			}
			\tabularnewline
			
			\midrule
			
			{\black\centering
			\(\Sigma^\dagger\Sigma\Sigma^\dagger \Phi\)
			\qquad
			\((\#\,2)\)
			}
			\tabularnewline
			
			\midrule
			
			{\black
			\(
			(\Sigma^\dagger T_L^A \hQ_{qQ}\tau_L^I \delmu H)
			(\Sigma^\dagger \langle \hQ_{QQ}^\dagger \tau_L^I\rangle
			T_L^A \delmu\Sigma)
			\)
			}
			\\[0.6ex]
			
			{\black
			\(
			(\Sigma^\dagger T_L^A T_R^B \hQ_{qQ}\tau_L^I \delmu H)
			(\Sigma^\dagger \langle \hQ_{QQ}^\dagger \tau_L^I\rangle
			T_L^A T_R^B \delmu\Sigma)
			\)
			}
			\\[0.8ex]
			
			\midrule
			
			{\black\centering
			\(\Sigma^\dagger\Sigma\Phi_c^\dagger\Sigma_H\)
			\qquad
			\((\#\,2)\)
			}
			\tabularnewline
			
			\midrule
			
			{\black
			\(
			(\Sigma^\dagger \hQ_{qQ}\tau_L^I \delmu\Sigma_H)
			(\Phi_c^\dagger \langle \hQ_{QQ}^\dagger \tau_L^I\rangle
			\delmu\Sigma)
			\)
			}
			\\[0.6ex]
			
			{\black
			\(
			(\Sigma^\dagger T_L^A \hQ_{qQ}\tau_L^I \delmu\Sigma_H)
			(\Phi_c^\dagger \langle \hQ_{QQ}^\dagger \tau_L^I\rangle
			T_L^A \delmu\Sigma)
			\)
			}
			\\[0.8ex]
			
			\midrule
			
			{\black\centering
			\(\Sigma^\dagger\Phi\Phi_c^\dagger\Phi_c\)
			\qquad
			\((\#\,2)\)
			}
			\tabularnewline
\bottomrule
		\end{tabular}
	\end{minipage}
	\hfill
	\begin{minipage}[t]{0.49\textwidth}
		\vspace{0pt}
		\centering
		\begin{tabular}{p{0.94\linewidth}}
			\toprule
			
			{\purp\centering
				\(\{\hQ_{QQ},\hQ_{qQ}^{\dagger}\}\)
			}
			\tabularnewline
			
			\midrule
			
			{\purp\centering
				\(\Sigma\Sigma^\dagger\Sigma H^\dagger\)
				\qquad
				\((\#\,2)\)
			}
			\tabularnewline
			
			\midrule
			
			{\purp
				\(
				(H^\dagger \tau_L^I \hQ_{qQ}^{\dagger}T_L^A \delmu\Sigma)
				(\Sigma^\dagger T_L^A
				\, \tau_L^I\hQ_{QQ}\, \delmu\Sigma)
				\)
			}
			\\[0.6ex]
			
			{\purp
				\(
				(H^\dagger \tau_L^I \hQ_{qQ}^{\dagger}T_L^A T_R^B \delmu\Sigma)
				(\Sigma^\dagger T_L^A T_R^B
				\, \tau_L^I\hQ_{QQ}\, \delmu\Sigma)
				\)
			}
			\\[0.8ex]
			
			\midrule
			
			{\purp\centering
				\(\Sigma^\dagger\Sigma\Phi_c\Sigma_H^\dagger\)
				\qquad
				\((\#\,2)\)
			}
			\tabularnewline
			
			\midrule
			
			{\purp
				\(
				(\Sigma_H^\dagger \tau_L^I \hQ_{qQ}^{\dagger}\delmu\Sigma)
				(\Sigma^\dagger
				\, \tau_L^I\hQ_{QQ}\, \delmu\Phi_c)
				\)
			}
			\\[0.6ex]
			
			{\purp
				\(
				(\Sigma_H^\dagger \tau_L^I \hQ_{qQ}^{\dagger}T_L^A\delmu\Sigma)
				(\Sigma^\dagger T_L^A
				\, \tau_L^I\hQ_{QQ}\, \delmu\Phi_c)
				\)
			}
			\\[0.8ex]
			
			\midrule
			
			{\purp\centering
				\(\Sigma\Phi^\dagger\Phi_c\Phi_c^\dagger\)
				\qquad
				\((\#\,2)\)
			}
			\tabularnewline
\bottomrule
		\end{tabular}
	\end{minipage}
	
	\caption{Explicit operator forms for the \(\widehat Q_{qQ}\widehat Q_{QQ}^{\dagger}D^2\) class and its hermitian conjugate.}
\end{table}

\begin{table}[H]
	\centering
	\renewcommand{\arraystretch}{0.90}
	\setlength{\tabcolsep}{2pt}
	\tiny
	
	\begin{minipage}[t]{0.49\textwidth}
		\vspace{0pt}
		\centering
		\begin{tabular}{p{0.94\linewidth}}
			\toprule
			
			{\black\centering
			\(\{\hQ_{Qq},\hQ_{QQ}^{\dagger}\}\)
			}
			\tabularnewline
			
			\midrule
			
			{\black\centering
			\(\Sigma^\dagger\Sigma\Sigma^\dagger H\)
			\qquad
			\((\#\,2)\)
			}
			\tabularnewline
			
			\midrule
			
			{\black
			\(
			(\Sigma^\dagger T_L^A \hQ_{Qq}\tau_L^I \delmu H)
			(\Sigma^\dagger \langle \hQ_{QQ}^\dagger \tau_L^I\rangle
			T_L^A \delmu\Sigma)
			\)
			}
			\\[0.6ex]
			
			{\black
			\(
			(\Sigma^\dagger T_L^A T_R^B \hQ_{Qq}\tau_L^I \delmu H)
			(\Sigma^\dagger \langle \hQ_{QQ}^\dagger \tau_L^I\rangle
			T_L^A T_R^B \delmu\Sigma)
			\)
			}
			\\[0.8ex]
			
			\midrule
			
			{\black\centering
			\(\Sigma^\dagger\Sigma\Phi_c^\dagger\Sigma_H\)
			\qquad
			\((\#\,2)\)
			}
			\tabularnewline
			
			\midrule
			
			{\black
			\(
			(\Sigma^\dagger \hQ_{Qq}\tau_L^I \delmu\Sigma_H)
			(\Phi_c^\dagger \langle \hQ_{QQ}^\dagger \tau_L^I\rangle
			\delmu\Sigma)
			\)
			}
			\\[0.6ex]
			
			{\black
			\(
			(\Sigma^\dagger T_L^A \hQ_{Qq}\tau_L^I \delmu\Sigma_H)
			(\Phi_c^\dagger \langle \hQ_{QQ}^\dagger \tau_L^I\rangle
			T_L^A \delmu\Sigma)
			\)
			}
			\\[0.8ex]
			
			\midrule
			
			{\black\centering
			\(\Sigma^\dagger\Phi\Phi_c^\dagger\Phi_c\)
			\qquad
			\((\#\,2)\)
			}
			\tabularnewline
\bottomrule
		\end{tabular}
	\end{minipage}
	\hfill
	\begin{minipage}[t]{0.49\textwidth}
		\vspace{0pt}
		\centering
		\begin{tabular}{p{0.94\linewidth}}
			\toprule
			
			{\purp\centering
				\(\{\hQ_{QQ},\hQ_{Qq}^{\dagger}\}\)
			}
			\tabularnewline
			
			\midrule
			
			{\purp\centering
				\(\Sigma\Sigma^\dagger\Sigma H^\dagger\)
				\qquad
				\((\#\,2)\)
			}
			\tabularnewline
			
			\midrule
			
			{\purp
				\(
				(H^\dagger \tau_L^I \hQ_{Qq}^{\dagger}T_L^A \delmu\Sigma)
				(\Sigma^\dagger T_L^A
				\, \tau_L^I\hQ_{QQ}\, \delmu\Sigma)
				\)
			}
			\\[0.6ex]
			
			{\purp
				\(
				(H^\dagger \tau_L^I \hQ_{Qq}^{\dagger}T_L^A T_R^B \delmu\Sigma)
				(\Sigma^\dagger T_L^A T_R^B
				\, \tau_L^I\hQ_{QQ}\, \delmu\Sigma)
				\)
			}
			\\[0.8ex]
			
			\midrule
			
			{\purp\centering
				\(\Sigma^\dagger\Sigma\Phi_c\Sigma_H^\dagger\)
				\qquad
				\((\#\,2)\)
			}
			\tabularnewline
			
			\midrule
			
			{\purp
				\(
				(\Sigma_H^\dagger \tau_L^I \hQ_{Qq}^{\dagger}\delmu\Sigma)
				(\Sigma^\dagger
				\, \tau_L^I\hQ_{QQ}\, \delmu\Phi_c)
				\)
			}
			\\[0.6ex]
			
			{\purp
				\(
				(\Sigma_H^\dagger \tau_L^I \hQ_{Qq}^{\dagger}T_L^A\delmu\Sigma)
				(\Sigma^\dagger T_L^A
				\, \tau_L^I\hQ_{QQ}\, \delmu\Phi_c)
				\)
			}
			\\[0.8ex]
			
			\midrule
			
			{\purp\centering
				\(\Sigma\Phi^\dagger\Phi_c\Phi_c^\dagger\)
				\qquad
				\((\#\,2)\)
			}
			\tabularnewline
\bottomrule
		\end{tabular}
	\end{minipage}
	
	\caption{Explicit operator forms for the \(\widehat{Q}_{Qq}\widehat{Q}_{QQ}^{\dagger}D^2\) class and its hermitian conjugate.}
\end{table}

%%%%%%%%%%%%%%%%%%%%%%%%%

\begin{table}[H]
	\centering
	\renewcommand{\arraystretch}{1.00}
	\setlength{\tabcolsep}{2pt}
	\tiny
	
\begin{minipage}[t]{0.49\textwidth}
		\vspace{0pt}
		\centering
		\begin{tabular}{>{\black}p{0.94\linewidth}}
			\toprule
			
			\centering
			\(\{\hQ_{qq},\hQ_{Qq}^{\dagger}\}\)
			\tabularnewline
			
			\midrule
			
			\centering
			\(\Sigma_H\Phi_c^\dagger\Sigma\Sigma^\dagger\)
			\qquad
			\((\#\,4)\)
			\tabularnewline
			
			\midrule
			
			\(
			(\Sigma \, \hQ_{qq}T_L^A\,
			T_L^A\delmu\Sigma^\dagger)
			(\Sigma_H\hQ_{Qq}^{\dagger}\delmu\Phi_c^\dagger)
			\)
			\\[1.2ex]
			
			\(
			(\Sigma \, \hQ_{qq}T_L^A\,
			\delmu\Sigma^\dagger)
			(\Sigma_H\hQ_{Qq}^{\dagger}
			T_L^A\delmu\Phi_c^\dagger)
			\)
			\\[1.2ex]
			
			\(
			f^{ABC}
			(\Sigma \, \hQ_{qq}T_L^A\,
			T_L^B\delmu\Sigma^\dagger)
			(\Sigma_H\hQ_{Qq}^{\dagger}
			T_L^C\delmu\Phi_c^\dagger)
			\)
			\\[1.2ex]
			
			\(
			d^{ABC}
			(\Sigma \, \hQ_{qq}T_L^A\,
			T_L^B\delmu\Sigma^\dagger)
			(\Sigma_H\hQ_{Qq}^{\dagger}
			T_L^C\delmu\Phi_c^\dagger)
			\)
			\\[1.5ex]
			
			\midrule
			
			\centering
			\(\Sigma_H\Phi_c^\dagger\Phi_c\Phi_c^\dagger\)
			\qquad
			\((\#\,4)\)
			\tabularnewline
			
			\midrule
			
			\(
			(\Phi_c \, \hQ_{qq}T_L^A\,
			T_L^A\delmu\Phi_c^\dagger)
			(\Sigma_H\hQ_{Qq}^{\dagger}\delmu\Phi_c^\dagger)
			\)
			\\
\bottomrule
		\end{tabular}
	\end{minipage}
	\hfill
	\begin{minipage}[t]{0.49\textwidth}
		\vspace{0pt}
		\centering
		\begin{tabular}{>{\purp}p{0.94\linewidth}}
			\toprule
			
			\centering
			\(\{\hQ_{Qq},\hQ_{qq}^{\dagger}\}\)
			\tabularnewline
			
			\midrule
			
			\centering
			\(\Sigma_H^\dagger\Phi_c\Sigma^\dagger\Sigma\)
			\qquad
			\((\#\,4)\)
			\tabularnewline
			
			\midrule
			
			\(
			(\Sigma^\dagger \,\hQ_{qq}^\dagger T_L^A\,
			T_L^A \delmu\Sigma)
			(\Sigma_H^\dagger \hQ_{Qq}\delmu\Phi_c)
			\)
			\\[1.2ex]
			
			\(
			(\Sigma^\dagger \,\hQ_{qq}^\dagger T_L^A\,
			\delmu\Sigma)
			(\Sigma_H^\dagger \hQ_{Qq}
			T_L^A \delmu\Phi_c)
			\)
			\\[1.2ex]
			
			\(
			f^{ABC}
			(\Sigma^\dagger \,\hQ_{qq}^\dagger T_L^A\,
			T_L^B \delmu\Sigma)
			(\Sigma_H^\dagger \hQ_{Qq}
			T_L^C \delmu\Phi_c)
			\)
			\\[1.2ex]
			
			\(
			d^{ABC}
			(\Sigma^\dagger \,\hQ_{qq}^\dagger T_L^A\,
			T_L^B \delmu\Sigma)
			(\Sigma_H^\dagger \hQ_{Qq}
			T_L^C \delmu\Phi_c)
			\)
			\\[1.5ex]
			
			\midrule
			
			\centering
			\(\Sigma_H^\dagger\Phi_c\Phi_c^\dagger\Phi_c\)
			\qquad
			\((\#\,4)\)
			\tabularnewline
			
			\midrule
			
			\(
			(\Phi_c^\dagger \,\hQ_{qq}^\dagger T_L^A\,
			T_L^A \delmu\Phi_c)
			(\Sigma_H^\dagger \hQ_{Qq}\delmu\Phi_c)
			\)
			\\
\bottomrule
		\end{tabular}
	\end{minipage}
	
	\caption{Explicit operator forms for the first 18 operators of the $\widehat{Q}_{qq}\widehat{Q}_{Qq}^{\dagger}D^2$ class and its hermitian conjugate.}
\end{table}

\begin{table}[H]
	\centering
	\renewcommand{\arraystretch}{1.00}
	\setlength{\tabcolsep}{2pt}
	\tiny
	
\begin{minipage}[t]{0.49\textwidth}
		\vspace{0pt}
		\centering
		\begin{tabular}{>{\black}p{0.94\linewidth}}
			\toprule
			
			\centering
			\(\{\hQ_{qq},\hQ_{Qq}^{\dagger}\}\)
			\tabularnewline
			
			\midrule
			
			\centering
			\(\Phi\Sigma^\dagger\Phi_c\Phi_c^\dagger\)
			\qquad
			\((\#\,4)\)
			\tabularnewline
			
			\midrule
			
			\(
			(\Phi \hQ_{Qq}^{\dagger}\delmu\Sigma^\dagger)
			(\Phi_c
			\, \hQ_{qq}T_L^A\,
			T_L^A\delmu\Phi_c^\dagger)
			\)
			\\[1.2ex]
			
			\(
			(\Phi T_L^A\hQ_{Qq}^{\dagger}\delmu\Sigma^\dagger)
			(\Phi_c
			\, \hQ_{qq}T_L^A\,
			\delmu\Phi_c^\dagger)
			\)
			\\[1.2ex]
			
			\(
			f^{ABC}
			(\Phi T_L^B\hQ_{Qq}^{\dagger}\delmu\Sigma^\dagger)
			(\Phi_c
			\, \hQ_{qq}T_L^A\,
			T_L^C\delmu\Phi_c^\dagger)
			\)
			\\[1.2ex]
			
			\(
			d^{ABC}
			(\Phi T_L^B\hQ_{Qq}^{\dagger}\delmu\Sigma^\dagger)
			(\Phi_c
			\, \hQ_{qq}T_L^A\,
			T_L^C\delmu\Phi_c^\dagger)
			\)
			\\[1.5ex]
			
			\midrule
			
			\centering
			\(\Sigma_H\Sigma^\dagger\Sigma\Phi_c^\dagger\)
			\qquad
			\((\#\,4)\)
			\tabularnewline
			
			\midrule
			
			\(
			(\Sigma
			\, \hQ_{qq}T_L^A\,
			T_L^A\delmu\Sigma^\dagger)
			(\Sigma_H\hQ_{Qq}^{\dagger}\delmu\Phi_c^\dagger)
			\)
			\\
\bottomrule
		\end{tabular}
	\end{minipage}
	\hfill
	\begin{minipage}[t]{0.49\textwidth}
		\vspace{0pt}
		\centering
		\begin{tabular}{>{\purp}p{0.94\linewidth}}
			\toprule
			
			\centering
			\(\{\hQ_{Qq},\hQ_{qq}^{\dagger}\}\)
			\tabularnewline
			
			\midrule
			
			\centering
			\(\Phi^\dagger\Sigma\Phi_c^\dagger\Phi_c\)
			\qquad
			\((\#\,4)\)
			\tabularnewline
			
			\midrule
			
			\(
			(\Phi^\dagger \hQ_{Qq}\delmu\Sigma)
			(\Phi_c^\dagger
			\,\hQ_{qq}^\dagger T_L^A\,
			T_L^A\delmu\Phi_c)
			\)
			\\[1.2ex]
			
			\(
			(\Phi^\dagger T_L^A\hQ_{Qq}\delmu\Sigma)
			(\Phi_c^\dagger
			\,\hQ_{qq}^\dagger T_L^A\,
			\delmu\Phi_c)
			\)
			\\[1.2ex]
			
			\(
			f^{ABC}
			(\Phi^\dagger T_L^B\hQ_{Qq}\delmu\Sigma)
			(\Phi_c^\dagger
			\,\hQ_{qq}^\dagger T_L^A\,
			T_L^C\delmu\Phi_c)
			\)
			\\[1.2ex]
			
			\(
			d^{ABC}
			(\Phi^\dagger T_L^B\hQ_{Qq}\delmu\Sigma)
			(\Phi_c^\dagger
			\,\hQ_{qq}^\dagger T_L^A\,
			T_L^C\delmu\Phi_c)
			\)
			\\[1.5ex]
			
			\midrule
			
			\centering
			\(\Sigma_H^\dagger\Sigma\Sigma^\dagger\Phi_c\)
			\qquad
			\((\#\,4)\)
			\tabularnewline
			
			\midrule
			
			\(
			(\Sigma^\dagger
			\,\hQ_{qq}^\dagger T_L^A\,
			T_L^A\delmu\Sigma)
			(\Sigma_H^\dagger \hQ_{Qq}\delmu\Phi_c)
			\)
			\\
\bottomrule
		\end{tabular}
	\end{minipage}
	
	\caption{Explicit operator forms for the remaining 18 operators of the $\widehat{Q}_{qq}\widehat{Q}_{Qq}^{\dagger}D^2$ class and its hermitian conjugate.}
\end{table}
%%%%%%%%%%%%%%%%%%%%%%%%%%%

\begin{table}[H]
	\centering
	\renewcommand{\arraystretch}{1.00}
	\setlength{\tabcolsep}{2pt}
	\tiny
	
\begin{minipage}[t]{0.49\textwidth}
		\vspace{0pt}
		\centering
		\begin{tabular}{>{\black}p{0.94\linewidth}}
			\toprule
			
			\centering
			\(\{\hQ_{qq},\hQ_{qQ}^{\dagger}\}\)
			\tabularnewline
			
			\midrule
			
			\centering
			\(\Sigma_H\Phi_c^\dagger\Sigma\Sigma^\dagger\)
			\qquad
			\((\#\,4)\)
			\tabularnewline
			
			\midrule
			
			\(
			(\Sigma \, \hQ_{qq}T_L^A\,
			T_L^A\delmu\Sigma^\dagger)
			(\Sigma_H\hQ_{qQ}^{\dagger}\delmu\Phi_c^\dagger)
			\)
			\\[1.2ex]
			
			\(
			(\Sigma \, \hQ_{qq}T_L^A\,
			\delmu\Sigma^\dagger)
			(\Sigma_H\hQ_{qQ}^{\dagger}
			T_L^A\delmu\Phi_c^\dagger)
			\)
			\\[1.2ex]
			
			\(
			f^{ABC}
			(\Sigma \, \hQ_{qq}T_L^A\,
			T_L^B\delmu\Sigma^\dagger)
			(\Sigma_H\hQ_{qQ}^{\dagger}
			T_L^C\delmu\Phi_c^\dagger)
			\)
			\\[1.2ex]
			
			\(
			d^{ABC}
			(\Sigma \, \hQ_{qq}T_L^A\,
			T_L^B\delmu\Sigma^\dagger)
			(\Sigma_H\hQ_{qQ}^{\dagger}
			T_L^C\delmu\Phi_c^\dagger)
			\)
			\\[1.5ex]
			
			\midrule
			
			\centering
			\(\Sigma_H\Phi_c^\dagger\Phi_c\Phi_c^\dagger\)
			\qquad
			\((\#\,4)\)
			\tabularnewline
			
			\midrule
			
			\(
			(\Phi_c \, \hQ_{qq}T_L^A\,
			T_L^A\delmu\Phi_c^\dagger)
			(\Sigma_H\hQ_{qQ}^{\dagger}\delmu\Phi_c^\dagger)
			\)
			\\
\bottomrule
		\end{tabular}
	\end{minipage}
	\hfill
	\begin{minipage}[t]{0.49\textwidth}
		\vspace{0pt}
		\centering
		\begin{tabular}{>{\purp}p{0.94\linewidth}}
			\toprule
			
			\centering
			\(\{\hQ_{qQ},\hQ_{qq}^{\dagger}\}\)
			\tabularnewline
			
			\midrule
			
			\centering
			\(\Sigma_H^\dagger\Phi_c\Sigma^\dagger\Sigma\)
			\qquad
			\((\#\,4)\)
			\tabularnewline
			
			\midrule
			
			\(
			(\Sigma^\dagger \,\hQ_{qq}^\dagger T_L^A\,
			T_L^A \delmu\Sigma)
			(\Sigma_H^\dagger \hQ_{qQ}\delmu\Phi_c)
			\)
			\\[1.2ex]
			
			\(
			(\Sigma^\dagger \,\hQ_{qq}^\dagger T_L^A\,
			\delmu\Sigma)
			(\Sigma_H^\dagger \hQ_{qQ}
			T_L^A \delmu\Phi_c)
			\)
			\\[1.2ex]
			
			\(
			f^{ABC}
			(\Sigma^\dagger \,\hQ_{qq}^\dagger T_L^A\,
			T_L^B \delmu\Sigma)
			(\Sigma_H^\dagger \hQ_{qQ}
			T_L^C \delmu\Phi_c)
			\)
			\\[1.2ex]
			
			\(
			d^{ABC}
			(\Sigma^\dagger \,\hQ_{qq}^\dagger T_L^A\,
			T_L^B \delmu\Sigma)
			(\Sigma_H^\dagger \hQ_{qQ}
			T_L^C \delmu\Phi_c)
			\)
			\\[1.5ex]
			
			\midrule
			
			\centering
			\(\Sigma_H^\dagger\Phi_c\Phi_c^\dagger\Phi_c\)
			\qquad
			\((\#\,4)\)
			\tabularnewline
			
			\midrule
			
			\(
			(\Phi_c^\dagger \,\hQ_{qq}^\dagger T_L^A\,
			T_L^A \delmu\Phi_c)
			(\Sigma_H^\dagger \hQ_{qQ}\delmu\Phi_c)
			\)
			\\
\bottomrule
		\end{tabular}
	\end{minipage}
	
	\caption{Explicit operator forms for the first 18 operators of the $\widehat{Q}_{qq}\widehat{Q}_{qQ}^{\dagger}D^2$ class and its hermitian conjugate.}
\end{table}

\begin{table}[H]
	\centering
	\renewcommand{\arraystretch}{1.00}
	\setlength{\tabcolsep}{2pt}
	\tiny
	
\begin{minipage}[t]{0.49\textwidth}
		\vspace{0pt}
		\centering
		\begin{tabular}{>{\black}p{0.94\linewidth}}
			\toprule
			
			\centering
			\(\{\hQ_{qq},\hQ_{qQ}^{\dagger}\}\)
			\tabularnewline
			
			\midrule
			
			\centering
			\(\Phi\Sigma^\dagger\Phi_c\Phi_c^\dagger\)
			\qquad
			\((\#\,4)\)
			\tabularnewline
			
			\midrule
			
			\(
			(\Phi \hQ_{qQ}^{\dagger}\delmu\Sigma^\dagger)
			(\Phi_c
			\, \hQ_{qq}T_L^A\,
			T_L^A\delmu\Phi_c^\dagger)
			\)
			\\[1.2ex]
			
			\(
			(\Phi T_L^A\hQ_{qQ}^{\dagger}\delmu\Sigma^\dagger)
			(\Phi_c
			\, \hQ_{qq}T_L^A\,
			\delmu\Phi_c^\dagger)
			\)
			\\[1.2ex]
			
			\(
			f^{ABC}
			(\Phi T_L^B\hQ_{qQ}^{\dagger}\delmu\Sigma^\dagger)
			(\Phi_c
			\, \hQ_{qq}T_L^A\,
			T_L^C\delmu\Phi_c^\dagger)
			\)
			\\[1.2ex]
			
			\(
			d^{ABC}
			(\Phi T_L^B\hQ_{qQ}^{\dagger}\delmu\Sigma^\dagger)
			(\Phi_c
			\, \hQ_{qq}T_L^A\,
			T_L^C\delmu\Phi_c^\dagger)
			\)
			\\[1.5ex]
			
			\midrule
			
			\centering
			\(\Sigma_H\Sigma^\dagger\Sigma\Phi_c^\dagger\)
			\qquad
			\((\#\,4)\)
			\tabularnewline
			
			\midrule
			
			\(
			(\Sigma
			\, \hQ_{qq}T_L^A\,
			T_L^A\delmu\Sigma^\dagger)
			(\Sigma_H\hQ_{qQ}^{\dagger}\delmu\Phi_c^\dagger)
			\)
			\\
\bottomrule
		\end{tabular}
	\end{minipage}
	\hfill
	\begin{minipage}[t]{0.49\textwidth}
		\vspace{0pt}
		\centering
		\begin{tabular}{>{\purp}p{0.94\linewidth}}
			\toprule
			
			\centering
			\(\{\hQ_{qQ},\hQ_{qq}^{\dagger}\}\)
			\tabularnewline
			
			\midrule
			
			\centering
			\(\Phi^\dagger\Sigma\Phi_c^\dagger\Phi_c\)
			\qquad
			\((\#\,4)\)
			\tabularnewline
			
			\midrule
			
			\(
			(\Phi^\dagger \hQ_{qQ}\delmu\Sigma)
			(\Phi_c^\dagger
			\,\hQ_{qq}^\dagger T_L^A\,
			T_L^A\delmu\Phi_c)
			\)
			\\[1.2ex]
			
			\(
			(\Phi^\dagger T_L^A\hQ_{qQ}\delmu\Sigma)
			(\Phi_c^\dagger
			\,\hQ_{qq}^\dagger T_L^A\,
			\delmu\Phi_c)
			\)
			\\[1.2ex]
			
			\(
			f^{ABC}
			(\Phi^\dagger T_L^B\hQ_{qQ}\delmu\Sigma)
			(\Phi_c^\dagger
			\,\hQ_{qq}^\dagger T_L^A\,
			T_L^C\delmu\Phi_c)
			\)
			\\[1.2ex]
			
			\(
			d^{ABC}
			(\Phi^\dagger T_L^B\hQ_{qQ}\delmu\Sigma)
			(\Phi_c^\dagger
			\,\hQ_{qq}^\dagger T_L^A\,
			T_L^C\delmu\Phi_c)
			\)
			\\[1.5ex]
			
			\midrule
			
			\centering
			\(\Sigma_H^\dagger\Sigma\Sigma^\dagger\Phi_c\)
			\qquad
			\((\#\,4)\)
			\tabularnewline
			
			\midrule
			
			\(
			(\Sigma^\dagger
			\,\hQ_{qq}^\dagger T_L^A\,
			T_L^A\delmu\Sigma)
			(\Sigma_H^\dagger \hQ_{qQ}\delmu\Phi_c)
			\)
			\\
\bottomrule
		\end{tabular}
	\end{minipage}
	
	\caption{Explicit operator forms for the remaining 18 operators of the $\widehat{Q}_{qq}\widehat{Q}_{qQ}^{\dagger}D^2$ class and its hermitian conjugate.}
\end{table}

\begin{table}[H]
	\centering
	\renewcommand{\arraystretch}{1.00}
	\setlength{\tabcolsep}{2pt}
	\tiny
	
	\begin{minipage}[t]{0.49\textwidth}
		\vspace{0pt}
		\centering
		\begin{tabular}{>{\black}p{0.94\linewidth}}
			\toprule
			
			\centering
			\(\{\hQ_{qq},\hQ_{QQ}^{\dagger}\}\)
			\tabularnewline
			
			\midrule
			
			\centering
			\(\Sigma\Sigma^\dagger\Phi\Phi^\dagger\)
			\qquad
			\((\#\,2)\)
			\tabularnewline
			
			\midrule
			
			\(
			(\Sigma^\dagger \, \hQ_{qq} T_L^A\,
			T_L^A\delmu\Sigma)
			(\Phi^\dagger \, \hQ_{QQ}^{\dagger}\tau_L^I\,
			\tau_L^I\delmu\Phi)
			\)
			\\[1.2ex]
			
			\(
			(\Sigma^\dagger T_R^B\, \hQ_{qq} T_L^A\,
			T_L^A\delmu\Sigma)
			(\Phi^\dagger T_R^B
			\, \hQ_{QQ}^{\dagger}\tau_L^I\,
			\tau_L^I\delmu\Phi)
			\)
			\\[1.5ex]
			
			\midrule
			
			\centering
			\(\Sigma\Sigma^\dagger\Sigma_H\Sigma_H^\dagger\)
			\qquad
			\((\#\,1)\)
			\tabularnewline
			
			\midrule
			
			\(
			(\Sigma^\dagger \, \hQ_{qq}T_L^A\,
			T_L^A\delmu\Sigma)
			(\Sigma_H^\dagger
			\, \hQ_{QQ}^{\dagger}\tau_L^I\,
			\tau_L^I\delmu\Sigma_H)
			\)
			\\[1.5ex]
			
			\midrule
			
			\centering
			\(\Phi_c\Phi_c^\dagger\Phi\Phi^\dagger\)
			\qquad
			\((\#\,1)\)
			\tabularnewline
			
			\midrule
			
			\(
			(\Phi_c^\dagger \, \hQ_{qq}T_L^A\,
			T_L^A\delmu\Phi_c)
			(\Phi^\dagger
			\, \hQ_{QQ}^{\dagger}\tau_L^I\,
			\tau_L^I\delmu\Phi)
			\)
			\\
\bottomrule
		\end{tabular}
	\end{minipage}
	\hfill
	\begin{minipage}[t]{0.49\textwidth}
		\vspace{0pt}
		\centering
		\begin{tabular}{>{\purp}p{0.94\linewidth}}
			\toprule
			
			\centering
			\(\{\hQ_{QQ},\hQ_{qq}^{\dagger}\}\)
			\tabularnewline
			
			\midrule
			
			\centering
			\(\Sigma^\dagger\Sigma\Phi^\dagger\Phi\)
			\qquad
			\((\#\,2)\)
			\tabularnewline
			
			\midrule
			
			\(
			(\Sigma \, \hQ_{qq}^{\dagger} T_L^A\,
			T_L^A\delmu\Sigma^\dagger)
			(\Phi \, \hQ_{QQ}\tau_L^I\,
			\tau_L^I\delmu\Phi^\dagger)
			\)
			\\[1.2ex]
			
			\(
			(\Sigma T_R^B\, \hQ_{qq}^{\dagger} T_L^A\,
			T_L^A\delmu\Sigma^\dagger)
			(\Phi T_R^B
			\, \hQ_{QQ}\tau_L^I\,
			\tau_L^I\delmu\Phi^\dagger)
			\)
			\\[1.5ex]
			
			\midrule
			
			\centering
			\(\Sigma^\dagger\Sigma\Sigma_H^\dagger\Sigma_H\)
			\qquad
			\((\#\,1)\)
			\tabularnewline
			
			\midrule
			
			\(
			(\Sigma \, \hQ_{qq}^{\dagger}T_L^A\,
			T_L^A\delmu\Sigma^\dagger)
			(\Sigma_H
			\, \hQ_{QQ}\tau_L^I\,
			\tau_L^I\delmu\Sigma_H^\dagger)
			\)
			\\[1.5ex]
			
			\midrule
			
			\centering
			\(\Phi_c^\dagger\Phi_c\Phi^\dagger\Phi\)
			\qquad
			\((\#\,1)\)
			\tabularnewline
			
			\midrule
			
			\(
			(\Phi_c \, \hQ_{qq}^{\dagger}T_L^A\,
			T_L^A\delmu\Phi_c^\dagger)
			(\Phi
			\, \hQ_{QQ}\tau_L^I\,
			\tau_L^I\delmu\Phi^\dagger)
			\)
			\\
\bottomrule
		\end{tabular}
	\end{minipage}
	
	\caption{Explicit operator forms for the \(\widehat Q_{qq} \widehat Q_{QQ}^{\dagger}D^2\) class and its hermitian conjugate \(\widehat Q_{QQ} \widehat Q_{qq}^{\dagger}D^2\).}
\end{table}

\begin{table}[H]
	\centering
	\renewcommand{\arraystretch}{1.15}
	\setlength{\tabcolsep}{3pt}
	\scriptsize
	
	\begin{minipage}[t]{0.49\textwidth}
		\vspace{0pt}
		\centering
		\begin{tabular}{>{\black}p{0.94\linewidth}}
			\toprule
			
			\centering
			\(\{\hQ_{qQ},\hQ_{Qq}^{\dagger}\}\)
			\tabularnewline
			
			\midrule
			
			\centering
			\((\Sigma^\dagger)^2\Phi^2\)
			\qquad
			\((\#\,2)\)
			\tabularnewline
			
			\midrule
			
			\(
			(\Sigma^\dagger \hQ_{qQ}\delmu\Phi)
			(\Sigma^\dagger \hQ_{Qq}^{\dagger}\delmu\Phi)
			\)
			\\[1.2ex]
			
			\(
			(\Sigma^\dagger T_L^A\hQ_{qQ}\tau_L^I T_R^B\delmu\Phi)
			(\Sigma^\dagger T_L^A\hQ_{Qq}^{\dagger}\tau_L^I T_R^B\delmu\Phi)
			\)
			\\[1.5ex]
			
			\midrule
			
			\centering
			\((\Phi_c^\dagger)^2\Sigma_H^2\)
			\qquad
			\((\#\,2)\)
			\tabularnewline
			
			\midrule
			
			\(
			(\Phi_c^\dagger \hQ_{qQ}\delmu\Sigma_H)
			(\Phi_c^\dagger \hQ_{Qq}^{\dagger}\delmu\Sigma_H)
			\)
			\\[1.2ex]
			
			\(
			(\Phi_c^\dagger T_L^A\hQ_{qQ}\tau_L^I \tau_R^J\delmu\Sigma_H)
			(\Phi_c^\dagger T_L^A\hQ_{Qq}^{\dagger}\tau_L^I \tau_R^J\delmu\Sigma_H)
			\)
			\\
			
			\bottomrule
		\end{tabular}
	\end{minipage}
	\hfill
	\begin{minipage}[t]{0.49\textwidth}
		\vspace{0pt}
		\centering
		\begin{tabular}{>{\purp}p{0.94\linewidth}}
			\toprule
			
			\centering
			\(\{\hQ_{Qq},\hQ_{qQ}^{\dagger}\}\)
			\tabularnewline
			
			\midrule
			
			\centering
			\(\Sigma^2(\Phi^\dagger)^2\)
			\qquad
			\((\#\,2)\)
			\tabularnewline
			
			\midrule
			
			\(
			(\Phi^\dagger \hQ_{qQ}^{\dagger}\delmu\Sigma)
			(\Phi^\dagger \hQ_{Qq}\delmu\Sigma)
			\)
			\\[1.2ex]
			
			\(
			(\Phi^\dagger T_R^B\tau_L^I\hQ_{qQ}^{\dagger}T_L^A\delmu\Sigma)
			(\Phi^\dagger T_R^B\tau_L^I\hQ_{Qq}T_L^A\delmu\Sigma)
			\)
			\\[1.5ex]
			
			\midrule
			
			\centering
			\(\Phi_c^2(\Sigma_H^\dagger)^2\)
			\qquad
			\((\#\,2)\)
			\tabularnewline
			
			\midrule
			
			\(
			(\Sigma_H^\dagger \hQ_{qQ}^{\dagger}\delmu\Phi_c)
			(\Sigma_H^\dagger \hQ_{Qq}\delmu\Phi_c)
			\)
			\\[1.2ex]
			
			\(
			(\Sigma_H^\dagger \tau_R^J\tau_L^I\hQ_{qQ}^{\dagger}T_L^A\delmu\Phi_c)
			(\Sigma_H^\dagger \tau_R^J\tau_L^I\hQ_{Qq}T_L^A\delmu\Phi_c)
			\)
			\\
			
			\bottomrule
		\end{tabular}
	\end{minipage}
	
	\caption{Explicit operator forms for the \(\{\hQ_{qQ},\hQ_{Qq}^{\dagger}\}\) class and its hermitian conjugate.}
\end{table}

\section{Explicit form of the two-body decay amplitudes}\label{app:explcit_amp}
\subsection{$B\to P_1 P_2$ all operator expansion}\label{app:BPP}
%%%%%%%%%%%%%%%%%%%%%%%%%

\begin{table}
\centering
\rotatebox{90}{%
\renewcommand{\arraystretch}{1}
\resizebox{.9\textheight}{!}{$
\begin{array}{l|c|c|c|c|c|c}
\toprule
 & & \multicolumn{2}{c|}{\alpha=c} & \multicolumn{3}{c}{\alpha=u} \\
\text{Mode} & \text{CKM} & O_{1b} & O_{2b}\,(\times\lambda_b) & O_{4a} & O_{4c}\,(\times i) & O_{4d} \\
\midrule
{\cal A}(B^-\!\to\pi^-\pi^0) & V_{d\bar\alpha}V_{\bar b\alpha} &
 0 & \dfrac{2\sqrt2}{3}\,p_B(p_{\pi^-}-p_{\pi^0}) &
 -\dfrac{2\sqrt2}{3}\,p_{\pi^0}(p_B+p_{\pi^-}) & -\dfrac{2\sqrt2}{3}\,p_{\pi^0}(p_B+p_{\pi^-}) & -\dfrac{2\sqrt2}{9}\,p_{\pi^0}(p_B+p_{\pi^-}) \\
{\cal A}(\bar B_d^0\!\to\pi^+\pi^-) & V_{d\bar\alpha}V_{\bar b\alpha} &
 -4\big(2p_{\pi^-}p_{\pi^+}+p_B(p_{\pi^-}+p_{\pi^+})\big) & -\dfrac{4}{3}\big(2p_Bp_{\pi^-}+(3p_B+5p_{\pi^-})p_{\pi^+}\big) &
 -\dfrac{4}{3}\,p_{\pi^-}(p_B+p_{\pi^+}) & \dfrac{4}{3}\,(p_B+p_{\pi^-})p_{\pi^+} & -\dfrac{4}{9}\big(4p_Bp_{\pi^-}+(3p_B+7p_{\pi^-})p_{\pi^+}\big) \\
{\cal A}(\bar B_d^0\!\to\pi^0\pi^0) & V_{d\bar\alpha}V_{\bar b\alpha} &
 -4\sqrt2\,p_{\pi^0}(p_B+p_{\pi^0}) & -\dfrac{10\sqrt2}{3}\,p_{\pi^0}(p_B+p_{\pi^0}) &
 0 & \dfrac{4\sqrt2}{3}\,p_{\pi^0}(p_B+p_{\pi^0}) & -\dfrac{4\sqrt2}{3}\,p_{\pi^0}(p_B+p_{\pi^0}) \\
{\cal A}(\bar B_d^0\!\to K^+K^-) & V_{d\bar\alpha}V_{\bar b\alpha} &
 0 & -4\big(2p_{K^-}p_{K^+}+p_B(p_{K^-}+p_{K^+})\big) &
 -4\,p_{K^-}(p_B+p_{K^+}) & \dfrac{8}{3}\big(2p_Bp_{K^-}+(p_B+3p_{K^-})p_{K^+}\big) & -\dfrac{16}{3}\,(p_B+p_{K^-})p_{K^+} \\
{\cal A}(\bar B_d^0\!\to K^0\bar K^0) & V_{d\bar\alpha}V_{\bar b\alpha} &
 -4\big(2p_{K^0}p_{\bar K^0}+(p_{K^0}+p_{\bar K^0})p_B\big) & -\dfrac{4}{3}\big(3p_{\bar K^0}p_B+p_{K^0}(3p_{\bar K^0}+2p_B)\big) &
 0 & \dfrac{8}{3}\big(2p_{K^0}p_{\bar K^0}+(p_{K^0}+p_{\bar K^0})p_B\big) & 0 \\
 {\cal A}(\bar B_s^0\!\to K^+\pi^-) & V_{d\bar\alpha}V_{\bar b\alpha} &
 -4\big(2p_{\pi^-}p_{K^+}+p_B(p_{\pi^-}+p_{K^+})\big) & \dfrac{4}{3}\,p_{\pi^-}(p_B+p_{K^+}) &
 \dfrac{8}{3}\,p_{\pi^-}(p_B+p_{K^+}) & -\dfrac{4}{3}\big(4p_Bp_{\pi^-}+(p_B+5p_{\pi^-})p_{K^+}\big) & \dfrac{4}{9}\big(-4p_Bp_{\pi^-}+(9p_B+5p_{\pi^-})p_{K^+}\big) \\
\bottomrule
\end{array}
$}
}
\caption{Explicit amplitude expressions for $B\to P_1 P_2$ modes.}
\label{tab:BPPamps24mom}
\end{table}
\clearpage

\subsection{$B\to D P$ all operator expansion}
%%%%%%%%%%%%%%%%%%%%%%%%%

\begin{table}[H]
\centering
\renewcommand{\arraystretch}{1.35}
\setlength{\tabcolsep}{4pt}
\resizebox{\textwidth}{!}{
\begin{tabular}{c c c c}
\toprule
\textbf{Process} & $O_{5a}$ & $O_{5b}$ & $O_{6a}$ \\
\midrule

\({\cal A}\!\left(B_u^- \to D_u\,\pi^-\right)\)
&
\(
8i\,V_{d\bar u}V_{\bar b c}
\left[
p_{D_u}\!\cdot p_{\pi^-}
-
p_{B_u}\!\cdot p_{\pi^-}
\right]
\)
&
\(
\frac{8i}{3}\,V_{d\bar u}V_{\bar b c}
\left[
p_{D_u}\!\cdot p_{\pi^-}
-
p_{B_u}\!\cdot p_{\pi^-}
\right]
\)
&
\(0\)
\\[1.2em]

\({\cal A}\!\left(\bar B_d^0 \to D_d^+\,\pi^-\right)\)
&
\(
8i\,V_{d\bar u}V_{\bar b c}
\left[
p_{D_d^+}\!\cdot p_{\pi^-}
-
p_{B_d}\!\cdot p_{\pi^-}
\right]
\)
&
\(
\frac{8i}{3}\,V_{d\bar u}V_{\bar b c}
\left[
-
p_{B_d}\!\cdot q_{\pi^-}
+
p_{D_d^+}\!\cdot q_{\pi^-}
\right]
\)
&
\(
4i\,V_{d\bar u}V_{\bar b c}\frac{\lambda_b}{f_{\pi}}
\left[
p_{D_d^+}\!\cdot q_{\pi^-}
-
p_{B_d}\!\cdot p_{\pi^-}
\right]
\)
\\[1.2em]

\({\cal A}\!\left(\bar B_d^0 \to D_u\,\pi^0\right)\)
&
\(0\)
&
\(0\)
&
\(
2\sqrt{2}\,i\,V_{d\bar u}V_{\bar b c}\frac{\lambda_b}{f_{\pi}}
\left[
p_{D_u}\!\cdot q_{\pi^0}
-
p_{B_d}\!\cdot p_{\pi^0}
\right]
\)
\\[1.2em]

\({\cal A}\!\left(\bar B_s^0 \to D_s^+\,\pi^-\right)\)
&
\(
8i\,V_{d\bar u}V_{\bar b c}
\left[
p_{D_s^+}\!\cdot p_{\pi^-}
-
p_{B_s}\!\cdot p_{\pi^-}
\right]
\)
&
\(
\frac{16i}{3}\,V_{d\bar u}V_{\bar b c}
\left[
p_{B_s}\!\cdot p_{\pi^-}
-
p_{D_s^+}\!\cdot p_{\pi^-}
\right]
\)
&
\(0\)
\\[1.2em]

\({\cal A}\!\left(\bar B_s^0 \to D_u\,\pi^0\right)\)
&
\(0\)
&
\(
4i\sqrt{2}\,V_{\bar b c}V_{s\bar u}
\left[
-p_{B_s}\!\cdot q_{\pi^0}
+
p_{D_u}\!\cdot q_{\pi^0}
\right]
\)
&
\(
2\sqrt{2}\,i\,V_{\bar b c}V_{s\bar u}\frac{\lambda_b}{f_{\pi}}
\left[
p_{D_u}\!\cdot q_{\pi^0}
-
p_{B_s}\!\cdot p_{\pi^0}
\right]
\)
\\[1.2em]

\({\cal A}\!\left(\bar B_s^0 \to D_d^+\,\pi^-\right)\)
&
\(0\)
&
\(
8i\,V_{\bar b c}V_{s\bar u}
\left[
-p_{B_s}\!\cdot q_{\pi^-}
+
p_{D_d^+}\!\cdot q_{\pi^-}
\right]
\)
&
\(
4i\,V_{\bar b c}V_{s\bar u}\frac{\lambda_b}{f_{\pi}}
\left[
p_{D_d^+}\!\cdot q_{\pi^-}
-
p_{B_s}\!\cdot p_{\pi^-}
\right]
\)
\\
\bottomrule
\end{tabular}}
\caption{Explicit amplitude expressions for \(B \to DP\) modes mediated via $b\to c \bar u q$}
\label{tab:BtoDP_op5a_op5b_op6a_combined}
\end{table}
%%%%%%%%%%%%%%%%%%%%%%%%%%%%%%%%%%%%%%%%%%%%%%%%%%%%%%%

%%%%%%%%%%%%%%%%%%%%%%%%%%%%%%%%%%%%%%%%%%%%%%%%%%%%%%
\begin{table}
	\centering
	\begin{minipage}[c]{0.8\textwidth}
		\rotatebox{90}{%
			\renewcommand{\arraystretch}{2}
			\setlength{\tabcolsep}{3pt}
			\resizebox{\textheight}{!}{
				\begin{tabular}{l|c|llllll}
					\toprule
					\textbf{Amplitude}
					& \textbf{CKM}
					& \(\boldsymbol{O_{7a}}\)
					& \(\boldsymbol{O_{7b}}\)
					& \(\boldsymbol{O_{8a}}\)
					& \(\boldsymbol{O_{8b}}\)
					& \(\boldsymbol{O_{8c}}\)
					& \(\boldsymbol{O_{8d}}\) \\
					\midrule
					
					\(\cA\!\left(B^- \to \bar D_u\pi^-\right)\)
					& \(V_{\bar b u}V_{d\bar c}\)
					& \(i\!\left[-\frac{16}{9}p_Bp_D+\frac{56}{9}p_Bp_\pi-\frac{40}{9}p_Dp_\pi\right]\)
					& \(i\!\left[-p_Bp_D+p_Bp_\pi\right]\)
					& \(0\)
					& \(0\)
					& \(-\dfrac{2i\lambda_b}{f_\pi}\!\left[p_Bp_\pi-p_Dp_\pi\right]\)
					& \(i\!\left[\dfrac{2\lambda_b}{f_\pi}(p_Bp_\pi-p_Dp_\pi)+4\dfrac{\lambda_c}{f_\pi} p_Bp_\pi\right]\)
					\\
					
					\(\cA\!\left(B^- \to \bar D_u\pi^0\right)\)
					& \(V_{\bar b u}V_{d\bar c}\)
					& \(i\!\left[-\frac{16\sqrt2}{9}p_Bp_\pi+\frac{16\sqrt2}{9}p_Dp_\pi\right]\)
					& \(\dfrac{i}{\sqrt2}\!\left[-p_Bp_\pi+p_Dp_\pi\right]\)
					& \(0\)
					& \(i\!\left[\sqrt2\dfrac{\lambda_c}{f_\pi} p_Bp_\pi\right]\)
					& \(\dfrac{i\sqrt2\lambda_b}{f_\pi}\!\left[p_Bp_\pi-p_Dp_\pi\right]\)
					& \(i\!\left[-\dfrac{\sqrt2\lambda_b}{f_\pi}(p_Bp_\pi-p_Dp_\pi)-\dfrac{2\sqrt2}{3}\dfrac{\lambda_c}{f_\pi} p_Bp_\pi\right]\)
					\\
					
					\(\cA\!\left(B^- \to D_s^-\pi^0\right)\)
					& \(V_{\bar b u}V_{s\bar c}\)
					& \(i\!\left[\frac{44\sqrt2}{9}p_Bp_D-\frac{10\sqrt2}{3}p_Bp_\pi-\frac{14\sqrt2}{9}p_Dp_\pi\right]\)
					& \(\dfrac{i}{\sqrt2}\!\left[-p_Bp_D+p_Dp_\pi\right]\)
					& \(0\)
					& \(i\!\left[\sqrt2\dfrac{\lambda_c}{f_\pi} p_Bp_\pi\right]\)
					& \(0\)
					& \(-\dfrac{2i\sqrt2}{3}\dfrac{\lambda_c}{f_\pi} p_Bp_\pi\)
					\\
\bottomrule
		\end{tabular}}}
	\end{minipage}\hfill
	% --- upright caption on the right ---
	\begin{minipage}[c]{0.18\textwidth}
		\caption{Explicit amplitude expressions for $B\to D P$ modes mediated via $b\to u \bar c q$.}
		\label{tab:BDPexplicit}
	\end{minipage}
\end{table}
\clearpage

%%%%%%%%%%%%%%%%%%%
\bibliographystyle{JHEP}
\bibliography{HeavyLightMesons}
%%%%%%%%%%%%%%%%%%%

\end{document}